\documentclass[fleqn,usenatbib]{mnras}
\usepackage{array, multirow, bigdelim, makecell, booktabs} 
\usepackage{newtxtext,newtxmath}
\usepackage[T1]{fontenc}
\usepackage{ae,aecompl}
\usepackage{graphicx}	
\usepackage{amsmath}	
\usepackage{topcapt}
\usepackage{supertabular}
\usepackage{adjustbox}
\usepackage{hyperref}
\usepackage{stfloats}
\renewcommand{\floatpagefraction}{0.9}

\hypersetup{
    colorlinks=true,
    linkcolor=blue,
    filecolor=magenta,      
    urlcolor=cyan,
}

\title[X-ray luminous clusters at $z>0.5$]{Beyond MACS: Physical properties of extremely X-ray luminous clusters at \emph{z} > 0.5}
\author[Ebeling et al.]{
H.\ Ebeling$^1$, J.\ Richard$^2$, B.\ Beauchesne$^3$, Q.\ Basto$^2$, A.C.\ Edge$^4$\thanks{E-mail: alastair.edge@durham.ac.uk} \& I.\ Smail$^4$
\\ \\
$^1$ Institute for Astronomy, University of Hawaii, 640 N.\ Aohoku Place, Hilo, HI 96720, USA\\
$^2$ Univ Lyon, Univ Lyon1, Ens de Lyon, CNRS, Centre de Recherche Astrophysique de Lyon UMR5574, F-69230, Saint-Genis-Laval, France\\
$^3$ Institute of Physics, Laboratory of Astrophysics, École Polytechnique Fédérale de Lausanne, Observatoire de Sauverny, CH-1290 Versoix, Switzerland\\
$^4$ Centre for Extragalactic Astronomy, Department of Physics, Durham University, South Road, Durham DH1 3LE, UK
}
\date{Accepted ---. Received ---; in original form ---}
\pubyear{2024}
\begin{document}
\renewcommand{\floatpagefraction}{1}

\label{firstpage}
\pagerange{\pageref{firstpage}--\pageref{lastpage}}
\maketitle

\begin{abstract}
We present a sample of over 100 highly X-ray luminous galaxy clusters at $z\approx0.5$--$0.9$, discovered by the extended Massive Cluster Survey (eMACS) in {\it ROSAT} All-Sky Survey (RASS) data. Follow-up observations of a subset at higher resolution and greater depth with the {\it Chandra X-ray Observatory} are used to map the gaseous intra-cluster medium, while strong-gravitational-lensing features identified in {\it Hubble Space Telescope} imaging allow us to constrain the total mass distribution. We present evidence of the exceptional gravitational-lensing power of these massive systems, search for substructure along the line of sight by mapping the radial velocities of cluster members obtained through extensive ground-based spectroscopy, and identify dramatic cases of galaxy evolution in high-density cluster environments. The available observations of the eMACS sample presented here provide a wealth of insights into the properties of very massive clusters ($\gtrsim 10^{15}$ M$_\odot$) at $z>0.5$, which act as powerful lenses to study galaxies in the very distant Universe. 
We also discuss the evolutionary state, galaxy population, and large-scale environment of eMACS clusters and release to the community all data and science products to further the understanding of the first generation of truly massive clusters to have formed in the Universe. 
\end{abstract}


\section{Introduction}
\label{sec:intro}
Occupying the top spot in the mass-ranked hierarchy of gravitationally collapsed structures in the Universe, massive galaxy clusters play a unique role in numerous areas of astrophysical and cosmological research: they allow us to study the evolution and interaction of baryonic and non-baryonic matter across a wide range of environments and redshifts \citep[e.g.,][]{Vikhlinin2001,Markevitch2002,Harvey2015,Ebeling2017,vanWeeren2019}, act as powerful gravitational telescopes that magnify faint background sources out to the very highest redshifts \citep[e.g.,][and references therein]{Kneib2011}, and provide independent constraints on fundamental cosmological parameters \citep[e.g.,][]{Vikhlinin2009,Allen2011,Planck2014,Mantz2015,Bocquet2019}. Importantly, they also provide valuable opportunities to observe and understand the evolution of galaxies in high-density environments, from ram-pressure stripping of infalling galaxies \citep[e.g.,][]{1972ApJ...176....1G,2001ApJ...561..708V,2007MNRAS.376..157C,ebeling2014,2022A&ARv..30....3B} to the mechanisms driving the growth of central cluster galaxies \citep[e.g.,][]{1976ApJ...209..693O,1985ApJ...289...18M,1998ApJ...502..141D,2007MNRAS.375....2D,2016ApJ...817...86M}.

The ability to quantify changes in the ensemble properties of the population of massive clusters with lookback time is crucial for many of the aforementioned science applications. However, attempts to constrain the physical properties of massive clusters, primarily their mass and relaxation state, at ever-increasing redshifts face the obvious challenge that the most massive clusters ($M\gtrsim 10^{15}$ M$_\odot$) have not had time to form in significant numbers beyond their typical redshift of formation, commonly thought to be $z\sim 1$. As a result, observational studies of such systems beyond the local Universe have, so far, been limited to extreme clusters at the median redshift of $z\sim 0.45$ probed by the X-ray-based\footnote{An X-ray luminosity of $10^{45}$ erg s$^{-1}$ (0.1--2.4 keV) corresponds to a cluster mass of $10^{15}$ M$_\odot$ \citep[e.g.][and references therein]{2020ApJ...892..102L}.} Massive Cluster Survey \citep[MACS;][]{Ebeling2001,Ebeling2007,Ebeling2010,Mann2012,Repp2018}. At significantly higher redshifts, studies frequently use the term ``massive cluster'' loosely, sampling systems of average mass ($\sim10^{14}$ M$_\odot$), such as those detected in surveys exploiting the Sunyaev-Zel'dovich \citep[SZ;][]{1980ARA&A..18..537S} effect \citep{2020ApJS..247...25B,2021ApJS..253....3H} or in all-sky infrared surveys \citep{2019ApJS..240...33G}, or targeting the precursors of today's massive clusters (in their infancy) at redshifts well beyond unity \citep[e.g.,][]{1998ApJ...492..428S,2010ApJ...716.1503P,2011A&A...526A.133G,2014ApJ...788...51N,2016ApJ...828...56W,2023ApJ...947L..24M}. 

The extended Massive Cluster Survey \citep[eMACS;][]{Ebeling2013}, described below, was designed to  push X-ray searches for very massive clusters to the highest possible redshifts without lowering the absolute mass threshold of $\sim10^{15}$ M$_\odot$ set at $z>0.3$ by MACS \citep{2014MNRAS.439...48A}. We here present the eMACS sample as it stands, including an overview of optical, radio, and X-ray follow-up observations of eMACS clusters performed to date. Based on the available data, we give first assessments of the physical properties (mass, evolutionary state, large-scale environment etc) of all eMACS clusters for which sufficient high-quality data, from both ground- and space-based follow-up observations, are in hand.

This paper is structured as follows. After this brief introduction, we provide an overview of the eMACS project in Section~\ref{sec:emacs}, summarize follow-up observations in Section~\ref{sec:obs}, and describe the methods used to analyze the reduced data in Section~\ref{sec:methods}. Section~\ref{sec:results} presents the results obtained for the physical properties of eMACS clusters, followed by our conclusions in Section~\ref{sec:wrapup}. More technical details about the compilation of the eMACS sample from X-ray and optical survey data can be found in Appendix \ref{sec:app-rass}, while Appendix~\ref{sec:app-img} serves as a repository for summaries of key observational findings for each cluster.

We assume the $\Lambda$CDM concordance cosmology, i.e., $\Omega_{m} = 0.3$, $\Omega_{\Lambda} = 0.7$, and $H_{0} = 100 h$ km s$^{-1}$ Mpc$^{-1}$, with $h = 0.7$.

\section{eMACS}
\label{sec:emacs}

The distant clusters presented and discussed here were all discovered (or, in a few cases, re-discovered) in the course of the extended Massive Cluster Survey, eMACS. An X-ray selected cluster survey based on {\it ROSAT} All-Sky Survey data \citep[RASS;][]{Voges1999} within a solid angle of over 20,000 deg$^2$, defined by $|b|>20^\circ$, $\delta>-30^\circ$, eMACS extends similar work at lower redshifts \citep[e.g.,][]{Ebeling1997,degrandi1999,Ebeling2000,Ebeling2001,Reiprich2001,Ebeling2002,Bohringer2004,Ebeling2007,Ebeling2010,Ebeling2013,Bohringer2013} to $z>0.5$ in an attempt to find the first generation of truly massive clusters out to redshifts approaching unity.

\begin{figure*}
    \centering
    \includegraphics[width=0.98\textwidth]{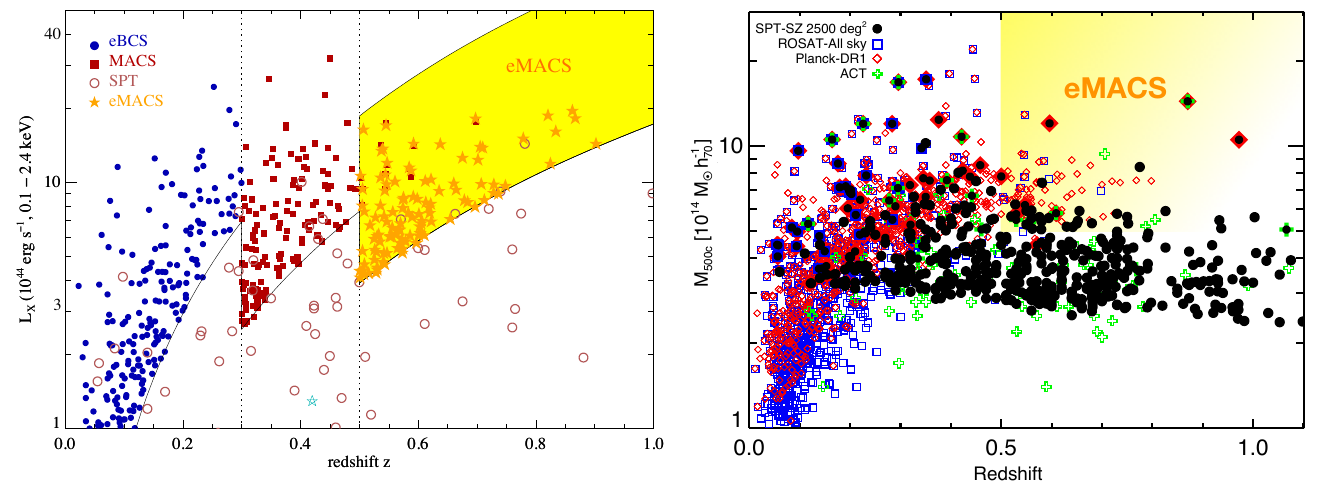}
    \caption{Left: Distribution of X-ray luminosity vs redshift for cluster samples compiled from RASS data, as well as for SZ-selected samples. Right: Mass estimates versus redshift for galaxy clusters from SZ and and X-ray selected samples \citep[adapted from][]{2015ApJS..216...27B}. eMACS complements existing surveys by focusing on the high-mass, high-redshift regime. Although total masses are not available for all clusters shown in the left-hand plot, the equivalence of an X-ray luminosity of $10^{45}$ erg s$^{-1}$ (0.1--2.4 keV) and a total mass of $10^{15}$ M$_\odot$ is well established for galaxy clusters \citep[e.g.,][]{2014MNRAS.439...48A}. } 
    \label{fig:lxmz}
\end{figure*}

Fig.~\ref{fig:lxmz} shows the X-ray luminosity, mass, and redshift regime explored by eMACS in comparison with other cluster surveys. The RASS flux limits of the eBCS \citep{Ebeling2000}, MACS, and eMACS surveys are $3\times 10^{-12}$, $10^{-12}$, and $0.5\times 10^{-12}$ erg cm$^{-2}$ s$^{-1}$ (0.1--2.4 keV), respectively. By design, eMACS finds clusters that are as X-ray luminous as the most extreme systems discovered in the eBCS and MACS surveys, but at yet higher redshift. Solid angle is crucial in this context; while the South Pole Telescope \citep[SPT;][]{2011PASP..123..568C}, a facility designed for the detection of the SZ effect in clusters, is capable of  detecting clusters with constant efficiency at all redshifts, the much smaller solid angle covered by the SPT surveys prevents it from finding exceptionally massive clusters in significant numbers. The commonalities and differences between SZ- and X-ray cluster surveys in mass-redshift space are summarized in Fig.~\ref{fig:lxmz} which also shows the cluster samples compiled by the SPT, the Atacama Cosmology Telescope \citep[ACT;][]{2011ApJS..194...41S}, and the Planck satellite. Although Planck covered the entire sky, the mission’s limited angular resolution results in a selection that is biased against clusters even at modest redshift \citep{2007MNRAS.377..253M,2015ApJS..216...27B}, creating a strongly redshift-dependent selection function more similar to that of RASS-based cluster samples at moderate redshifts than those of high-resolution SZ surveys such as those conducted with the SPT and ACT. As illustrated in Fig.~\ref{fig:lxmz}, eMACS does not compete with, but instead complements, SZ cluster surveys.

In a nutshell, the eMACS sample was compiled as follows. Candidates were selected from the RASS source catalogue\footnote{Note that eMACS, unlike MACS, draws not only from the RASS Bright Source Catalogue but also from the Faint Source Catalogue.} based on very few, simple X-ray selection criteria (beyond the mentioned cuts in Galactic and equatorial latitude, cuts in the spectral X-ray hardness ratio and in the RASS flux in the 0.1--2.4 keV band were applied; see Appendix~\ref{sec:app-rass} for details). Next, visual inspection of optical images covering a $5'\times 5'$ area around each source in the publicly available data obtained by the PanSTARRS (PS) facility \citep{Kaiser2002}, specifically the PS1 3$\pi$ survey \citep{Chambers2016}, was used to identify plausible clusters as either overdensities of faint galaxies of similar colour or fields that appear blank to the depth of the PS 3$\pi$ data. We refer the reader to Appendix~\ref{sec:app-rass} for more details. As a final step, spectroscopic follow-up observations of this tentative cluster sample were conducted to limit the sample to clusters at a redshift of $z>0.5$. Results from an eMACS pilot project were presented in \citet{Ebeling2013}, and an assessment of the lensing power of eMACS clusters is provided by \citet{Zalesky2020}.

References to the ``current'' eMACS sample are motivated by the sample's incompleteness, both with regard to the optical screening of all cluster candidates identified in the X-ray selection process described in Appendix~\ref{sec:app-rass} and with regard to the spectroscopic confirmation of these candidates. However, in light of the progress made in the compilation of cluster samples from the RASS' successor, the next-generation X-ray all-sky survey {\it eROSITA} \citep[e.g.,][]{2023MNRAS.522.1601C}, we opt for releasing the current eMACS sample (together with all supporting follow-up data) to prevent unnecessary investments of observing time into the confirmation and characterization of {\it eROSITA} detections that merely duplicate prior eMACS discoveries and results.

\begin{figure}
    \centering
    \includegraphics[width=0.47\textwidth]{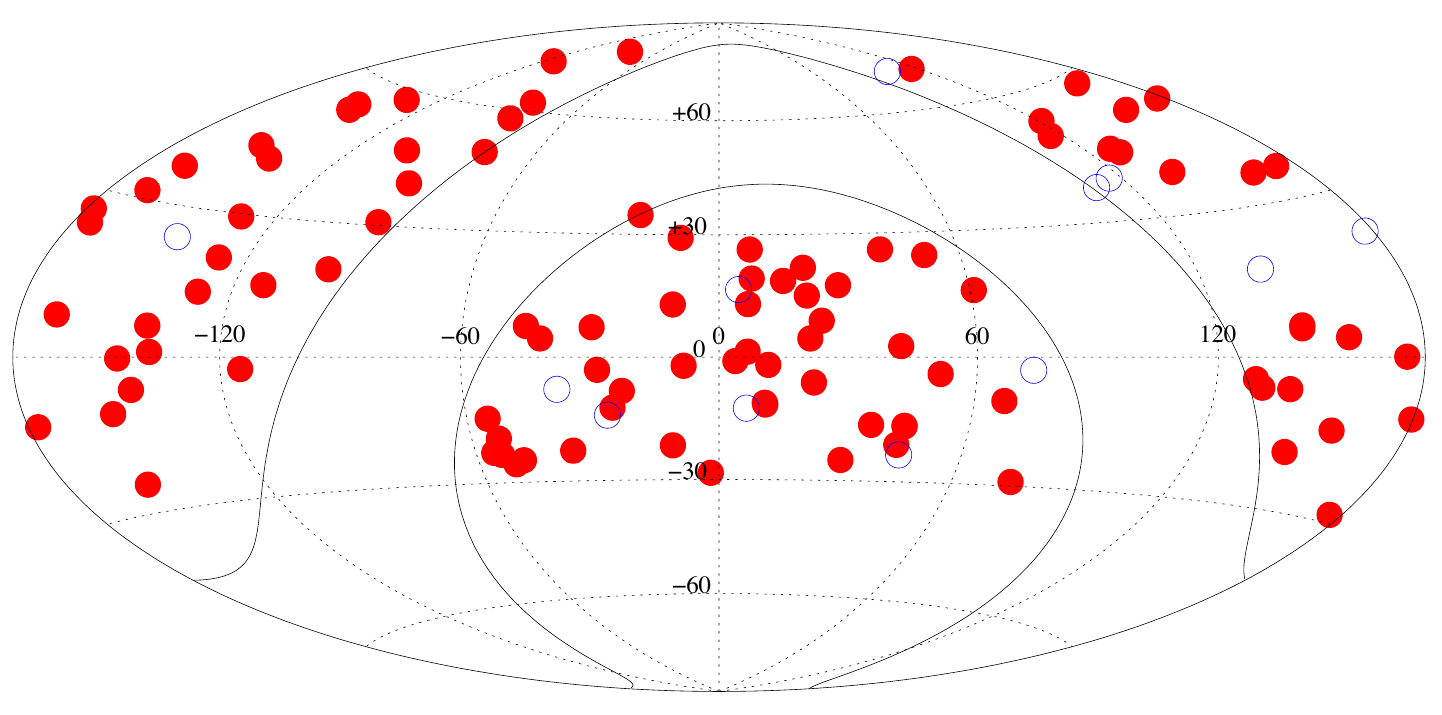}
    \caption{The distribution of eMACS clusters in the equatorial sky. The 12 clusters also included in the MACS sample and presented previously in \citet{Ebeling2007} are shown as open symbols. Solid lines delineate the 40-deg wide exclusion zone around the Galactic plane. While the statistical completeness of eMACS has not been quantified, the distribution of clusters across the surveyed solid angle is approximately isotropic.
    } 
    \label{fig:skymap}
\end{figure}

Table~\ref{tab:sample} lists all 111 entries in the current eMACS sample and provides an overview of selected follow-up observations performed to date (see Sections~\ref{sec:obs-hst} and \ref{sec:obs-cxo} for details). Fig.~\ref{fig:skymap} shows the location of all clusters on the sky and demonstrates that the incompleteness of the sample presented here is not a strong function of celestial position, either in Galactic or equatorial coordinates.
\begin{table*}
\begin{scriptsize}
  \centering
  \begin{tabular}{lcrrcl@{\hspace{1mm}}l@{\hspace{1mm}}l@{\hspace{1mm}}l@{\hspace{1mm}}l@{\hspace{1mm}}l@{\hspace{1mm}}l@{\hspace{1mm}}l}
name & z & $n_z$ & \multicolumn{1}{c}{t$_{\rm ACIS-I}$} & \multicolumn{1}{c}{\rm JVLA} &\multicolumn{8}{c}{HST passbands}\\
     &   &       & \multicolumn{1}{c}{(ks)}  & observed & \multicolumn{8}{c}{}\\[1mm] \hline\\[-1mm]
eMACSJ0014.9$-$0056 & 0.5330 &  18 &  30 & Y &        & F606W &       &       & F110W &       & F140W &       \\                                   
\hspace*{0.9mm}MACSJ0018.5$+$1626 & 0.5473 &  -   &  67 & Y &  F435W & F606W & F814W & F105W &       & F125W & F140W & F160W \\ 
\hspace*{0.9mm}MACSJ0025.4$-$1222 & 0.5848 &  -   & 158 & Y &  F435W &       & F814W & F105W &       & F125W & F140W & F160W \\ 
eMACSJ0026.2$+$0120 & 0.6357 &  22 &     & Y &        & F606W & F814W &       & F110W &       & F140W &       \\                                   
eMACSJ0027.0$+$1252 & 0.6063 &  22 &     & - &        &       &       &       & F110W &       & F140W &       \\                                   
eMACSJ0030.5$+$2618 & 0.4970 &  20 &  18 & Y &        & F606W & F814W &       & F110W &       & F140W &       \\                                   
eMACSJ0031.2$+$1907 & 0.5304 &  20 &     & - &        & F606W &       &       & F110W &       & F140W &       \\                                   
eMACSJ0042.5$-$1103 & 0.5701 &  13 &  12 & - &        &       & F814W &       & F110W &       & F140W &       \\                                   
eMACSJ0043.1$-$1129 & 0.7282 &  19 &     & - &        & F606W & F814W &       & F110W &       & F140W &       \\                                   
eMACSJ0045.2$-$0151 & 0.5471 &  15 &  49 & Y &        &       &       &       & F110W &       & F140W &       \\                                   
eMACSJ0101.1$+$1831 & 0.6533 &  23 &     & Y &        & F606W & F814W &       & F110W &       & F140W &       \\                                   
eMACSJ0121.4$+$2143 & 0.6011 &  17 &     & - &        & F606W & F814W &       & F110W &       & F140W &       \\                                   
eMACSJ0122.7$+$1454 & 0.5469 &  15 &     & - &        &       &       &       & F110W &       & F140W &       \\                                   
eMACSJ0124.0$+$0430 & 0.5363 &  14 &     & - &        & F606W &       &       &       &       &       &       \\                                   
eMACSJ0127.5$-$0606 & 0.5055 &  14 &     & - &        &       &       &       & F110W &       & F140W &       \\                                   
eMACSJ0135.2$+$0847 & 0.6185 &  38 &  12 & Y &        &       &       &       & F110W &       & F140W &       \\                                   
eMACSJ0153.4$+$1722 & 0.5464 &  21 &     & - &        & F606W &       &       &       &       &       &       \\                                   
eMACSJ0200.3$-$2454 & 0.7126 &  25 &     & - &       &       &       &       &       &       &       &       \\ 
eMACSJ0224.5$-$1615 & 0.6223 &  15 &     & - &        &       &       &       & F110W &       & F140W &       \\                                   
eMACSJ0241.0$+$2557 & 0.5752 &  22 &     & - &        & F606W &       &       & F110W &       & F140W &       \\                                   
eMACSJ0248.2$+$0237 & 0.5561 &  19 &  45 & - &        &       &       &       & F110W &       & F140W &       \\                                   
eMACSJ0252.4$-$2100 & 0.7017 &  26 &     & - &        &       &       &       & F110W &       &       &       \\ 
eMACSJ0256.7$-$1623 & 0.8621 &  18 &  33 & Y &        & F606W & F814W &       & F110W &       & F140W &       \\                                   
eMACSJ0256.9$-$1631 & 0.8670 &  17 &  33 & Y &        & F606W & F814W &       & F110W &       & F140W &       \\                                   
\hspace*{0.9mm}MACSJ0257.1$-$2325 & 0.5056 &   -  &  38 & Y &  F435W & F606W & F814W & F105W &       & F125W & F140W & F160W \\ 
eMACSJ0324.0$+$2421 & 0.9023 &  26 &  31 & - &  F435W & F606W & F814W & F105W &       & F125W & F140W & F160W \\                                   
eMACSJ0325.4$-$0359 & 0.5721 &  12 &     & - &        &       &       &       & F110W &       & F140W &       \\                                   
eMACSJ0403.6$+$1544 & 0.5275 &  20 &     & - &        & F606W & F814W &       & F110W &       & F140W &       \\                                   
eMACSJ0429.0$-$1011 & 0.5418 &  10 &     & - &        &       &       &       & F110W &       & F140W &       \\                                   
\hspace*{0.9mm}MACSJ0454.1$-$0300 & 0.5376 &   -  &  14 & Y &        &       & F814W & F105W & F110W & F125W & F140W & F160W \\ 
eMACSJ0502.9$-$2902 & 0.6028 &  23 &  49 & - &        & F606W & F814W &       & F110W &       & F140W &       \\                                   
\hspace*{0.9mm}MACSJ0647.7$+$7015 & 0.5922 &  -   &  39 & Y &  F435W & F606W & F814W & F105W & F110W & F125W & F140W & F160W \\ 
\hspace*{0.9mm}MACSJ0717.5$+$3745 & 0.5454 &  -   & 243 & Y &  F435W & F606W & F814W & F105W & F110W & F125W & F140W & F160W \\ 
eMACSJ0742.9$+$5001 & 0.6013 &  15 &     & - &        & F606W & F814W &       & F110W &       & F140W &       \\                                   
\hspace*{0.9mm}MACSJ0744.8$+$3927 & 0.6976 &  -   &  90 & Y &  F435W & F606W & F814W & F105W & F110W & F125W & F140W & F160W \\ 
eMACSJ0747.0$+$6937 & 0.5696 &  13 &     & - &        & F606W & F814W &       & F110W &       & F140W &       \\                                   
eMACSJ0804.6$+$5325 & 0.5786 &  20 &   4 & - &        & F606W & F814W &       & F110W &       & F140W &       \\                                   
eMACSJ0834.2$+$4524 & 0.6606 &  32 &  33 & Y &        & F606W & F814W &       & F110W &       & F140W &       \\                                   
eMACSJ0840.2$+$4421 & 0.6377 &  39 &  32 & Y &        & F606W & F814W &       & F110W &       & F140W &       \\                                   
eMACSJ0841.8$-$0429 & 0.5359 &  11 &     & - &        & F606W & F814W &       & F110W &       & F140W &       \\                                   
eMACSJ0850.2$-$0611 & 0.5744 &  22 &     & - &        & F606W & F814W &       & F110W &       & F140W &       \\ 
eMACSJ0910.8$+$3850 & 0.5618 &  19 &     & - &        & F606W & F814W &       & F110W &       & F140W &       \\                                   
\hspace*{0.9mm}MACSJ0911.2$+$1746 & 0.5051 &  -   &  42 & Y &  F435W &       & F814W & F105W &       & F125W & F140W & F160W \\ 
eMACSJ0921.6$-$0621 & 0.6842 &  15 &     & Y &        & F606W & F814W &       & F110W &       & F140W &       \\                                   
eMACSJ0934.6$+$0540 & 0.5627 &  15 &     & - &        & F606W &       &       & F110W &       & F140W &       \\                                   
eMACSJ0935.1$+$0614 & 0.7787 &  21 &  30 & Y &        & F606W & F814W &       & F110W &       & F140W &       \\                                   
eMACSJ0943.3$-$1842 & 0.5687 &  12 &     & Y &        & F606W & F814W &       & F110W &       & F140W &       \\                      
eMACSJ1025.0$-$1354 & 0.6196 &  13 &     & - &        & F606W & F814W &       & F110W &       & F140W &       \\                                   
eMACSJ1027.2$+$0345 & 0.7482 &  10 &     & Y &        & F606W & F814W &       & F110W &       & F140W &       \\                                   
\end{tabular}
   \caption{The eMACS sample at the time of publication; we list the cluster redshift, the number of spectroscopic redshifts measured, and the HST passbands in which imaging data were obtained. We also indicate which clusters were observed with the JVLA at 5\,GHz to date. The 12 eMACS clusters released earlier as part of the MACS sample \citep{Ebeling2007} are listed for completeness' sake ($n_z$ entries are deliberately omitted for these 12 clusters).}
\label{tab:sample}
\end{scriptsize}
\end{table*}

\setcounter{table}{0}
\begin{table*}
\begin{scriptsize}
  \centering
  \begin{tabular}{lcrrcl@{\hspace{1mm}}l@{\hspace{1mm}}l@{\hspace{1mm}}l@{\hspace{1mm}}l@{\hspace{1mm}}l@{\hspace{1mm}}l@{\hspace{1mm}}l}
name & z & $n_z$ & \multicolumn{1}{c}{t$_{\rm ACIS-I}$} & \multicolumn{1}{c}{\rm JVLA} &\multicolumn{8}{c}{HST passbands}\\
     &   &       & \multicolumn{1}{c}{(ks)}  & observed & \multicolumn{8}{c}{}\\[1mm] \hline\\[-1mm]                                            
eMACSJ1030.5$+$5132 & 0.5182 &  23 &  20 & Y &        & F606W & F814W &       & F110W &       & F140W &       \\                                   
eMACSJ1050.6$+$3548 & 0.5077 &  39 &     & Y &        & F606W & F814W &       &       &       &       &       \\                                   
eMACSJ1057.5$+$5759 & 0.6027 &  12 &     & - &        & F606W & F814W &       & F110W &       & F140W &       \\                                   
eMACSJ1131.1$+$3553 & 0.5159 &  12 &     & Y &        & F606W & F814W &       & F110W &       & F140W &       \\                                   
eMACSJ1136.8$+$0005 & 0.5968 &  26 &   4 & Y &        & F606W & F814W &       & F110W &       & F140W &       \\                                   
eMACSJ1144.2$-$2836 & 0.5070 &  30 &  31 & Y &        & F606W & F814W &       & F110W &       & F140W &       \\                                   
eMACSJ1148.0$+$5116 & 0.5834 &  28 &     & Y &        & F606W & F814W &       & F110W &       & F140W &       \\                                   
\hspace*{0.9mm}MACSJ1149.5$+$2223 & 0.5445 &  - & 366 & Y &  F435W & F606W & F814W & F105W & F110W & F125W & F140W & F160W \\ 
eMACSJ1157.9$-$1046 & 0.5570 &  25 & 156 & - &  F435W &       & F814W &       & F110W &       & F140W &       \\          eMACSJ1209.4$+$2640 & 0.5553 &  22 &  16 & Y &        &       &       & F105W & F110W &       & F140W & F160W \\ 
eMACSJ1212.5$-$1216 & 0.6786 &  28 &     & Y &        &       & F814W &       & F110W &       & F140W &       \\                                   
eMACSJ1222.3$+$2418 & 0.5085 &  14 &     & Y &        & F606W &       &       & F110W &       & F140W &       \\                                   
eMACSJ1248.2$+$0743 & 0.5733 &  15 &     & - &        & F606W & F814W &       & F110W &       & F140W &       \\                                   
eMACSJ1251.3$+$3131 & 0.5069 &  16 &     & - &        & F606W & F814W &       & F110W &       & F140W &       \\                                   
eMACSJ1257.3$+$3654 & 0.5253 &  14 &     & - &        & F606W &       &       & F110W &       & F140W &       \\    
eMACSJ1341.9$-$2442 & 0.8339 &  26 &  34 & Y &        & F606W & F814W & F105W & F110W & F125W & F140W & F160W \\                                   
eMACSJ1350.7$-$1055 & 0.8247 &  12 &  33 & Y &        &       &       &       &       &       &       &       \\ 
eMACSJ1353.7$+$4329 & 0.7364 &  31 &  40 & Y &  F435W & F606W & F814W & F105W &       & F125W & F140W & F160W \\   
eMACSJ1407.0$-$0015 & 0.5520 &  14 &   4 & Y &        & F606W & F814W &       &       &       &       &       \\                                   
eMACSJ1414.7$+$5446 & 0.6121 &  26 &  33 & - &        & F606W & F814W &       & F110W & F125W & F140W & F160W \\ 
eMACSJ1419.1$-$0624 & 0.5432 &  22 &     & Y &        &       & F814W &       & F110W &       & F140W &       \\                                   
eMACSJ1419.2$+$5326 & 0.6384 &  32 &     & - &        &       & F814W &       & F110W &       & F140W &       \\                                   
\hspace*{0.9mm}MACSJ1423.8$+$2404 & 0.5434 &  -   &  19 & Y &  F435W & F606W & F814W & F105W & F110W & F125W & F140W & F160W \\ 
eMACSJ1430.0$+$4127 & 0.6646 &  13 &     & Y &        & F606W & F814W &       & F110W &       & F140W &       \\                                   
eMACSJ1437.8$+$0616 & 0.5350 &  19 &     & - &        & F606W &       &       & F110W &       & F140W &       \\                                   
eMACSJ1443.2$+$0102 & 0.5284 &  23 &     & Y &        & F606W & F814W &       & F110W &       & F140W &       \\  
eMACSJ1508.1$+$5755 & 0.5421 &  48 &  49 & Y &        & F606W & F814W &       & F110W &       & F140W &       \\ eMACSJ1518.3$+$2927 & 0.6050 &  33 &     & Y &        & F606W & F814W &       & F110W &       & F140W &       \\                                   
eMACSJ1523.1$+$1329 & 0.5010 &  18 &     & - &        & F606W & F814W &       & F110W &       & F140W &       \\                                   
eMACSJ1527.6$+$2044 & 0.6967 &  28 &  25 & Y &  F435W & F606W & F814W & F105W & F110W & F125W & F140W & F160W \\                                   
eMACSJ1621.2$-$0231 & 0.5014 &   9 &     & Y &        & F606W & F814W &       & F110W &       & F140W &       \\                                   
eMACSJ1621.4$+$7232 & 0.5863 &  34 &     & Y &        & F606W & F814W &       & F110W &       & F140W &       \\                                   
eMACSJ1631.1$+$1528 & 0.5087 &  11 &     & Y &        & F606W & F814W &       & F110W &       & F140W &       \\                                   
eMACSJ1709.5$+$4731 & 0.5527 &  40 &     & Y &        & F606W & F814W &       & F110W &       & F140W &       \\                                   
eMACSJ1732.4$+$1934 & 0.5406 &  41 &     & Y &        & F606W & F814W &       & F110W &       & F140W &       \\                                   
eMACSJ1756.8$+$4008 & 0.5743 & 117 &  89 & Y &  F435W & F606W & F814W & F105W & F110W & F125W & F140W & F160W \\                                   
eMACSJ1757.5$+$3045 & 0.6105 &  32 &     & Y &        & F606W & F814W &       & F110W &       & F140W &       \\                                   
eMACSJ1823.1$+$7822 & 0.6754 &  20 &  22 & Y &  F435W & F606W & F814W & F105W & F110W & F125W & F140W & F160W \\                                   
eMACSJ1831.1$+$6214 & 0.8207 &  25 &  23 & Y &        & F606W & F814W &       & F110W &       & F140W &       \\                                   
eMACSJ1831.9$+$5746 & 0.7072 &  15 &     & Y &        & F606W & F814W &       & F110W &       & F140W &       \\                                   
eMACSJ1852.0$+$4900 & 0.6035 &  32 &  20 & Y &        & F606W & F814W &       & F110W &       & F140W &       \\                                   
eMACSJ2018.3$-$2242 & 0.5302 &  20 &     & Y &        & F606W & F814W &       & F110W &       & F140W &       \\                                   
eMACSJ2020.1$-$1432 & 0.5653 &  22 &     & - &        & F606W & F814W &       & F110W &       & F140W &       \\                                   
eMACSJ2025.5$-$2313 & 0.7098 &  19 &     & Y &        & F606W & F814W &       & F110W &       & F140W &       \\                                   
eMACSJ2026.7$-$1920 & 0.6219 &  40 &  32 & Y &        & F606W & F814W &       & F110W &       & F140W &       \\                                   
eMACSJ2037.1$-$2534 & 0.5263 &  36 &     & Y &        & F606W & F814W &       & F110W &       & F140W &       \\                                   
eMACSJ2045.8$-$2438 & 0.5917 &  20 &     & - &        &       & F814W &       & F110W &       & F140W &       \\                                   
eMACSJ2100.4$+$0724 & 0.5742 &  13 &     & - &        & F606W & F814W &       & F110W &       & F140W &       \\                                   
eMACSJ2114.4$+$0427 & 0.5898 &  20 &     & Y &        & F606W & F814W &       & F110W &       & F140W &       \\                                   
\hspace*{0.9mm}MACSJ2129.4$-$0741 & 0.5879 &  -   &  40 & Y &  F435W & F606W & F814W & F105W & F110W & F125W & F140W & F160W \\ 
eMACSJ2137.2$-$2232 & 0.6131 &  23 &     & Y &        & F606W & F814W &       & F110W &       & F140W &       \\                                   
eMACSJ2201.9$+$0711 & 0.5633 &  11 &     & - &        & F606W & F814W &       & F110W &       & F140W &       \\                                   
eMACSJ2207.5$-$0302 & 0.5304 &  21 &     & Y &        & F606W & F814W &       & F110W &       & F140W &       \\                                   
\hspace*{0.9mm}MACSJ2214.9$-$1359 & 0.5022 &  -   &  38 & Y &        &       & F814W & F105W &       & F125W &       & F160W \\ 
eMACSJ2220.3$-$1211 & 0.5292 &  35 &  12 & - &        & F606W & F814W &       & F110W &       & F140W &       \\                                   
eMACSJ2229.9$-$0808 & 0.6214 &  65 &     & Y &        & F606W & F814W &       & F110W &       & F140W &       \\                                   
eMACSJ2236.0$+$3451 & 0.7287 &  25 &     & - &        & F606W & F814W &       & F110W &       & F140W &       \\                                   
eMACSJ2315.3$-$2128 & 0.5400 &  43 &  31 & - &        & F606W & F814W &       & F110W &       & F140W &       \\                                   
eMACSJ2316.6$+$1246 & 0.5258 &  32 &     & Y &        & F606W & F814W &       & F110W &       & F140W &       \\                                   
eMACSJ2320.9$+$2912 & 0.4960 &  45 &  10 & - &        &       & F814W &       & F110W &       & F140W &       \\                                   
eMACSJ2327.4$-$0204 & 0.7055 &  24 & 146 & Y &  F435W & F606W & F814W &       &       & F125W &       & F160W \\ 
eMACSJ2351.6$-$2818 & 0.5679 &  18 &     & - &        & F606W &       &       & F110W &       & F140W &       \\                                   
\end{tabular}
\caption{continued.}
\end{scriptsize}
\end{table*}

\section{Observations and data reduction}
\label{sec:obs}

\subsection{Groundbased imaging}
By design, all eMACS clusters are covered in five optical passbands ($g^{\prime}r^{\prime}i^{\prime}z^{\prime}y^{\prime}$) in the PS1 3$\pi$ survey \citep{Chambers2016}. Although crucial for the identification of clusters among the many faint X-ray sources in the RASS, the PS1 3$\pi$ imaging data are too shallow to serve more ambitious purposes. 

Consequently, we obtained short ($\sim$300s) observations of all eMACS clusters to a limiting magnitude of approximately 23 (AB) through the $g^{\prime},r^{\prime}$, and $i^{\prime}$ filters\footnote{For a small number of clusters, imaging with GMOS was also obtained in the $z^{\prime}$ passband.} with the GMOS imaging spectrograph on the 8m Gemini-North telescope on Maunakea. Extending to over 1 Mpc in radius from the centre of a cluster at $z>0.5$, the  5.5\arcmin\ field of view of GMOS offers a comprehensive view of the projected spatial distribution of likely cluster galaxies (as identified from their characteristic red-sequence colours) and allowed the efficient selection of targets for the spectroscopic follow-up observations described in Section~\ref{sec:obs-spec}. A more detailed description of the GMOS imaging survey of eMACS clusters, as well as of the associated data reduction techniques, is provided in the context of the mass-traces-light algorithm \textsc{AStroLens} that provided a first assessment of the gravitational-lensing strength of eMACS clusters \citep{Zalesky2020}. 

\subsection{\emph{HST} imaging}
\label{sec:obs-hst}

Optical and near-infrared imaging with high angular resolution and low sky background is critical for several key science goals of the eMACS project, most notably the detection of strong-gravitational-lensing features, as well as dust and star formation in brightest cluster galaxies (BCGs), and ram-pressure stripping of ``jellyfish'' galaxies. 

To achieve these objectives, eMACS clusters were observed with the Advanced Camera for Surveys (ACS) and the Wide-Field Camera 3 (WFC3) aboard the {\it Hubble Space Telescope} ({\it HST}) in four SNAPshot programs (GO-13671, GO-14098, GO-15132, and GO-15843). In addition, eMACS clusters of special interest were targeted by us in dedicated {\it HST} imaging campaigns (GO-15466, GO-15608, GO-15844, GO-16428) that added either depth or area on the sky to existing SNAPshot data. A number of eMACS clusters were also observed independently by other teams. 

The listed SNAPshot programmes requested imaging observations in four passbands: F606W and F814W (ACS; 1200s each), as well as F110W and F140W (WFC3; 700s each), although the random nature of SNAPshot surveys meant that not all clusters were observed in all passbands. Imaging was also performed in the F435W passband of ACS, as well as the WFC3 near-infrared filters F105W, F125W, and F160W, in targeted GO observation awarded to us or other teams. An overview of the available {\it HST} data in these eight passbands is provided in Table~\ref{tab:sample}. Except for eMACSJ0200.3 which has no {\it HST} imaging, all eMACS clusters were observed in at least one passband\footnote{Scheduled {\it HST} observations of eMACSJ1350.7 were canceled by the proposers after {\it CXO} observations of the field failed to detect diffuse X-ray emission, thereby ruling out the presence of a galaxy cluster (see Section~\ref{sec:misid}).}, 

All {\it HST} data were reduced using standard procedures. Specifically, all available {\it HST} images for a given field were corrected for geometric distortions and mapped to a common coordinate frame tied to the Gaia astrometric reference frame, using the \textsc{TweakReg} and \textsc{AstroDrizzle} packages \citep{fruchter2010}. False-colour RGB images\footnote{Whenever possible (see Table~\ref{tab:sample}), we use the F606W, F814W, and F110W images to populate the B, G, and R channels, respectively.} were then created using the same colour table, scaling, and limits for all fields. 
We used {\sc SExtractor} \citep{bertin1996} to generate source catalogues and to obtain photometry in the three chosen passbands. Cluster members were tentatively identified as members of the red sequence \citep[e.g.,][]{visvanathan1977,1992MNRAS.254..601B} based on colour-magnitude diagrams. We refer the reader to Basto et al.\ (in preparation) for additional details.

We note that, since the vast majority of the {\it HST} images were obtained in randomly selected SNAPshot observations, the resulting {\it HST} imaging dataset can be considered unbiased and representative of the eMACS sample as a whole. A mild bias is introduced, however, by our efforts to secure multi-passband coverage (or greater observational depth) for clusters showing exceptional strong-lensing features in the SNAPshot observations.

\subsection{Optical and near-infrared spectroscopy}
\label{sec:obs-spec}

Extensive spectroscopic follow-up observations were conducted to target presumed cluster members, potential strong-lensing features, and likely counterparts of X-ray point sources (detected in the {\it CXO} observations described in Section~\ref{sec:obs-cxo}) in the eMACS cluster fields. These observations were performed with the DEep Imaging Multi-Object Spectrograph \citep[DEIMOS;][]{DEIMOS2003}, the Low-Resolution Imaging Spectrometer \citep[LRIS;][]{LRIS1995}, and the Multi-Object Spectrometer For Infra-Red Exploration \citep[MOSFIRE;][]{MOSFIRE2010,MOSFIRE2012} on the Keck-1 and -2 10m telescopes on Maunakea, Hawai'i. The individual multi-object spectroscopy targets were selected from {\it HST}, GMOS, PS1/3$\pi$, and {\it CXO} observations. Whenever possible (i.e., depending on the availability and quality of {\it HST} and {\it CXO} data), priority was given to redshifts measurements of likely strong-lensing features and X-ray point sources, with the remainder of the mask being used to target probable cluster members (selected by their colours). In the following we briefly summarize the instrumental setup for the three spectrographs used.

\begin{description}
\item[{\it DEIMOS}:] 1\arcsec\ slit width; 600 line mm$^{-1}$ grating; GG455 order-blocking filter; central wavelength 6300\AA; 1.2\AA\ pixel$^{-1}$ sampling;
\item[{\it LRIS}:] 1\arcsec\ slit width; D680 dichroic; 300/5000 grism (8.8\AA\ FWHM resolution, red arm); 600/7500 grating and central wavelength 8100\AA\ (4.7\AA\ resolution, blue arm);
\item[{\it MOSFIRE}:] 0.7\arcsec\ slit width; $I$, $J$, $K$, and / or $H$ band, depending on the expected wavelength of emission-line features; ABBA dither pattern with (typically) 1.5\arcsec\ offset; 2.8\AA\ pixel$^{-1}$ sampling.
\end{description}
Exposure times ranged from approximately 1800s to several hours, depending on atmospheric conditions, airmass, and the faintness of the primary targets on a given MOS mask. We typically observed all targets first in the optical window and then added near-infrared spectroscopy with MOSFIRE only for strong-lensing features for which no identifiable features were detected in prior observations with LRIS or DEIMOS.

Most of the DEIMOS spectra were reduced using a modified version of the DEEP2 pipeline \citep{2012ascl.soft03003C,2013ApJ...765...25N}, with more recently acquired data having been processed with the {\sc PypeIt} package \citep{pypeit:joss_pub}. 
An early version of {\sc PypeIt} was also used to reduce the LRIS data, while the MOSFIRE data reduction was mostly performed with the pipeline designed by the MOSFIRE commissioning team and written by Nick Konidaris with extensive checking and feedback from Chuck Steidel and other MOSFIRE team members \citep{2014ApJ...795..165S} . 

Although we aimed at obtaining, for each eMACS cluster, at least 20 spectroscopic redshifts of cluster members as well as spectroscopic confirmation of at least one gravitational-lensing feature, weather patterns on Maunakea and the vagaries of the telescope-time allocation process introduced seasonal biases. We also deliberately devoted extra resources to the spectroscopic follow up of eMACS clusters of particular interest, e.g., the most distant ones, the most obvious and most spectacular gravitational lenses, or the most extreme mergers. As a result, the spectroscopic coverage of the sample presented here is uneven, as is apparent from the number of redshifts obtained for each cluster listed in Table~\ref{tab:sample}.

\subsection{X-ray imaging spectroscopy}
\label{sec:obs-cxo}

We searched the public archive CHASER\footnote{\url{https://cda.harvard.edu/chaser/}} for data collected for eMACS clusters with the Advanced CCD Imaging Spectrometer (ACIS) aboard the {\it Chandra X-ray Observatory} ({\it CXO}); for consistency, we retrieved and analyzed only observations performed with the instrument's ACIS-I array in VFAINT mode. Of the resulting list of 47 eMACS clusters observed (see Table~\ref{tab:sample}), the majority were targeted by the eMACS team (PI Ebeling), with most of the remainder having been selected as part of follow-up observations of detections of the Sunyaev-Zeldovich effect. Excluding the 12 systems at $z>0.5$ discovered previously by the MACS project \citep{Ebeling2007} and discussed extensively in the literature \citep[e.g.,][]{2003ApJ...583..559L,Ebeling2007,2008ApJ...687..959B,2011MNRAS.410.1939Z,2012A&A...544A..71L,2016ApJ...819..113O,2018MNRAS.476.3415W,2018MNRAS.481.2901J}, we are left with 35 new systems. For each cluster we reduce and then merge all available observational datasets.

We emphasize that the resulting subset of eMACS clusters with {\it CXO}/ACIS-I data is not representative of the full sample but reflects the scientific interests and target selection criteria put forward by the authors of the underlying {\it CXO} proposals. As a result, biases in favour of the most X-ray luminous, the most distant, and the most disturbed clusters are introduced; additional biases may be present but remain unknown to us.

Our data reduction extensively uses routines from the \textsc{CIAO} software environment \citep{fruscione2006} and largely follows established procedures, beginning with the reprocessing and reprojecting of the archival data, before removing time intervals affected by background flares (we first examine the high-energy 9--12 keV band, then the entire ACIS-I passband). We subsequently identify X-ray point sources as detected by the {\sc wavdetect} source-detection algorithm \citep{freeman2002} but employ a visual check to ensure the validity of all point sources in the automatically generated source list.

Since our analysis is similar to the process described in detail in the literature \citep[e.g.,][]{mahler2023,beauchesne2024}; we here provide only a brief summary. For each observation, we retrieve the associated ACIS-I blank-sky background from the public archive and compute the response of the instruments within a circle of radius $1$~Mpc, initially centred on the cluster position in the RASS source catalogue. To model the background emission (both instrumental and sky backgrounds), we use an empirical model consisting of B-spline functions. For each observation, we build this model by first fitting the blank-sky spectrum for each CCD in the 0.1--13 keV band. We then use a linear combination of these models to define a background model per observation as a linear combination of the CCD models. This linear combination is optimised on the blank-sky spectrum of the chosen cluster region by keeping all coefficients frozen for the spectral fit of the cluster and optimizing only one global amplitude per observation. This procedure allows us to estimate the scaling between the science and the blank-sky observations without relying on the observing time or the particle background. More details about the background estimation can be found in Appendix A of \citet{beauchesne2024}. 

\subsection{Radio observations}
\label{sec:obs-radio}

The majority of the clusters presented here (see Table~\ref{tab:sample}) were observed with the Jansky VLA in B-array at 5~GHz in 2016 as part of a larger campaign to observe MACS and eMACS clusters to search for activity in the BCG and other cluster members (PI Edge, 16A-167). All targets were observed in two visits of 300s each, separated by 60--80~mins, with an associated, nearby phase calibrator. These observations reach a consistent noise level of $\sim30\ \mu$Jy per beam and a resolution of $\sim$2.0$''$ that depends weakly on declination. 

\section{Analysis}
\label{sec:methods}

\subsection{Galaxy spectroscopy}

Redshifts were measured and corrected to the heliocentric frame using an adaptation of the {\sc SpecPro} package \citep{2011PASP..123..638M}. We determined cluster velocity dispersions with the \textsc{ROSTAT} software developed by \citet{1990AJ....100...32B}.

\subsection{X-ray imaging spectroscopy}

To determine fundamental physical parameters of the ICM emission recorded with {\it CXO}, we proceed as follows. The overall spectral model is the combination of the previously defined empirical background with a physical model of the cluster emission; we adopt a single-temperature plasma, represented by a single APEC model (Astrophysical Plasma Emission Code, {\sc xsapec}) with solar abundance ratios from \citet{asplund2009} and photoelectric absorption parameterized by the column density of neutral hydrogen ($n_{\rm H}$) along the line of sight ({\sc xsphabs}). We fit the combined spectrum to the data in the 0.5--7 keV range using the nested-sampling algorithm provided in the \textsc{Multinest} package \citep{feroz2019} through its Python wrapper \citep{buchner2014}. During this fit, the density, temperature (k$T$), and metallicity ($Z$) of the intracluster medium (ICM) are allowed to vary, in addition to the background normalisation. Since $n_{\rm H}$ is poorly constrained for most clusters, we set it to the Galactic value provided by the HI4PI project \citep{HI4PI2016}.

We perform the spectral fitting within a circular area that is determined iteratively. We start by fitting the data within a circle of radius $1$~Mpc around the RASS position of each eMACS cluster. From the posterior distribution of the ICM temperature and the cluster redshift, we derive the distributions of $r_{\rm 1000}$ and $M_{\rm 1000}$ using the relation provided in \citet{arnaud2002}. We also compute the centroid of the X-ray emission within the initial fitting area of $1$~Mpc with the python package {\sc pyproffit} \citep{eckert2020} and then define a new extraction region around this centroid position, with a radius given by the median of the $r_{\rm 1000}$ distribution. We repeat this process until the new $r_{\rm 1000}$ median is included in the median-centred interval of $68$ per cent of the previous distribution, and the associated emission centroids are similar. We do not quantitatively monitor the convergence of the centre position, but require the previous centre position to be within $6$ arcsec of the new one. In most cases, the differences in the location of the centroid between fits are of the order of $1$ arcsec or less. The resulting values of $r_{\rm 1000}$ range approximately from $500$ to $1000$~kpc for our sample.

The spatial distribution of the X-ray emission from eMACS clusters observed with {\it CXO} is, in all figures in this paper, represented by the isointensity contours of the X-ray surface brightness after adaptive smoothing to $3\sigma$ significance with the \textsc{asmooth} algorithm \citep{2006MNRAS.368...65E}. 

\subsection{Strong gravitational lensing}
\label{sec:SL}

We use the multi-band high-resolution images obtained with {\it HST} (all but four 
eMACS clusters were observed with {\it HST} in at least two passbands; see Section~\ref{sec:obs-hst} and Table~\ref{tab:sample}) to identify potential strongly lensed images based on their similarity in colours and morphologies. The corresponding sources were given high priority for the spectroscopic follow-up observations described in Section~\ref{sec:obs-spec}. For the strong-lensing analysis described below we consider the subset of 25 
eMACS clusters with at least one multiple-image system confirmed by spectroscopic redshifts.

We use the \textsc{Lenstool} software \citep{2007NJPh....9..447J} to create lens models by parameterising the mass distribution of each cluster as a combination of double Pseudo Isothermal Elliptical (dPIE) mass profiles at both large (cluster) and small (galaxy) scales. Following similar work on large cluster samples \citep[e.g.,][]{Richard2010,Richard2021}, we optimise the set of parameters of large-scale dPIE components to best reproduce the locations of the confirmed multiple-image systems in each cluster. To account for the mass in small-scale components, we include each cluster member (selected from the red sequence) as a galaxy-scale perturber of the overall mass distribution. We fix these galaxies' geometric parameters (centre, ellipticity, position angle) at the values in the {\it HST} photometric catalogue, while the remaining parameters of these small-scale dPIEs (velocity dispersion, cut and core radii) are assumed to follow a light-traces-mass relation with their luminosity.

\section{Results and Discussion}
\label{sec:results}

In the following we present our findings on the gravitational-lensing strength (as estimated from our strong-lensing analysis), the properties of the ICM, and the  galaxy population of eMACS clusters. 

We stress again that the eMACS sample presented here is not statistically complete; in addition, the sample's coverage in follow-up observations (be it galaxy spectroscopy, {\it HST} imaging, or X-ray imaging spectroscopy with {\it CXO}) is currently inhomogeneous. Given these limitations, the goal of this paper is not a comprehensive characterization of the population of X-ray luminous clusters at $z>0.5$. Rather, we aim to highlight subsets of the sample as well as individual systems that were found to be exceptional, warrant further study, or yield insights into the selection biases of cluster samples compiled at different wavelengths and using different survey techniques.

The statistical completeness (or lack thereof) of eMACS and any eMACS subsample is strongly affected by both statistical and systematic uncertainties in the RASS data from which cluster candidates were selected (see Section~\ref{sec:emacs} and Appendix~\ref{sec:app-rass} for details). The most important contributors are shot noise (some RASS detections of eMACS clusters are comprised of a dozen or fewer net X-ray photons) and contamination by X-ray point sources, a systematic issue that we describe and assess in the following section.

\subsection{\emph{CXO} versus RASS}
\label{sec:xps}

Since {\it ROSAT}'s effective point-spread function (PSF) is significantly broadened during the RASS compared to the on-axis PSF of the system \citep[and can not be easily modeled,][]{1997ApJ...486..738D}, the standard source-detection algorithms used in the RASS reliably recognize only relatively nearby and X-ray bright clusters as intrinsically extended sources \citep{1996MNRAS.281..799E}. Moreover, the RASS X-ray fluxes and luminosities of clusters even at $z<0.3$ are prone to contamination from unresolved X-ray point sources, be they embedded in the cluster or physically unrelated fore- or background sources. At the far higher redshifts and dramatically poorer photon statistics probed by eMACS, the impact of X-ray point sources on our RASS-selected sample is \textit{a priori} unquantifiable but can be measured through follow-up observations with {\it CXO} that allow point sources to be resolved and removed.

\begin{figure}
    \centering
    \hspace*{-5mm}\includegraphics[width=0.5\textwidth]{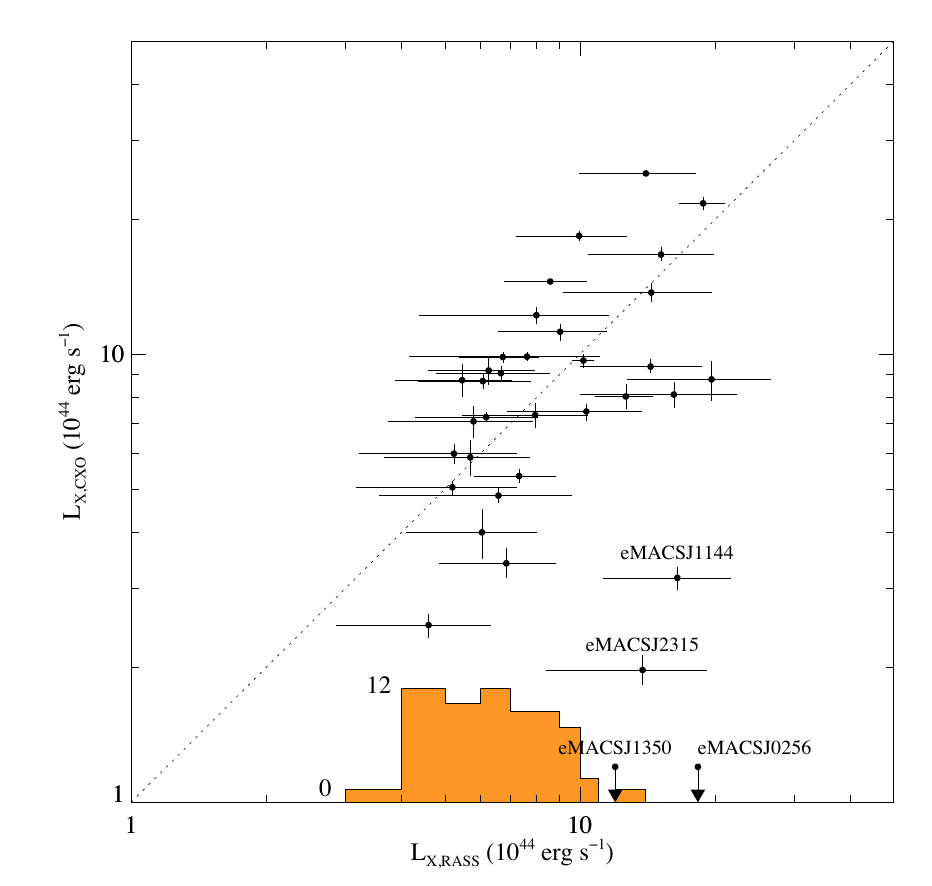}
    \caption{Comparison of eMACS cluster luminosities in the 0.1--2.4 keV band, as derived from RASS and {\it CXO} observations. The most extreme outliers are labelled and discussed in Section~\ref{sec:xcont}. Two {\it CXO} non-detections (see Section~\ref{sec:misid}) are plotted as upper limits at fiducial {\it CXO} luminosities of $1.2\times 10^{44}$ erg s$^{-1}$. The distribution in RASS X-ray luminosity of the eMACS clusters without {\it CXO}/ACIS-I data is shown as a histogram. Note that X-ray point sources were removed before measuring cluster luminosities in the high-resolution {\it CXO} data, whereas the luminosities observed in the RASS include X-ray point sources (where present). } 
    \label{fig:lxlx}
\end{figure}

Fig.~\ref{fig:lxlx} demonstrates that eMACS delivers on the promise of Fig.~\ref{fig:lxmz} by showing a comparison of the cluster luminosities as observed in the RASS and in the dedicated follow-up observations performed with {\it CXO} of selected eMACS clusters. Here, we have removed all point sources from the {\it CXO} data prior to measuring the luminosity of the diffuse ICM emission, whereas (by necessity) no such correction was applied to the RASS luminosities. We find that, in general, contamination from X-ray point sources is a minor effect\footnote{For the sample presented here, the median flux from point sources within the RASS detect cell is less than $5\times 10^{-15}$ erg cm$^{-2}$ s$^{-1}$ (0.1--2.4 keV).}, as no significant offset from the identity relation (shown as a dotted line in Fig.~\ref{fig:lxlx}) is seen in the correlation, and the scatter about it is largely consistent with the statistical uncertainties in the RASS-based measurements, with the exception of the four systems labeled in Fig.~\ref{fig:lxlx} and discussed below.

\subsubsection{Flux contamination}
\label{sec:xcont}

We highlight in Fig.~\ref{fig:lxlx} two clusters (eMACSJ1144.2 and eMACSJ2315.3) for which the point-source corrected X-ray luminosity as measured from {\it CXO} observations falls below the RASS estimate by over a factor of five. Even for such outliers, however, the cause of the discrepancy is not always readily apparent. As shown in Figs.~B20  
and B43,  
several X-ray point sources contribute to the X-ray flux measured for these two clusters at the low angular resolution of the RASS; however, the total flux from all X-ray point sources out to 2\arcmin\ radius from the peak of the ICM emission amounts to less than 10\% of the reported RASS fluxes in Fig.~\ref{fig:lxlx}. Then again, since many types of X-ray point sources are time variable, any of the point sources apparent in Figs.~B20  
and B43  
could have been substantially brighter in the early 1990s when the RASS was conducted.

The degree of point-source contamination in X-ray selected cluster surveys may also depend on the mass and evolutionary state of the systems under study. Low-mass clusters and groups of galaxies are less dense and dynamically younger; as a result, late-type galaxies tend to be more prevalent and nuclear activity (in cluster members as well as in galaxies in the cluster environment) may be stronger and more common. Anecdotally, the point source contamination appears indeed to be more pronounced in the RASS-selected MACS and BCS cluster samples which, probing lower redshifts, sample a broader range of cluster masses than eMACS \citep[e.g.,][]{Ebeling2010}.\\

\subsubsection{Misidentifications}
\label{sec:misid}

While the relative flux contributions from the diffuse ICM and from related or unrelated point sources can be difficult to disentangle for individual clusters (but does not invalidate the presence and discovery of the respective cluster), a total absence of ICM emission renders the respective entry in our catalogue a plain misidentification. 

Based on the current {\it CXO} follow-up observations, we are aware of two such cases: as noted in Fig.~\ref{fig:lxlx}, eMACSJ0256.9 ($z=0.8670$) and eMACSJ1350.7 ($z=0.8247$) show no significant extended X-ray emission in 33ks observations with {\it CXO}. We show the {\it HST} image of the former (an already awarded {\it HST} observation of the latter was cancelled by us after the {\it CXO} observation failed to detect ICM emission) as well as the histogram of radial velocities of galaxies in this region in Fig.~B6.  
Note the absence of ICM emission from eMACSJ0256.9 (Fig.~B6)  
in spite of an obvious overdensity of galaxies at the catalogued RASS source position and the high velocity dispersion of 1170$^{+140}_{-240}$ km s$^{-1}$ measured for the structure (based on 17 concordant redshifts). eMACSJ0256.9 thus underlines the critical importance of moderately deep follow-up observations with a high-resolution X-ray facility like {\it CXO} to identify non-clusters: as is apparent from Fig.~B6, 
the X-ray flux detected in the RASS is in fact generated by several X-ray point sources, the brightest of which was spectroscopically identified by us as a QSO at $z=0.8562$, i.e., at the same redshift of the purported cluster. We discuss eMACSJ0256.9 further in Section~\ref{sec:results-lss} in the context of large-scale structure.

\subsection{Strong lensing}
\label{sec:results-SL}

Given the size of the eMACS sample (and the wealth of strong-lensing features present), a comprehensive strong-lensing study of all eMACS clusters not only requires significant observational and analytical resources but also faces systematic challenges. Although compelling signs of strong gravitational lensing are  found in the majority of eMACS clusters for which the prerequisite {\it HST} imaging is available (Fig.~\ref{fig:sl}), the limitations of the data available for some clusters (e.g., no colour information; no spectroscopic redshifts; clipping of the strong-lensing area by the WFC3 aperture) hampers the secure identification of the most powerful lensing constraints: multiple-image systems. Nevertheless, for clusters with at least one spectroscopically confirmed multiple-image system, we are able to create tentative lens models as described in Section~\ref{sec:SL}. 

\subsubsection{Lensed sources}

Strong lensing by massive galaxy clusters offers unique insights into the properties of background galaxies in two complementary ways: by probing the background population in a statistical manner, thereby providing valuable constraints on, for instance, the galaxy luminosity function at  redshifts that are not observationally accessible in other ways (e.g., \citealt{Bouwens2022}), and by enabling detailed studies of the physical properties of individual, intrinsically faint background objects at more modest redshifts (e.g., \citealt{Alavi2014}). If quantitative insights are aimed for, both of these avenues of research require credible models of the mass distribution within the cluster lenses, and hence the identification of unambiguous strong-lensing features to constrain these lens models (see Sect.~\ref{sec:SL}). While exploiting the former, statistical goal requires much deeper follow-up observations (such as those of the Frontier Fields project \citealt{Lotz2017},  or the JWST UNCOVER  program \citealt{Bezanson2022}) than have been performed for eMACS to date, first steps toward the latter goal can be taken with the data at hand.

\begin{figure}
    \centering
    \includegraphics[width=0.48\textwidth]{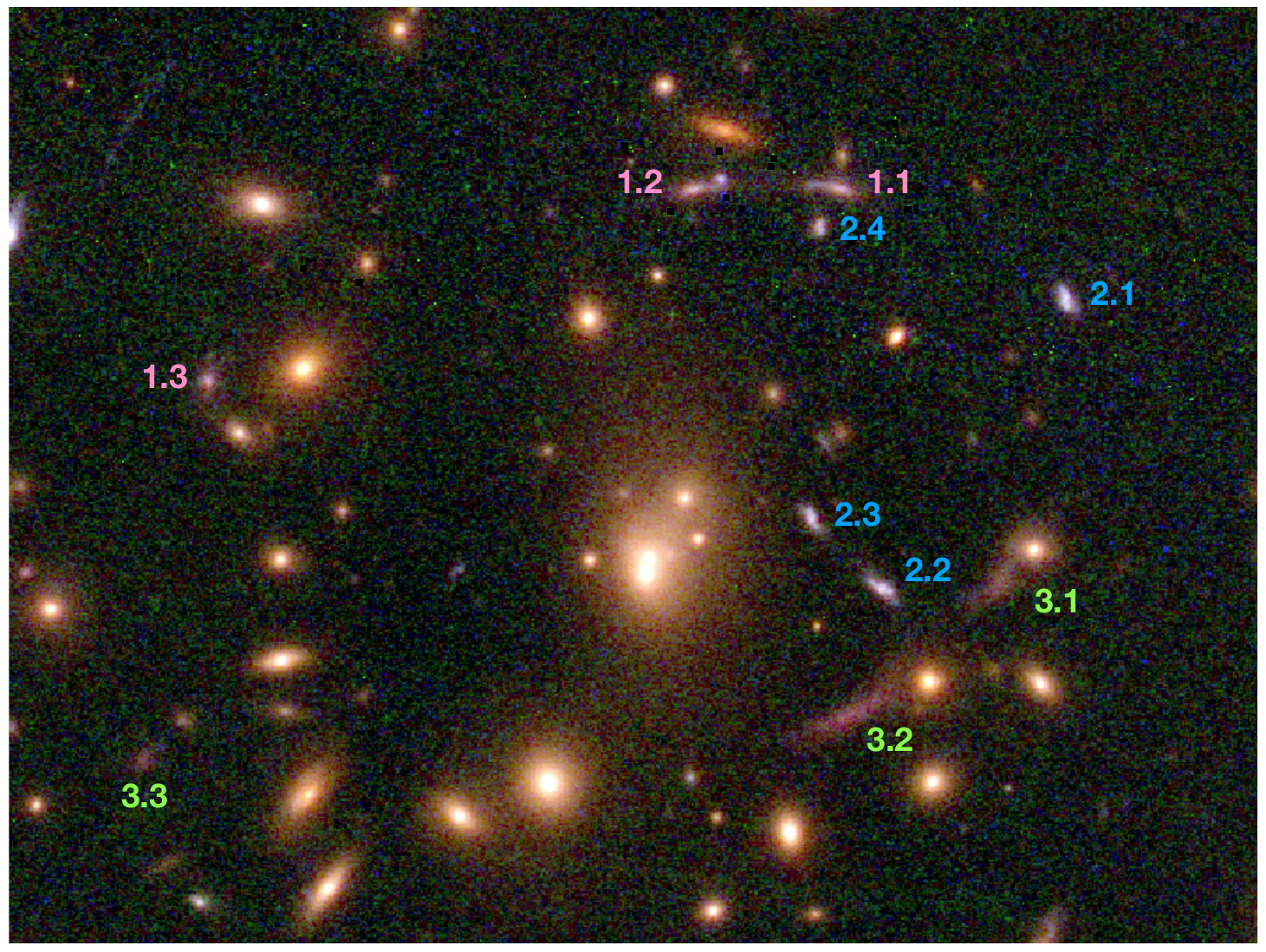}
    \caption{\textit{HST} image of the core of eMACSJ1248.2 at $z=0.573$ ($40\arcsec\times 30\arcsec$, see also Fig.~B25). 
      The components of three multiple-image systems are marked, among them a rare hyperbolic-umbilical configuration (System 2, with four images: 2.1--2.4), although unfortunately this currently lacks a spectroscopic redshift.}
    \label{fig:e1248HU}
\end{figure}

\begin{figure*}
    \includegraphics[width=\textwidth]{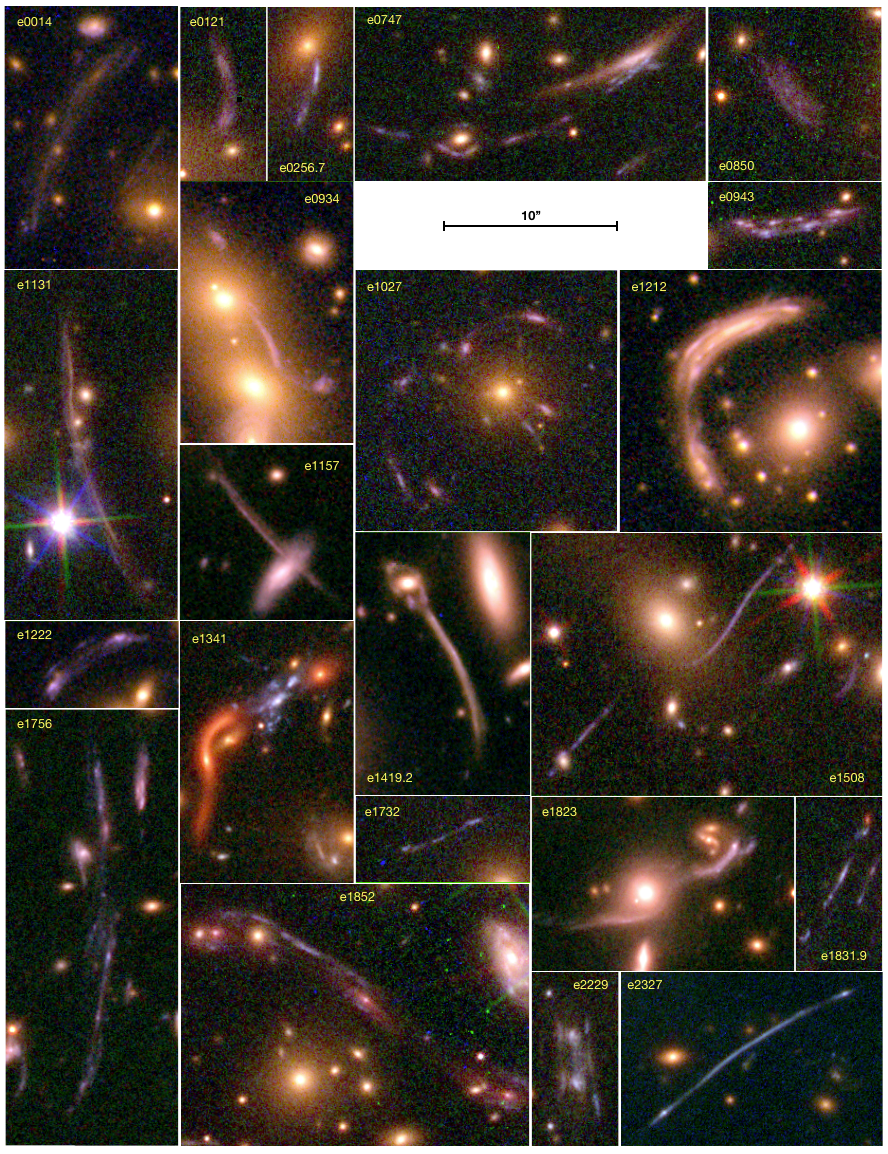}
\caption{Examples of strong-lensing features in \textit{HST} images of eMACS clusters. We use a shorthand notation for the cluster names.}
\label{fig:sl}
\end{figure*}

For 106 eMACS clusters existing multi-band {\it HST} imaging enables us to visually survey the core regions of the clusters for potential strongly lensed background sources, identifiable as highly sheared images (arcs and giant arcs) and multiply imaged sources. Many of these systems are cusp arcs, producing three images, the brightest two of which display mirror symmetry (e.g., eMACSJ0121.4 and eMACSJ1222.3 in Fig.~\ref{fig:sl}). However, higher multiplicity and more complex configurations are seen, such as quads (e.g., eMACSJ1027.2 in Fig.~\ref{fig:sl}) or rare hyperbolic umbilical lensing events as in eMACSJ1248.2 (System 2 in Fig.\ref{fig:e1248HU}; see also Fig.~B25). 
The significance of such exotic lenses is discussed, e.g., by \citet{Lagattuta23}.

\begin{figure}
    \centering
    \includegraphics[width=0.48\textwidth]{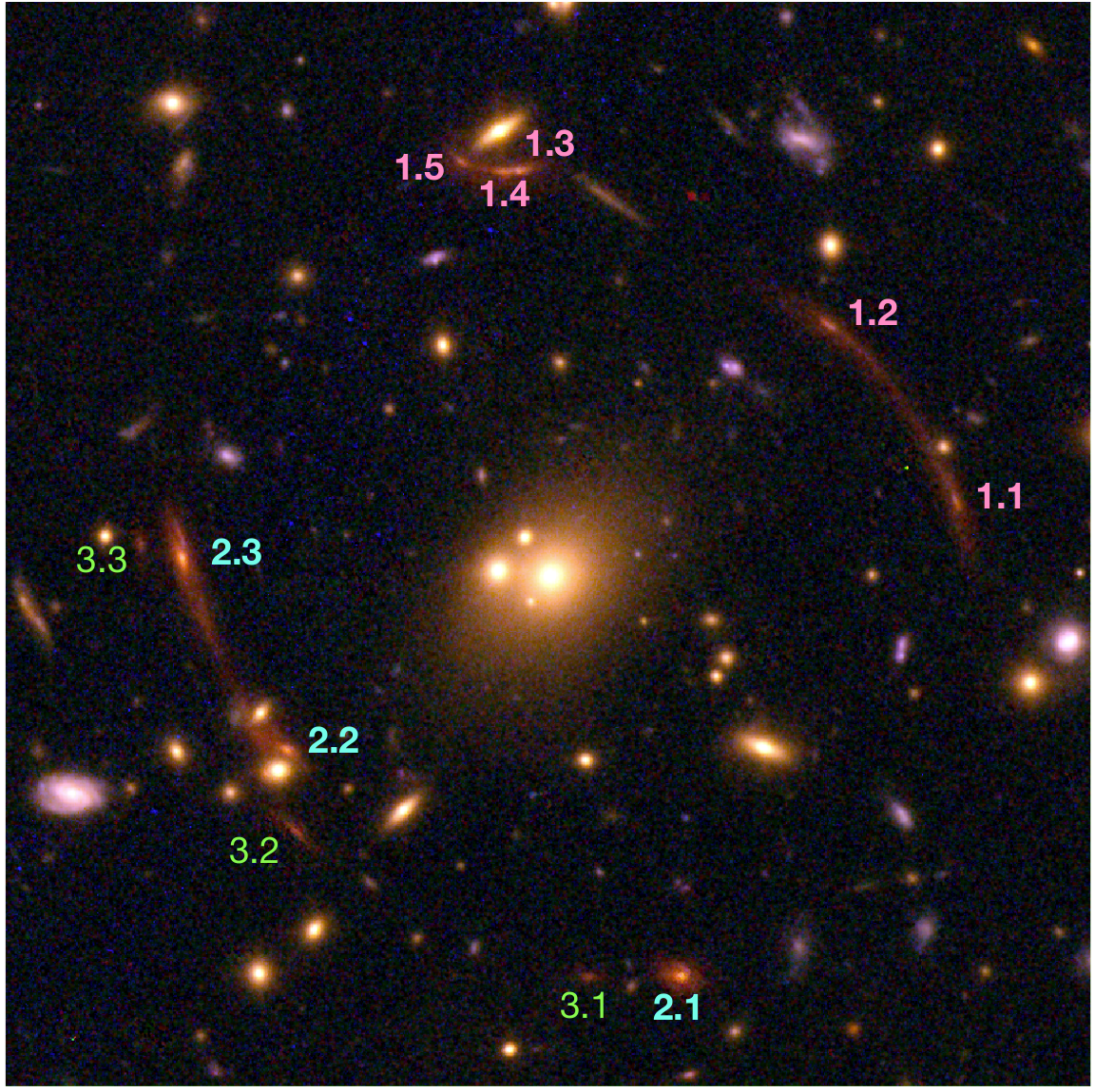}
    \caption{\textit{HST} image of the core of eMACSJ1437.8 at $z=0.535$ ($50\arcsec\times 50\arcsec$, see also Fig.~B30), 
      revealing a group of background galaxies at $z=2.55$ gravitationally lensed by the foreground cluster. The components of three multiple-image systems (two of them, Systems 1 and 2, spectroscopically confirmed at $z=2.546$ and $z=2.556$) are marked.}
    \label{fig:e1437core}
\end{figure}

The colours and morphologies of the strongly lensed features showcase the diversity of high-redshift sources seen through the cores of massive clusters at $z>0.5$ and, in rare cases, can even reveal extended background structures, such as the galaxy group at $z=2.55$ lensed by eMACSJ1437.8 (Figs.~\ref{fig:e1437core} and B30). 
For background sources at modest redshifts, typically at $z\leq 2$, the larger apparent sizes and in general brighter  magnitudes allow complex internal structures of the sources to be readily apparent through the mirror symmetry and image shear that result from strong gravitational lensing; examples are the spiral galaxy at $z=1.195$ behind eMACSJ1212.5 (Figs.~\ref{fig:sl} and B23), 
or the quiescent galaxy at $z=1.594$ behind eMACSJ1341.9 (discussed in \citealt{2018ApJ...852L...7E}; see also Figs.~\ref{fig:sl} and B26),  
although resolved galaxy features can be seen out to higher redshifts (e.g., The Scream at $z=3.375$ behind eMACSJ222.9, Figs.~\ref{fig:sl} and B42). 
At higher redshifts, the combination of strong image shear and increasingly compact sources sizes (particularly in the restframe UV at $z\gtrsim 2$), as well as surface-brightness dimming, contribute to the apparent narrowness of some lensed features, such as the complex ``straight arc'' at $z=2.03$ in eMACSJ1756.8 (Figs.~\ref{fig:sl} and B35),  
or highly sheared examples like the giant arc in eMACSJ2327.4 at $z=2.983$ \citep[][see also Figs.~\ref{fig:sl} and B45]{2015ApJ...814...21S}.  

\subsubsection{Einstein radii and systematic effects}
\label{sec:results-re}

As a first indicator of central lensing strength, we list in Table~\ref{tab:re} the Einstein radius, $R_{\rm E}$\footnote{Following, e.g., \citet{2023A&A...678L...2M}, we compute the equivalent Einstein radius $R_{\rm E}$ as the radius of the circular aperture enclosing the same area as the tangential critical line at this redshift.}, of the 25 eMACS clusters modeled to date with {\sc Lenstool}, estimated for a fiducial source at $z=2$; in addition, we also list the total projected gravitational mass within that radius. Both quantities exhibit a large spread for reasons that include true diversity in the clusters' physical characteristics\footnote{Including the cluster environment, as strong lensing is sensitive to the total projected mass along the line of sight.}  as well as uncertainties in the model assumptions for cluster lenses currently constrained by only a single multiple-image system. 

\begin{table}
    \centering
    \begin{tabular}{lcccc}
Name & $R_{\rm E,PM}$ &  $M(r<R_{\rm E,PM})$ & $N_{\rm mi}$ & $N_{\rm mi,z}$\\
     &    (arcsec)    & ($10^{12}$ M$_\odot$) \\[1mm] \hline\\[-1mm]
 eMACSJ0030.5$+$2618  &  \;\,$9\pm2$      &  \;\,16  & 4 & 4 \\
 eMACSJ0121.4$+$2143  &     $12\pm1$      &  \;\,40  & 3 & 3 \\
 eMACSJ0252.4$-$2100  &     $15\pm2$      &  \;\,72  &12\;\, &12\;\, \\
 eMACSJ0256.7$-$1623  &     $11\pm1$      &  \;\,49  &12\;\, &12\;\, \\
 eMACSJ0324.0$+$2421  &     $11\pm2$      &  \;\,51  & 3 & 3 \\
 eMACSJ0502.9$-$2902  &     $25\pm3$      &     164  &20\;\, & 9 \\
 eMACSJ0834.2$+$4524  &     $32\pm4$      &     324  & 9 & 9 \\
 eMACSJ0840.2$+$4421  &     $14\pm1$      &  \;\,58  & 8 & 8 \\
 eNACSJ0934.6$+$0540  &     $16\pm2$      &  \;\,60  &10\;\, & 7 \\
 eMACSJ0943.3$-$1842  &     $>12$         &  $>37$\;\,  & 3 & 3 \\
 eMACSJ1057.5$+$5759  &     $20\pm10$\;\; &     119  & 4 & 2 \\
 eMACSJ1157.9$-$1046  &       $>15$       & $>44$\;\,  & 9 & 3\\
 eMACSJ1209.4$+$2640  &     $22\pm2$      &     131  &19\;\, & 7 \\
 eMACSJ1212.5$-$1216  &     $40\pm4$      &     542  & 3 & 3\\
 eMACSJ1248.2$+$0743  &     $17\pm2$      &  \;\,69  &11\;\, & 3\\
 eMACSJ1341.9$-$2442  &     $16\pm2$      &  \;\,79  & 5 & 5\\
 eMACSJ1353.7$+$4329  &     $38\pm5$      &     525  &13\;\, & 4\\
 eMACSJ1437.8$+$0616  &     $14\pm1$      &  \;\,47  &11\;\, & 8\\
 eMACSJ1527.6$+$2044  &     $21\pm2$      &     136  & 7 & 2\\
 eMACSJ1756.8$+$4008  &      $>19$        &  $>79$\;\, & 7 & 7\\
 eMACSJ1831.1$+$6214  &      $>18$        & $>133$\;\;\;  & 5 & 5\\
 eMACSJ1852.0$+$4900  &     $15\pm2$      &  \;\,54  & 5 & 5\\
 eMACSJ2229.9$-$0808  &      $>11$        &  $>27$\;\, & 9 & 6\\
 eMACSJ2316.6$+$1246  &     $22\pm3$      &     114  & 9 & 9\\
 eMACSJ2327.4$-$0204  &     $35\pm4$      &     420  &12\;\, &12\;\, \\
    \end{tabular}
    \caption{Einstein radii at $z_{\rm source}=2$ and enclosed masses from our strong-lensing analysis of all eMACS clusters with spectroscopically confirmed multiple-image systems (PM = Parametric Model). We show entries as lower limits where an obviously present second mass component is not included in our lens model due to a lack of strong-lensing constraints. Also listed is $N_{\rm mi}$, the number of multiple images used to constrain each model, and $N_{\rm mi,z}$, the number of such images belonging to spectroscopically confirmed multiple-image systems.}
    \label{tab:re}
\end{table}

The impact of modeling uncertainties (specifically fixing the core radius of cluster-scale mass components) on the Einstein radii derived for eMACS clusters with few strong-lensing constraints is particularly strong for models based on a single multiple-image system with a redshift much lower or higher than our fiducial redshift of $z=2$.

That both real physical cluster characteristics and the lens-modeling methodology affect the derived Einstein radii becomes apparent also when the Einstein radii from our {\sc Lenstool} analysis are compared to those obtained in a previous strong-lensing analysis of almost the entire eMACS cluster sample with the \textsc{AStroLens} package \citep{Zalesky2020}. Unlike {\sc Lenstool}, \textsc{AStroLens} uses a light-traces-mass approach that, although calibrated using a small number of clusters with spectroscopically confirmed strong-lensing features, relies entirely on the spatial distribution of light from the cluster galaxies (as measured from ground-based imaging) to derive mass models for eMACS clusters. As such, \textsc{AStroLens} provides an estimate of the global lensing power on scales determined by the extent of the distribution of cluster galaxies. {\sc Lenstool}, by contrast, models the mass distribution in much greater detail from actual strong-lensing features -- but only within the strong-lensing regime, i.e., the cluster core.

\begin{figure}
    \centering
    \hspace*{-5mm}\includegraphics[width=0.5\textwidth]{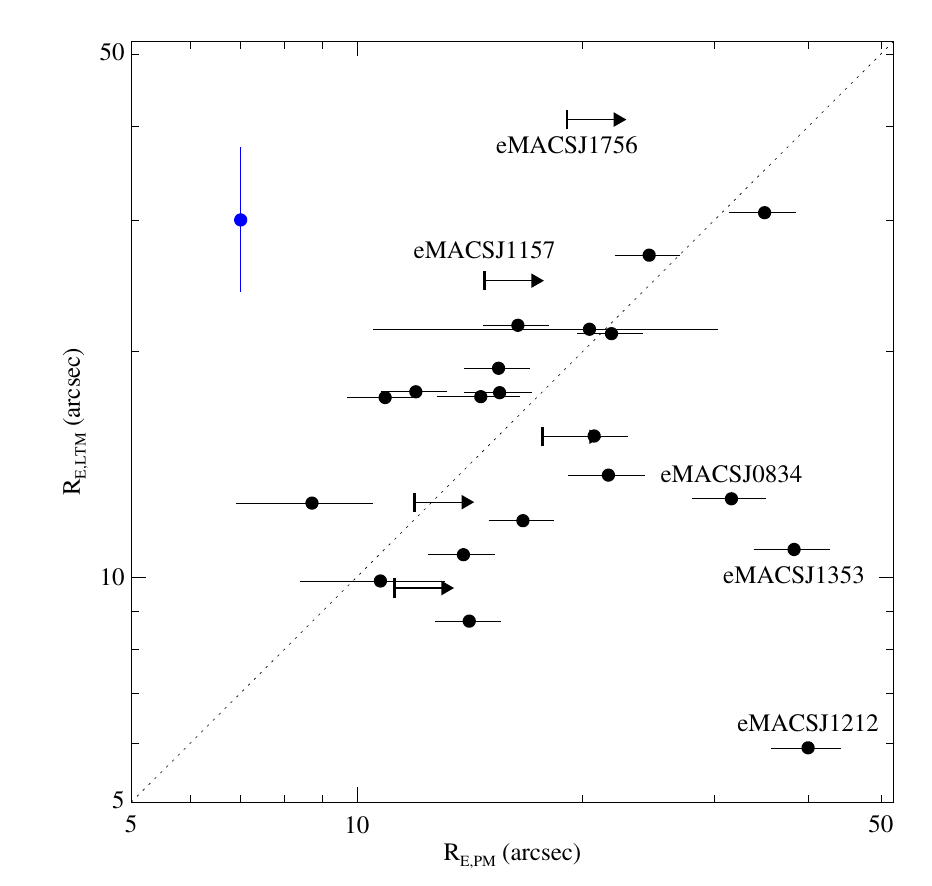}
    \caption{Comparison of Einstein radii estimated for a fiducial background source at $z=2$ for 25 eMACS cluster using the parametric-modeling software {\sc Lenstool} ($R_{\rm E,PM}$) and the light-traces-mass package AStroLens ($R_{\rm E,LTM}$). The dotted line marks the identity relation; the five most extreme outliers, labelled in this graph, are discussed in Section~\ref{sec:results-SL}. The blue error bar in the upper left indicates the approximate statistical uncertainty of 25\% of the $R_{\rm E,LTM}$ values. Note that all \textsc{AStroLens} measurements shown in this graph were scaled by a factor of 0.8 to remove a systematic bias relative to the \textsc{Lenstool} results. \label{fig:rere}
    }
\end{figure}

Comparing the Einstein radii estimated by the two methods for the eMACS subsample listed in Table~\ref{tab:re}, we find a systematic offset between the results from the two methods, in the sense that \textsc{AStroLens} on average overestimates the Einstein radius by a factor of 1.26$\pm$0.12. Once this bias is removed\footnote{We note that the subsample of Table~\ref{tab:re} includes nine of the ten eMACS clusters that were used in the calibration of \textsc{AStroLens} (the tenth one being eMACSJ0850.2, Fig.~B12), 
and that discrepancies for this subset were, in part, apparent and acknowledged in \citet{Zalesky2020}. }, the results show an overall correlation that is reassuring, given the differences in the methodologies of the two analyses and in the underlying observational data (Fig.~\ref{fig:rere}). The  scatter about the identity line (characterized by $R_{\rm E,LTM}/R_{\rm E,PM}=1.00\pm0.28$, where $R_{\rm E,LTM}$ denotes the rescaled Einstein radii obtained with \textsc{AStroLens}) is substantial though and indicative of small-scale structure in the majority of eMACS clusters, as expected for a population of clusters that are actively growing through mergers and accretion (see Sections~\ref{sec:results-cxo} and \ref{sec:hilx}). The comparison of \textsc{AStroLens} and \textsc{Lenstool} results also highlights dramatically different Einstein radii measured by the two algorithms for three clusters, labelled in Fig.~\ref{fig:rere} and reviewed in more detail below. The more sophisticated strong-lensing analysis of eMACS clusters presented in Basto et al.\ (in preparation) will improve upon the lens models used here and allow a more robust investigation of systematic effects.

\begin{figure}
    \centering
    \hspace*{-2mm}\includegraphics[width=0.5\textwidth]{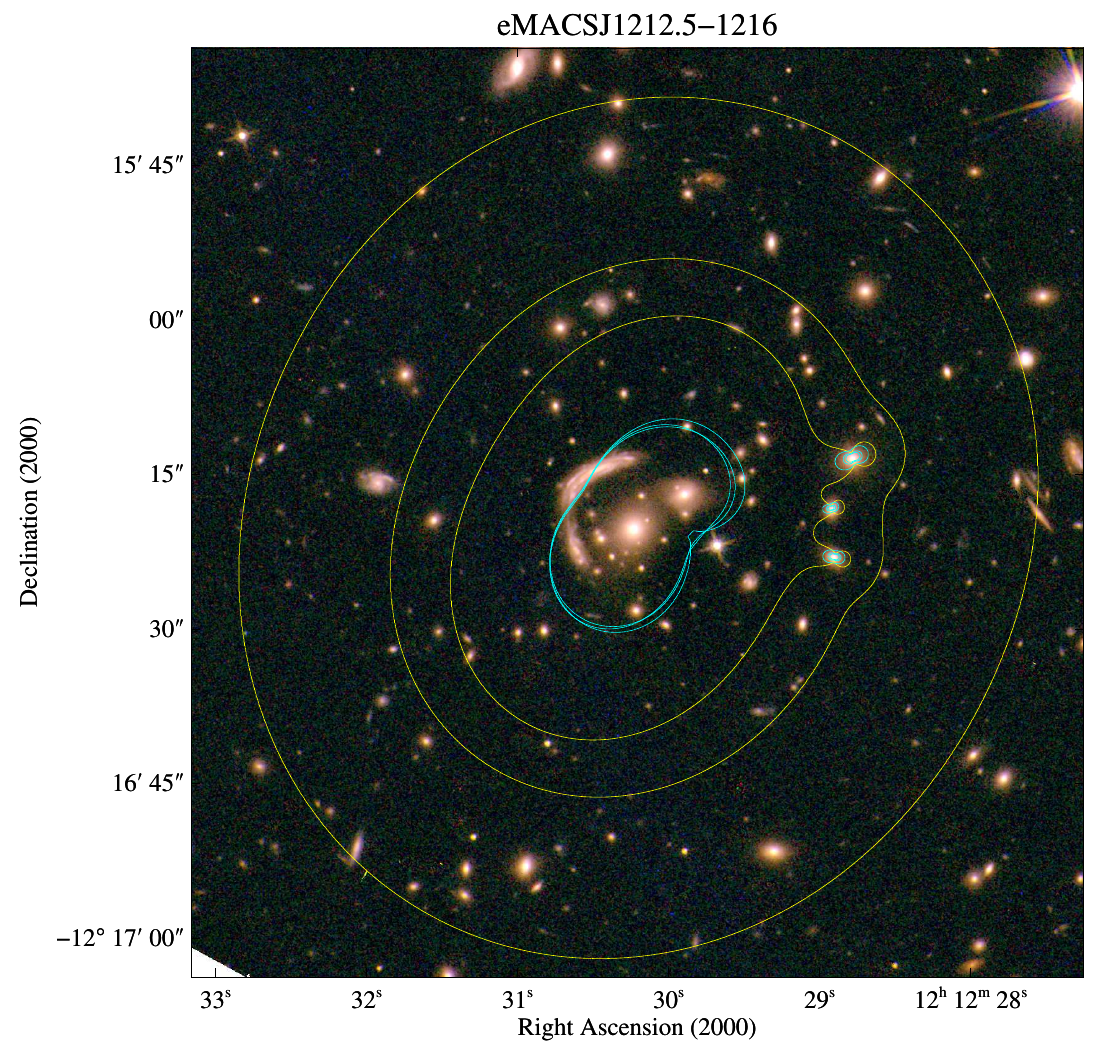}
    \caption{The core of eMACSJ1212.5 (cf.\ Fig.~B23) 
      with tangential critical lines for strong gravitational lensing of a background source at $z=1.195$ (the redshift of the observed multiple-image system) and $z=2$ overlaid in cyan and yellow, respectively. For each source redshift, the critical line is shown for cluster core radii of 60, 100, and 200 kpc. By design, the critical lines for $z_{\rm src} =1.195$  satisfy the strong-lensing constraints for any core radius. }
    \label{fig:e1212models}
\end{figure}

While the Einstein radii determined by the two methods are in good agreement once the mentioned systematic bias is removed, Fig.~\ref{fig:rere} reveals a handful of extreme outliers that warrant discussion. The most extreme one is eMACSJ1212.5, the cause being mostly the \textsc{Lenstool} assumption of a core radius of 200 kpc  and the significant extrapolation from the measured redshift of $z=1.2$ of the multiple-image system to the fiducial value of $z=2$, as illustrated in Fig.~\ref{fig:e1212models}. In addition, for compact clusters whose light distribution is dominated by only one or two galaxies with little contribution from other cluster members, the Einstein radius obtained with \textsc{AStroLens} depends sensitively on the assumed scaling relations for the BCG. Hence, while \textsc{Lenstool} likely overestimates the lensing strength of eMACSJ1212.5, the same quantity is likely to have been underestimated by \textsc{AStroLens}.

A different systematic effect is likely responsible for two slightly less extreme outliers in Fig.~\ref{fig:rere}, eMACSJ0834.2 (Fig.~B10) 
and eMACSJ1353.7 (Fig.~B27). 
Both systems are small-separation double clusters and active mergers with very high velocity dispersions of $1590^{+140}_{-200}$ km s$^{-1}$ and $1600^{+110}_{-160}$ km s$^{-1}$, respectively; both are also extremely X-ray luminous (see Section~\ref{sec:hilx}). In addition, the lens models of both clusters are constrained by robust strong-lensing constraints at 100-200\,kpc, i.e., at much larger radial distances than eMACSJ1212.5, and the redshifts of the lensed sources are closer to our fiducial value of $z=2$ than in the case of eMACSJ1212.5. As a result, the required extrapolations are much more modest, and the large Einstein radii presented in Table~\ref{tab:re} and Fig.~\ref{fig:rere} are not only less uncertain but also consistent with all other known physical properties of these clusters. The outlier status of eMACSJ0834.2 and eMACSJ1353.7 in Fig.~\ref{fig:rere} is thus likely to be due to \textsc{AStroLens} underestimating the lensing power of line-of-sight mergers whose galaxy distribution (and hence light distribution) is by necessity much less extended than that of similarly massive systems colliding along an axis that is closer to the plane of the sky.

The cause of the discrepancy in the $R_{\rm E}$ values of the remaining two clusters labeled in Fig.~\ref{fig:rere}, eMACSJ1157.9 (Fig~B21) 
and eMACSJ1756.8 (Fig.~B35) 
is less speculative and already hinted at by them being flagged as lower limits in Table~\ref{tab:re}. The reason for the small Einstein radius of eMACSJ1157.9, a system that is included in the list of the ten most powerful cluster lenses in the analysis by \citet{Zalesky2020}, is simply the lack of spectroscopic redshifts for several multiple-image systems readily identified in the {\it HST} images, which cause the western cluster component to be completely absent from the current, tentative \textsc{Lenstool}  model (Fig.~B21). 
Future spectroscopic follow-up observations are bound to secure additional redshifts for strong-lensing features in this area and help recover the full extent of the cluster mass distribution. An even more dramatic outlier (in the same sense as eMACSJ1157.9) is eMACSJ1756.8, shown in Fig.~B35). 
Again at least half of the mass distribution of this highly elongated system is missing from the parametric lens model. However, in this case the reason is not a lack of redshifts but an actual lack of strong-lensing features in the eastern half of the cluster. The cause of this cluster's outlier status in Fig.~\ref{fig:rere} thus appears to be a physical property of the system, namely an unusually flat mass profile that dramatically reduces the probability for strong gravitational lensing around the eastern cluster core. A weak-lensing analysis is needed to test this hypothesis and establish a robust total mass for this extreme merger. Weak-lensing follow-up studies might also benefit other eMACS clusters that appear to contain additional cluster-scale components that lack suitable strong-lensing features, such as eMACSJ0943.3 (Fig.~B15), 
eMACSJ1831.1 (Fig.~B38), 
or eMACSJ2229.9 (Fig.~B42). 

While the short list of eMACS clusters with \textsc{Lenstool} models presented here contains several exceptional gravitational lenses, many other eMACS clusters exhibit strong-lensing features that make them promising candidates for in-depth follow-up studies. Among the systems with existing {\it HST} images, such noteworthy clusters are eMACSJ0014.9 (Fig.~B1),  
eMACSJ0031.2 (Fig.~B3), 
eMACSJ0121.4 (Fig.~B4), 
eMACSJ0747.0 (Fig.~B9), 
eMACSJ0850.2 (Fig.~B12), 
eMACSJ0934.6 (Fig.~B13), 
eMACSJ\-1027.2 (Fig.~B16), 
eMACSJ1131.1 (Fig.~B19), 
eMACSJ1144.2 (Fig.~B20), 
eMACSJ1222.3 (Fig.~B24), 
eMACSJ1414.7 (Fig.~B28), 
eMACSJ1419.2 (Fig.~B29), 
eMACSJ1508.1 (Fig.~B31), 
eMACSJ\-1732.4 (Fig.~B34), 
and eMACSJ1823.1 (Fig.~B37). 

\subsection{ICM properties and relaxation state}
\label{sec:results-cxo}

An overview of the key cluster properties derived from the X-ray follow-up observations described in Section~\ref{sec:obs-cxo} is given in Table~\ref{tab:cxo}. Since most of the underlying {\it CXO} observations were obtained with the explicit (and modest) goal of establishing point-source-subtracted cluster luminosities and assessing the large-scale morphology of the ICM, the chosen ACIS-I exposure times are relatively short for such faint targets. 

\begin{table*}
\begin{scriptsize}
\centering
\begin{tabular}{lccr@{\hspace{5mm}}r@{\hspace{5mm}}r@{\hspace{2mm}}r@{\hspace{3mm}}r@{\hspace{3mm}}r@{\hspace{3mm}}r@{\hspace{3mm}}r@{\hspace{3mm}}r@{\hspace{3mm}}}
Name & $z$ & $n_{\rm H}$ & \multicolumn{1}{c}{$L_{\rm X, RASS}$} & \multicolumn{1}{c}{$t_{\rm ACIS-I}$} & \multicolumn{1}{c}{$r_{\rm 1000}$} 
& \multicolumn{1}{c}{$f_{\rm X, CXO}$}  & \multicolumn{1}{c}{$L_{\rm X, CXO}$} & \multicolumn{1}{c}{k$T_{\rm X, CXO}$} & \multicolumn{1}{c}{$M_{\rm X, CXO}(<r_{\rm 1000})$} & \multicolumn{1}{c}{Morph.}\\
     &   & ($10^{20}$ cm$^{-2}$) &  \multicolumn{1}{c}{[0.1--2.4 keV]} &  \multicolumn{1}{c}{(ks)} &  \multicolumn{1}{c}{(kpc)} 
& \multicolumn{1}{c}{[0.5--7 keV]} & \multicolumn{1}{c}{[0.5--7 keV]} & \multicolumn{1}{c}{(keV)} & \multicolumn{1}{c}{($10^{14}$ M$_{\sun}$)} & \multicolumn{1}{c}{class}\\[1mm] \hline\\[-1mm]
eMACSJ0014.9$-$0056 & 0.5330 &  2.95 &  6.2$\pm$1.9 & 29.3 &  884$^{+ 77}_{- 60}$ & 11.6$^{+0.6}_{-0.5}$ & 13.0$^{+0.6}_{-0.6}$ & 10.2$^{+1.7}_{-1.4}$ &  7.0$^{+2.0}_{-1.4}$ & 2 \\[2mm]
eMACSJ0030.5$+$2618 & 0.4970 &  3.47 &  4.6$\pm$1.7 & 15.6 &  626$^{+ 78}_{- 61}$ &  3.7$^{+0.4}_{-0.4}$ &  3.5$^{+0.4}_{-0.4}$ &  4.9$^{+1.3}_{-0.9}$ &  2.4$^{+1.0}_{-0.6}$ & 2 \\[2mm]
eMACSJ0042.5$-$1103 & 0.5701 &  2.54 &  6.9$\pm$2.0 & 11.9 &  865$^{+248}_{-193}$ &  4.6$^{+0.6}_{-0.6}$ &  6.0$^{+0.8}_{-0.8}$ & 10.1$^{+6.7}_{-3.9}$ &  6.9$^{+7.8}_{-3.7}$ & 4 \\[2mm]
eMACSJ0045.2$-$0151 & 0.5471 &  2.71 &  5.2$\pm$2.0 & 49.4 &  613$^{+ 38}_{- 35}$ &  5.9$^{+0.3}_{-0.3}$ &  7.1$^{+0.3}_{-0.4}$ &  5.0$^{+0.6}_{-0.6}$ &  2.4$^{+0.5}_{-0.4}$ & 3 \\[2mm]
eMACSJ0135.2$+$0847 & 0.6185 &  4.87 &  7.9$\pm$2.5 & 11.5 &  582$^{+ 82}_{- 60}$ &  6.3$^{+0.6}_{-0.5}$ & 10.1$^{+0.9}_{-0.9}$ &  4.9$^{+1.5}_{-1.0}$ &  2.2$^{+1.1}_{-0.6}$ & 4 \\[2mm]
eMACSJ0248.2$+$0237 & 0.5561 &  5.09 &  6.6$\pm$3.0 & 44.5 &  621$^{+ 45}_{- 38}$ &  5.6$^{+0.3}_{-0.3}$ &  6.9$^{+0.4}_{-0.3}$ &  5.2$^{+0.8}_{-0.6}$ &  2.5$^{+0.6}_{-0.4}$ & 4 \\[2mm]
eMACSJ0256.7$-$1623 & 0.8621 &  4.29 & 19.6$\pm$6.9 & 32.2 &  546$^{+127}_{- 94}$ &  3.3$^{+0.5}_{-0.4}$ & 11.9$^{+1.9}_{-1.5}$ &  5.7$^{+3.0}_{-1.8}$ &  2.5$^{+2.1}_{-1.1}$ & 3 \\[2mm]
eMACSJ0256.9$-$1631 & 0.8670 &  4.32 & 18.3$\pm$6.6 & 32.1 &   56$^{+ 13}_{- 16}$ &  0.0$^{+0.0}_{-0.0}$ &  0.0$^{+0.1}_{-0.0}$ &  0.1$^{+0.0}_{-0.0}$ &  0.0$^{+0.0}_{-0.0}$ & $-$ \\[2mm]
eMACSJ0324.0$+$2421 & 0.9023 &  8.48 & 14.4$\pm$5.3 & 30.8 &  787$^{+ 99}_{- 94}$ &  6.0$^{+0.4}_{-0.4}$ & 24.3$^{+1.8}_{-1.6}$ & 12.5$^{+3.4}_{-2.9}$ &  7.7$^{+3.3}_{-2.4}$ & 3 \\[2mm]
eMACSJ0502.9$-$2902 & 0.6028 &  1.25 &  6.7$\pm$1.9 & 47.5 &  753$^{+ 46}_{- 43}$ &  9.9$^{+0.4}_{-0.4}$ & 14.9$^{+0.6}_{-0.6}$ &  8.0$^{+1.0}_{-0.9}$ &  4.7$^{+0.9}_{-0.8}$ & 3 \\[2mm]
eMACSJ0804.6$+$5325 & 0.5786 &  3.81 &  5.8$\pm$2.1 &  3.9 &  888$^{+256}_{-174}$ &  9.3$^{+1.2}_{-1.2}$ & 12.7$^{+1.6}_{-1.7}$ & 10.8$^{+7.0}_{-3.8}$ &  7.5$^{+8.6}_{-3.6}$ & 4 \\[2mm]
eMACSJ0834.2$+$4524 & 0.6606 &  2.80 & 10.0$\pm$2.8 & 32.6 &  643$^{+ 32}_{- 30}$ & 14.6$^{+0.5}_{-0.5}$ & 27.5$^{+0.9}_{-1.0}$ &  6.2$^{+0.7}_{-0.6}$ &  3.2$^{+0.5}_{-0.4}$ & 4 \\[2mm]
eMACSJ0840.2$+$4421 & 0.6377 &  2.42 & 14.4$\pm$4.3 & 31.4 &  715$^{+ 28}_{- 30}$ &  8.8$^{+0.4}_{-0.3}$ & 15.2$^{+0.7}_{-0.5}$ &  7.5$^{+0.7}_{-0.6}$ &  4.2$^{+0.5}_{-0.5}$ & 2 \\[2mm]
eMACSJ0935.1$+$0614 & 0.7787 &  3.55 & 15.2$\pm$4.8 & 29.9 &  681$^{+ 59}_{- 47}$ &  9.3$^{+0.5}_{-0.5}$ & 26.3$^{+1.3}_{-1.4}$ &  8.1$^{+1.5}_{-1.1}$ &  4.3$^{+1.2}_{-0.8}$ & 3 \\[2mm]
eMACSJ1030.5$+$5132 & 0.5182 &  1.28 &  6.7$\pm$1.4 & 19.7 &  578$^{+ 22}_{- 21}$ & 12.7$^{+0.4}_{-0.4}$ & 13.3$^{+0.5}_{-0.5}$ &  4.3$^{+0.4}_{-0.3}$ &  1.9$^{+0.2}_{-0.2}$ & 1 \\[2mm]
eMACSJ1136.8$+$0005 & 0.5968 &  2.10 &  5.7$\pm$2.0 &  3.9 &  844$^{+261}_{-174}$ &  7.0$^{+0.9}_{-0.8}$ & 10.3$^{+1.4}_{-1.2}$ & 10.0$^{+7.2}_{-3.7}$ &  6.6$^{+8.2}_{-3.3}$ & 4 \\[2mm]
eMACSJ1144.2$-$2836 & 0.5070 &  5.52 & 16.5$\pm$5.2 & 30.8 &  675$^{+ 75}_{- 63}$ &  4.7$^{+0.4}_{-0.4}$ &  4.6$^{+0.4}_{-0.4}$ &  5.7$^{+1.3}_{-1.1}$ &  3.0$^{+1.1}_{-0.8}$ & 4 \\[2mm]
eMACSJ1157.9$-$1046 & 0.5570 &  2.91 &  7.6$\pm$3.4 &125.3 &  558$^{+ 17}_{- 17}$ & 10.3$^{+0.3}_{-0.3}$ & 12.9$^{+0.3}_{-0.4}$ &  4.2$^{+0.3}_{-0.3}$ &  1.8$^{+0.2}_{-0.2}$ & 4 \\[2mm]
eMACSJ1209.4$+$2640 & 0.5553 &  1.48 &  6.1$\pm$1.7 & 15.9 &  712$^{+ 67}_{- 50}$ & 11.2$^{+0.7}_{-0.6}$ & 13.9$^{+0.8}_{-0.8}$ &  6.8$^{+1.3}_{-0.9}$ &  3.8$^{+1.2}_{-0.7}$ & 3 \\[2mm]
eMACSJ1341.9$-$2442 & 0.8339 &  6.01 & 16.2$\pm$6.2 & 32.8 &  546$^{+ 69}_{- 53}$ &  3.3$^{+0.3}_{-0.3}$ & 11.0$^{+0.9}_{-0.9}$ &  5.6$^{+1.5}_{-1.1}$ &  2.4$^{+1.0}_{-0.6}$ & 4 \\[2mm]
eMACSJ1350.7$-$1055 & 0.8247 &  3.42 & 12.0$\pm$4.7 & 31.9 &   62$^{+ 10}_{- 18}$ &  0.0$^{+0.1}_{-0.0}$ &  0.0$^{+0.3}_{-0.0}$ &  0.1$^{+0.0}_{-0.0}$ &  0.0$^{+0.0}_{-0.0}$ & $-$ \\[2mm]
eMACSJ1353.7$+$4329 & 0.7364 &  1.14 &  9.0$\pm$2.5 & 40.1 &  694$^{+ 51}_{- 51}$ &  7.3$^{+0.4}_{-0.4}$ & 18.0$^{+0.9}_{-0.9}$ &  7.9$^{+1.2}_{-1.1}$ &  4.3$^{+1.0}_{-0.9}$ & 4 \\[2mm]
eMACSJ1407.0$-$0015 & 0.5520 &  3.57 &  6.0$\pm$2.0 &  3.9 &  849$^{+409}_{-291}$ &  5.6$^{+1.2}_{-1.4}$ &  6.8$^{+1.5}_{-1.7}$ & 10.0$^{+11.1}_{-5.8}$ &  6.4$^{+14.4}_{-4.6}$ & 3 \\[2mm]
eMACSJ1414.7$+$5446 & 0.6121 &  1.41 & 10.2$\pm$0.6 & 32.6 &  846$^{+ 60}_{- 52}$ & 10.9$^{+0.5}_{-0.5}$ & 17.1$^{+0.8}_{-0.7}$ & 10.2$^{+1.5}_{-1.2}$ &  6.8$^{+1.5}_{-1.2}$ & 3 \\[2mm]
eMACSJ1508.1$+$5755 & 0.5421 &  1.60 &  7.3$\pm$1.5 & 47.3 &  797$^{+ 52}_{- 55}$ &  7.7$^{+0.3}_{-0.3}$ &  9.0$^{+0.4}_{-0.4}$ &  8.3$^{+1.2}_{-1.1}$ &  5.2$^{+1.1}_{-1.0}$ & 4 \\[2mm]
eMACSJ1527.6$+$2044 & 0.6967 &  4.55 &  8.0$\pm$3.6 & 24.5 &  741$^{+ 78}_{- 59}$ &  9.4$^{+0.6}_{-0.5}$ & 20.2$^{+1.3}_{-1.1}$ &  8.7$^{+1.9}_{-1.3}$ &  5.0$^{+1.8}_{-1.1}$ & 3 \\[2mm]
eMACSJ1756.8$+$4008 & 0.5743 &  3.28 &  8.6$\pm$1.8 &126.1 &  799$^{+ 18}_{- 17}$ & 18.2$^{+0.2}_{-0.2}$ & 24.4$^{+0.3}_{-0.2}$ &  8.7$^{+0.4}_{-0.4}$ &  5.4$^{+0.4}_{-0.3}$ & 4 \\[2mm]
eMACSJ1823.1$+$7822-NE & 0.6803 &  3.02 & \textsuperscript{\textdagger}12.7$\pm$1.9 & 22.1 &  496$^{+ 41}_{- 33}$ &  5.0$^{+0.3}_{-0.3}$ &  9.9$^{+0.7}_{-0.6}$ &  3.8$^{+0.6}_{-0.5}$ &  1.5$^{+0.4}_{-0.3}$ & 1 \\[2mm]
eMACSJ1823.1$+$7822-SW & 0.6784 &  3.02 & \textsuperscript{\textdagger}12.7$\pm$1.9 & 22.1 &  534$^{+ 62}_{- 50}$ &  2.4$^{+0.2}_{-0.3}$ &  4.8$^{+0.5}_{-0.5}$ &  4.2$^{+1.0}_{-0.7}$ &  1.8$^{+0.7}_{-0.5}$ & 3 \\[2mm]
eMACSJ1831.1$+$6214 & 0.8207 &  4.15 & 18.8$\pm$2.2 & 22.0 &  718$^{+ 61}_{- 59}$ & 11.4$^{+0.6}_{-0.6}$ & 36.7$^{+1.9}_{-2.1}$ &  9.5$^{+1.7}_{-1.5}$ &  5.3$^{+1.5}_{-1.2}$ & 4 \\[2mm]
eMACSJ1852.0$+$4900 & 0.6035 &  3.90 &  6.3$\pm$1.7 & 19.8 &  596$^{+ 64}_{- 59}$ &  8.4$^{+0.6}_{-0.7}$ & 12.7$^{+1.0}_{-1.1}$ &  5.0$^{+1.1}_{-0.9}$ &  2.3$^{+0.8}_{-0.6}$ & 4 \\[2mm]
eMACSJ2026.7$-$1920 & 0.6219 &  4.14 & 10.3$\pm$3.4 & 31.8 &  541$^{+ 31}_{- 27}$ &  6.1$^{+0.3}_{-0.3}$ &  9.9$^{+0.5}_{-0.5}$ &  4.2$^{+0.5}_{-0.4}$ &  1.8$^{+0.3}_{-0.3}$ & 1 \\[2mm]
eMACSJ2220.3$-$1211 & 0.5292 &  3.78 &  5.2$\pm$2.0 & 12.4 &  934$^{+228}_{-136}$ & 10.0$^{+1.0}_{-0.9}$ & 11.0$^{+1.1}_{-1.0}$ & 11.3$^{+6.2}_{-3.1}$ &  8.2$^{+7.7}_{-3.1}$ & 2 \\[2mm]
eMACSJ2315.3$-$2128 & 0.5400 &  1.72 & 13.8$\pm$5.4 & 30.8 &  952$^{+276}_{-183}$ &  3.1$^{+0.4}_{-0.4}$ &  3.6$^{+0.4}_{-0.4}$ & 11.9$^{+7.9}_{-4.1}$ &  8.9$^{+10.2}_{-4.2}$ & 3 \\[2mm]
eMACSJ2320.9$+$2912 & 0.4960 &  7.35 &  5.5$\pm$1.6 &  9.6 &  709$^{+115}_{- 88}$ & 14.6$^{+1.3}_{-1.5}$ & 13.8$^{+1.3}_{-1.4}$ &  6.2$^{+2.2}_{-1.5}$ &  3.5$^{+2.0}_{-1.1}$ & 3 \\[2mm]
eMACSJ2327.4$-$0204 & 0.7055 &  4.62 & 14.0$\pm$4.1 &138.9 &  679$^{+ 16}_{- 16}$ & 17.8$^{+0.3}_{-0.4}$ & 39.4$^{+0.8}_{-0.8}$ &  7.3$^{+0.4}_{-0.4}$ &  3.9$^{+0.3}_{-0.3}$ & 2 \\[2mm]
\end{tabular}
\caption{Global physical characteristics of eMACS clusters from \textit{CXO}/ACIS-I data available at the time of publication; $n_{\rm H}$ values represent the Galactic neutral-hydrogen column density \citep{HI4PI2016}; all X-ray fluxes (unabsorbed) and luminosities are listed in units of $10^{-13}$ erg s$^{-1}$ cm$^{-2}$ and $10^{44}$ erg s$^{-1}$, respectively. Errors correspond to $1\sigma$ confidence. We list the properties of the two components of the double cluster eMACSJ1823 separately; note, however, that the listed RASS flux (marged with a \textdagger symbol) is the same and duplicated for both entries, since this source is not resolved in the RASS. See \citet{Ebeling2007} for a description of the morphological classification listed in the final column. An estimate of the total cluster mass within $r_{\rm 200}$ can be obtained by doubling $M_{\rm 1000}=M(<{\rm r_{1000}})$ (see Section~\ref{sec:results-cxo}).}
\label{tab:cxo}
\end{scriptsize}
\end{table*}

\subsubsection{ICM temperature and luminosity; cluster mass}

\begin{figure}
    \centering
    \includegraphics[width=0.47\textwidth]{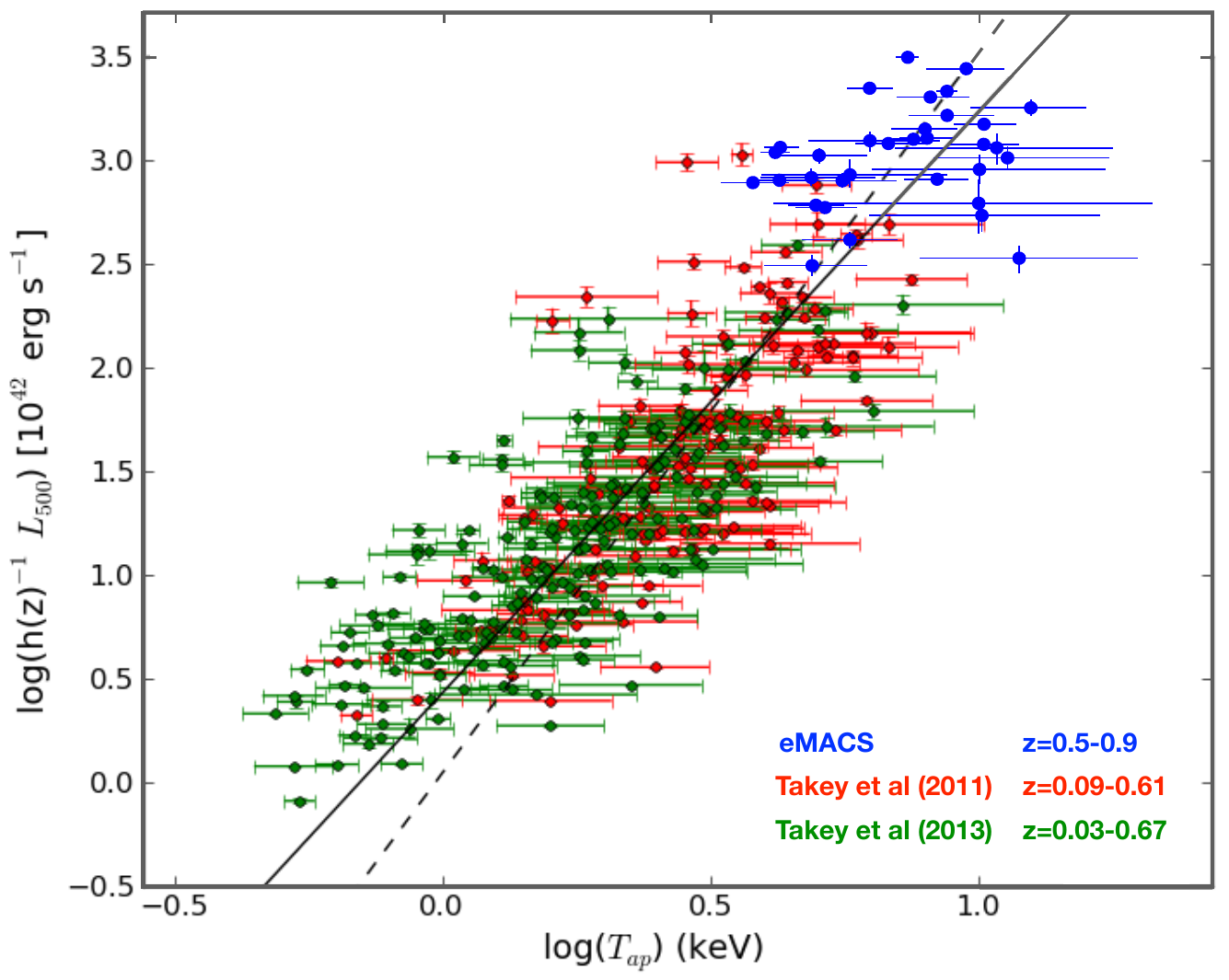}
    \caption{Bolometric luminosities and global ICM temperatures measured by us for the subsample of 35 eMACS clusters observed with {\it CXO}, overlaid on the $L_{\rm X}$--k$T$ relation determined by \citet{2013A&A...558A..75T} for a sample of 345 clusters detected with {\it XMM-Newton} \citep[reproduced from][]{2013A&A...558A..75T}. Note that, unlike Takey and collaborators, we measured both quantities within $r_{\rm 1000}$.} 
    \label{fig:lxkt}
\end{figure}

Although the global cluster luminosities and ICM temperatures presented here should be considered tentative (our analysis is unable to account for low-surface-brightness emission from the cluster outskirts, or for cool cores, shock-heated gas, and any other temperature variations), the results are in good agreement with the $L_{\rm X}$-k$T$ relation established for other X-ray selected cluster samples. As shown in Fig.~\ref{fig:lxkt}, eMACS extends the $L_{\rm X}$-k$T$ relation for galaxy clusters to higher luminosities and higher ICM temperatures than probed by existing X-ray selected cluster samples at $z>0.5$.

\begin{figure}
    \centering
    \includegraphics[width=0.48\textwidth]{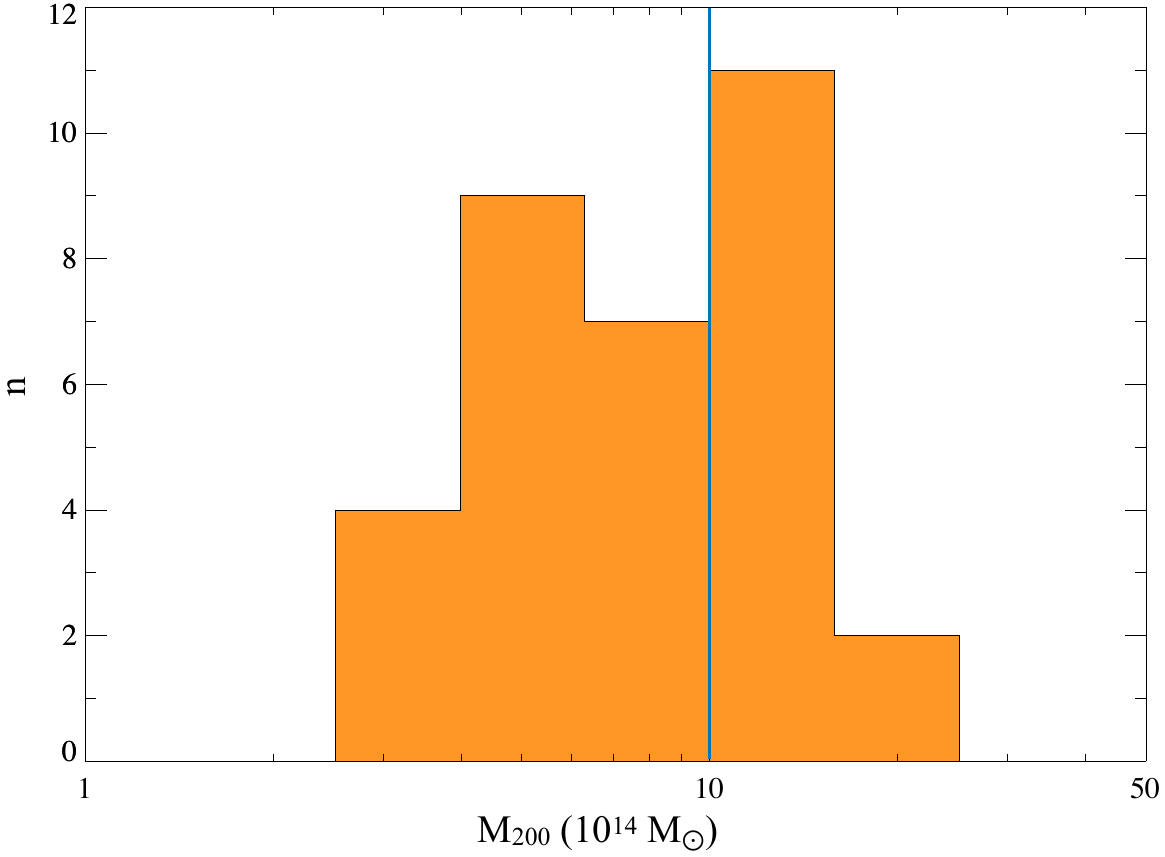}
    \caption{Distribution of total cluster masses ($M_{\rm 200}$) for the 33 eMACS clusters observed with \textit{CXO}/ACIS-I to date (see Table~\ref{tab:sample}). $M_{\rm 200}$ is estimated from $M_{\rm 1000}$ (Table~\ref{tab:cxo}) by extrapolating the universal cluster mass profile \citep[e.g.,][]{2005A&A...435....1P,2019A&A...621A..39E} from $r_{\rm 1000}$ to $r_{\rm 1000}$. The vertical line marks the  mass of $10^{15}$ M$_\odot$ referred to in Section~\ref{sec:intro} and highlighted as characteristic for the eMACS target regime in mass-redshift space (Fig.~\ref{fig:lxmz}, right). }
    \label{fig:m200} 
\end{figure}

The results of our X-ray analysis summarized in Table~\ref{tab:cxo} also unambiguously confirm that eMACS achieved its goal of probing both the redshift-X-ray luminosity and the redshift-mass regime highlighted in Fig.~\ref{fig:lxmz}. As shown in Figs.~\ref{fig:lxlx} and \ref{fig:lxkt}, the point-source corrected X-ray luminosities of eMACS clusters are of the order of $10^{45}$ erg s$^{-1}$, consistent with the eMACS target range shown in Fig.~\ref{fig:lxmz} (left). The same holds for the cluster masses. Although the existing \textit{CXO} observations allow direct measurements of the cluster mass (under the assumption of hydrostatic equilibrium) only to a radius of $r_{\rm 1000}$, $M_{\rm 200}$ can be estimated through extrapolation of the universal cluster mass profile to $r_{\rm 200}$ which approximately doubles these values \citep[e.g.,][]{2005A&A...435....1P,2019A&A...621A..39E}, yielding total masses in excess of $10^{15}$ M$_\odot$ for over one third of the subset for which \textit{CXO} data are currently available (Fig~\ref{fig:m200}). We note that this extrapolation assumes the NFW profile found to describe the mass distribution in relaxed clusters \citep{1997ApJ...490..493N}; for disturbed systems (of which eMACS found many; see below), the increase in the enclosed mass from $r_{\rm 1000}$ to $r_{\rm 200}$ would be even higher. Since our follow-up observations with \textit{CXO} targeted preferentially the most distant and most X-ray luminous clusters, the mass distribution shown in Fig~\ref{fig:m200} should, however, not be taken as representative of the eMACS sample as a whole.

\subsubsection{Relaxation state}

\begin{figure*}
    \centering
    \hspace*{2mm}\includegraphics[width=0.95\textwidth]{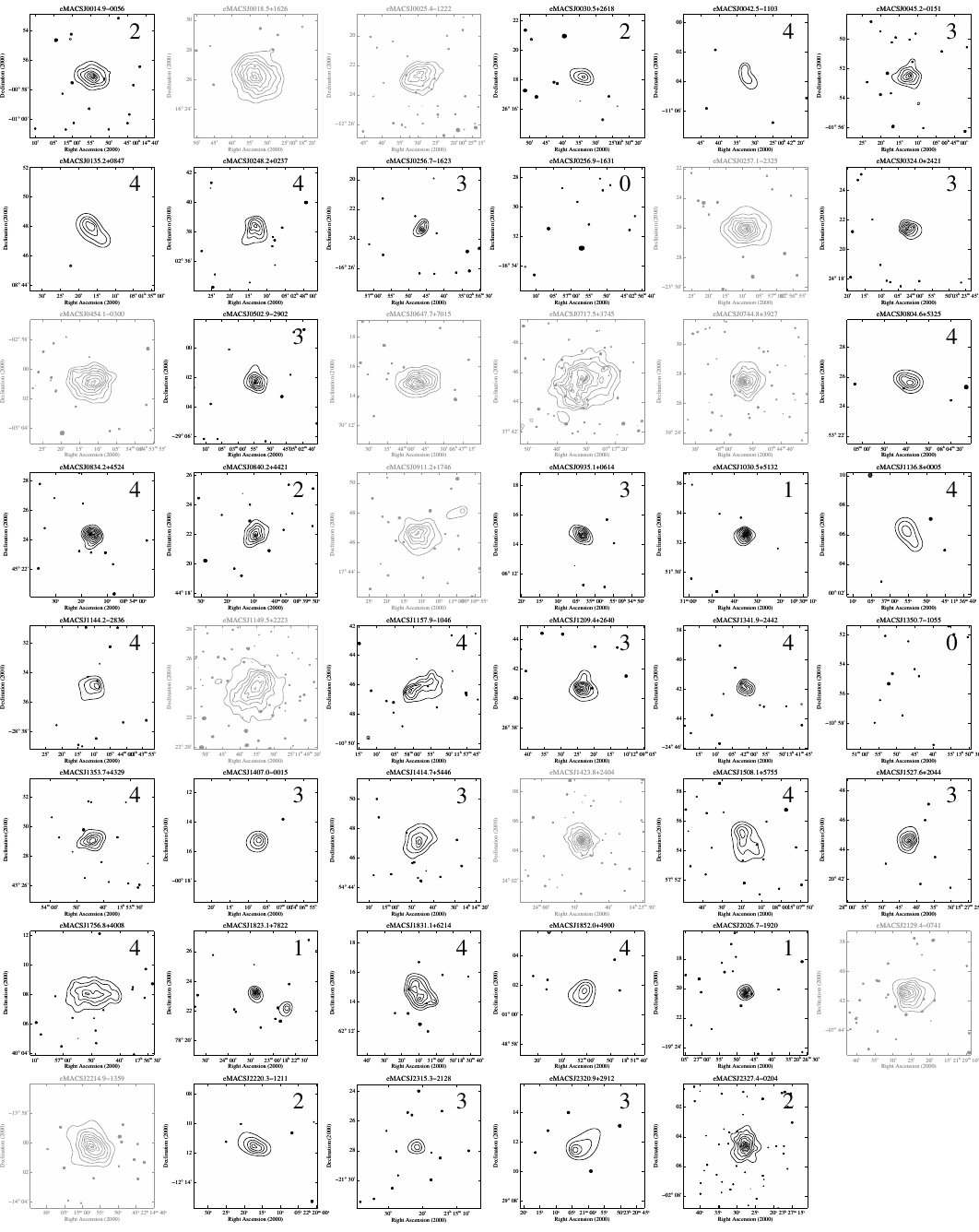}
    \caption{Contours of the adaptively smoothed X-ray surface brightness in the 0.5--7 keV band from the {\it CXO} observations described in Section~\ref{sec:obs-cxo}. Contours in this and all subsequent figures are logarithmically spaced by factors of 1.2, starting at three times the background level. The 12 clusters previously discovered in the MACS project \citep{Ebeling2007} are included only for completeness' sake and shown in gray. The morphological class assigned to each cluster is shown in the upper right corner of each panel. (Note that we again exclude the 12 MACS clusters that, by design, are also part of the eMACS sample.)
    } 
    \label{fig:cxocont}
\end{figure*}

We show in Fig.~\ref{fig:cxocont} the X-ray surface brightness for all eMACS clusters with {\it CXO}/ACIS-I data. Note the wide range of morphologies, from almost perfectly circular and strongly peaked (e.g., eMACSJ1030.5, eMACSJ2026.7) to extremely disturbed (e.g., eMACSJ1157.9, eMACSJ1508.1, eMACSJ1756.8). Also noteworthy are two fields which show no diffuse ICM emission at all, eMACSJ0256.9 and eMACSJ1350.7; we discussed these eMACS misidentification earlier in Section~\ref{sec:misid} within the larger context of contamination from X-ray point sources.

Recognizing the futility of a sophisticated ICM substructure analysis based on (typically) a few thousand photons, we adopt the simple but robust visual classification scheme of \citet{Ebeling2007} and assign each cluster a morphological classification from 1 (fully relaxed) through 4 (highly disturbed), based on X-ray and optical appearance, as well as X-ray/optical alignment. The resulting classification is indicated both in Fig.~\ref{fig:cxocont} and in Table~\ref{tab:cxo}. 

\begin{figure}
    \centering
    \hspace*{-3mm}\includegraphics[width=0.5\textwidth]{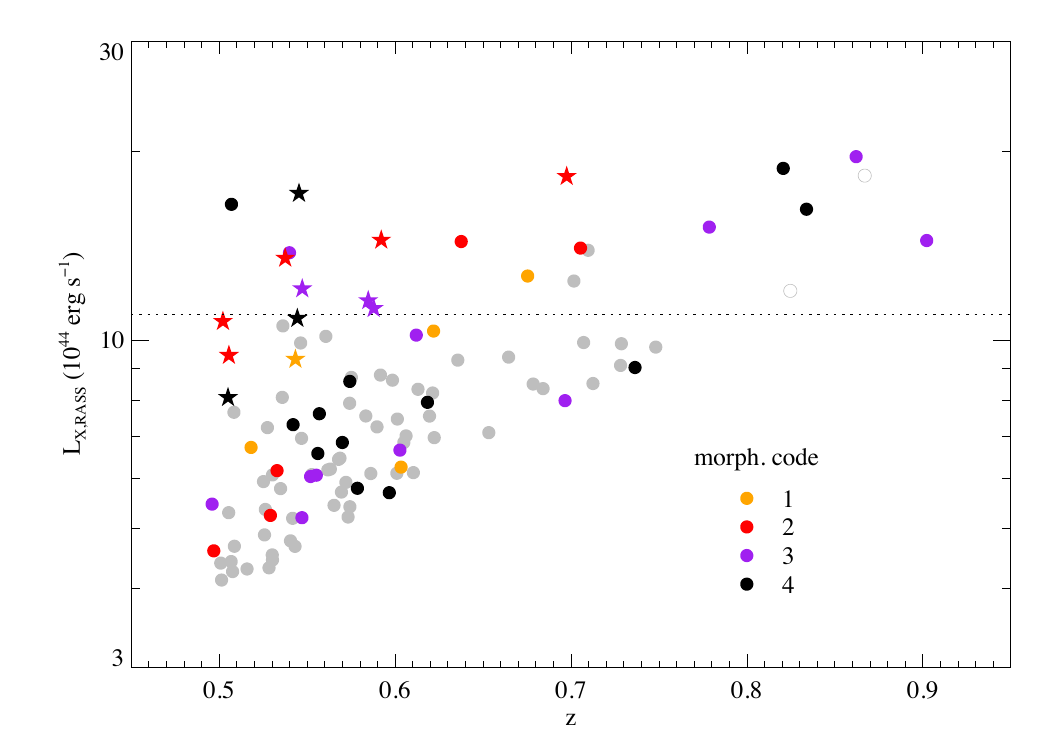}
    \caption{As Fig,~\ref{fig:lxmz} (left) but focusing solely on eMACS and highlighting the subset of eMACS clusters observed with {\it CXO}/ACIS-I (coloured symbols). The 19 clusters with $L_{\rm X,RASS}>1.1\times 10^{45}$ erg s$^{-1}$ (an additional two, shown as open circles, are the misidentifications discussed in Section~\ref{sec:misid}), form an almost complete subsample of which only two are lacking {\it CXO} data. The 12 MACS clusters at $z>0.5$ are shown as stars. Very few of the most X-ray luminous eMACS clusters exhibit a relaxed X-ray morphology.}
    \label{fig:lxz-morph}
\end{figure}

\begin{figure}
    \centering
    \includegraphics[width=0.45\textwidth]{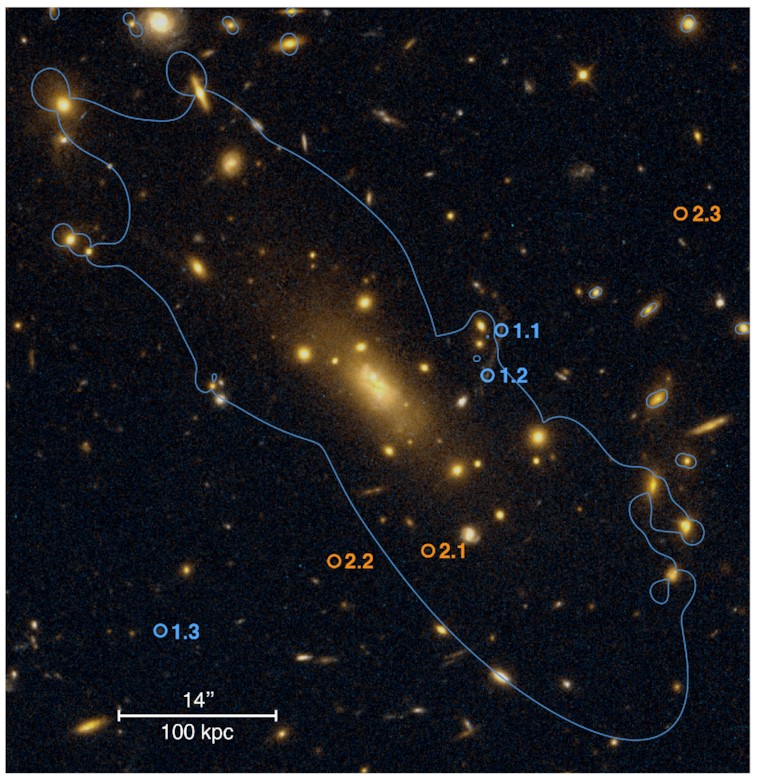}
    \caption{The core of eMACSJ0252.4$-$2100, as viewed with {\it HST}; overlaid is the critical line (for a source at $z=3.46$) from a lens model constrained by the two marked multiple-image systems \citep[figure reproduced from ][]{2021MNRAS.508.3663E}. Although no {\it CXO} data are available for eMACSJ0252 at present, the optical appearance and extreme dominance of the BCG suggest this system is very likely to have a morphological class of 1.}
    \label{fig:e0252}
\end{figure}

The low fraction of fully relaxed systems (morphological class 1) is immediately apparent, indicating that massive clusters at $z>0.5$ are undergoing vigorous growth through the accretion of surrounding structures. At face value, the dominance of disturbed systems seems to contradict claims that X-ray selected clusters samples suffer from a strong ``cool-core bias'' in favor of relaxed clusters at higher redshift \citep{2017MNRAS.468.1917R,2017ApJ...843...76A}. This claim is disputed though, as it may attribute erroneously to X-ray selection what is in fact caused by Malmquist bias: indeed, volume-complete (as opposed to flux limited) samples of X-ray selected clusters were found to be dominated not by cool-core clusters but by dynamically active, disturbed systems \citep{2017A&A...606L...4C}. As is, the data shown in Fig.~\ref{fig:cxocont} cannot be used to assess the validity of either claim, since eMACS is not volume complete and since the subset of eMACS clusters with existing {\it CXO} data is biased in favour of disturbed systems simply as a result of the scientific goals of the underlying {\it CXO} proposals. A subset with (almost) complete {\it CXO} coverage can, however, be defined  by considering only the most X-ray luminous eMACS clusters. As shown in Fig.~\ref{fig:lxz-morph}, only one of the 12 clusters in the almost volume-complete subsample of eMACS clusters with $L_{\rm X,RASS}> 1.1\times 10^{45}$ erg s$^{-1}$ is classified as fully relaxed (morphological class 1). Of the two systems without {\it CXO} data (both visible at $z\sim 0.7$ in Fig.~\ref{fig:lxz-morph}), one (eMACSJ0252.4, shown in Fig.~\ref{fig:e0252}) can be safely assumed to be highly relaxed (morphology class 1 or, at most, 2), based on the analysis of \citet{2021MNRAS.508.3663E} and the strong optical emission lines from its BCG, while the other (eMACSJ2025.5) is a very improbable candidate for the same classification, based on its optical appearance (see Fig.~B41). 

A conservative conclusion is thus that the fraction of cool-core clusters does not significantly exceed 10\% among the most X-ray luminous clusters at $z>0.5$ (2 of 19), in agreement with the findings of \citet{Mann2012}. A decrease in the fraction of relaxed systems as we approach the redshift of formation for massive clusters is not only plausible on evolutionary grounds, but also firmly expected from the impact of mergers in the rapidly evolving exponential tail of the most massive clusters. Such events boost, temporarily but dramatically, both the X-ray luminosity and the ICM temperature of the respective systems \citep{2002ApJ...577..579R}, thereby creating the opposite of a ``cool-core bias'' \citep{2007A&A...474..355F}.

\subsection{The most X-ray luminous eMACS clusters}
\label{sec:hilx}

We here provide a brief overview highlighting the ten most X-ray luminous systems\footnote{These systems fall precisely into the extreme regime of the $L_{\rm X}$-$z$ parameter space targeted by the eMACS project and highlighted in Fig.~\ref{fig:lxmz}.}  with $L_{\rm X,CXO} \geq 1.5\times 10^{45}$ erg s$^{-1}$, several of which were previously highlighted in \citet{Zalesky2020}, but note that an eleventh cluster, the massive merger eMACSJ0502.9 (Fig.~B8), 
makes the same luminosity cut, too. A visual summary of this subsample is shown in Fig.~\ref{fig:top10}; additional details for each cluster can be found in Appendix~\ref{sec:app-img}. A more comprehensive analysis and discussion of the properties of the most X-ray luminous eMACS clusters\footnote{Note that the subset of the most X-ray luminous eMACS clusters selected from RASS data  and highlighted in Fig.~\ref{fig:lxz-morph} differs from the one drawn from the {\it CXO}-based X-ray luminosities listed in Table~\ref{tab:cxo} and considered here and by Basto and collaborators. Although the X-ray fluxes recorded in the RASS and those derived from follow-up {\it CXO} observations are correlated, both photon noise and contamination from (possibly time variable) X-ray point sources introduce substantial scatter, as shown in Fig.~\ref{fig:lxlx}. Interestingly, the subset of the most X-ray luminous eMACS clusters drawn from RASS data (Fig.~\ref{fig:lxz-morph}) includes not only both of the eMACS misidentifications but also both of the most dramatic outliers from the mentioned correlation (see Fig.~\ref{fig:lxlx}).} is given by Basto et al.\ (in preparation).

\begin{figure*}
    \includegraphics[width=0.9\textwidth]{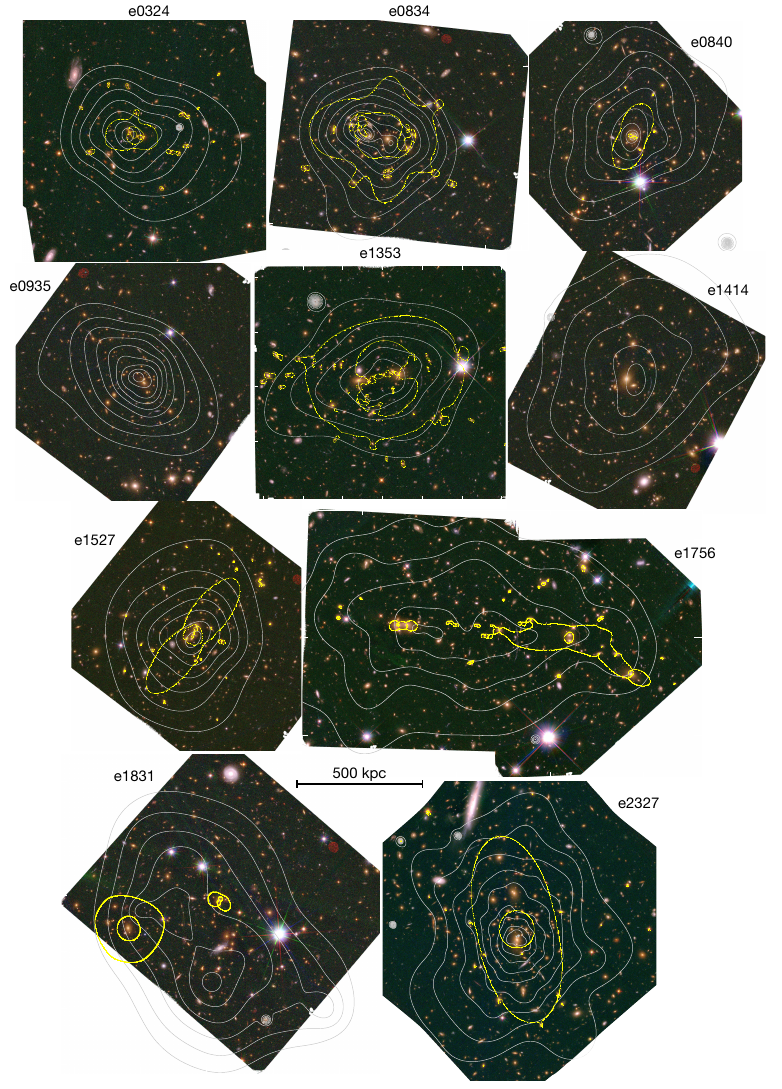}
    \caption{{\it HST} images of the ten most X-ray luminous eMACS clusters (based on point-source corrected fluxes measured from {\it CXO} observations). Overlaid in gray and yellow are isointensity contours of the adaptively smoothed X-ray emission and the critical line for gravitational lensing of a source at a fiducial redshift of $z=2$, respectively.}
    \label{fig:top10}
\end{figure*}

\subsubsection{eMACSJ0324.0$+$2421}
\label{sec:e0324}

eMACSJ0324.0 deserves special mention for being the only galaxy cluster at $z>0.9$ not merely detected, but in fact discovered in the RASS. Fig.~B7 
summarizes the observational evidence for this system (see also Fig.~\ref{fig:top10}). Two almost equally luminous BCG candidates, a disturbed ICM morphology that nonetheless features only a single X-ray peak, a very high global gas temperature of $(13\pm 3)$ keV, a very large Einstein radius and very high enclosed mass of $M(<240\,{\rm kpc})=4.6\times 10^{14}$ M$_\odot$ derived from a lens model constrained by two spectroscopically confirmed multiple-image systems (see Section~\ref{sec:results-SL}), and an exceptionally high velocity dispersion of $1520^{+150}_{-220}$ km s$^{-1}$ all suggest that eMACSJ0324.0 is an actively growing, massive cluster, observed in the post-collision phase (but still close to core passage) of a merger that proceeds largely along our line of sight. 

\subsubsection{eMACSJ0834.2$+$4524}

The appearance of eMACSJ0834.2 (Fig.~B10, 
see also Fig.~\ref{fig:top10} and Section~\ref{sec:results-SL}) provides ample and unambiguous evidence of an ongoing collision of two components of comparable mass. Although our strong-lensing mass model is unable to resolve multiple cluster-scale components, two well separated peaks are observed in the X-ray surface brightness. The offsets of these peaks in the ICM density from the BCGs of the two subclusters are indicative of a collision axis that lies at least partly in the plane of the sky. A line-of-sight component of the merger-induced peculiar velocities is, however, also likely, given the very high velocity dispersion of  $1590^{+140}_{-200}$ km s$^{-1}$. We note the quite modest ICM gas temperature of ($6.2\pm 0.7$) keV; such a low global temperature could be indicative of the presence of cool cores (the eastern component of e0834 exhibits a noticeably compact X-ray peak) and points to the absence of large amounts of shock-heated gas. We thus speculate that we either observe the merger well before first core passage or that the collision involves a large impact parameter, thereby leaving the cluster cores relatively unperturbed, even after several core passages.

Owing to the system's highly disturbed morphology, our best-fitting lens model is characterized by a large Einstein radius and a high enclosed mass of $M(<220\,{\rm kpc})=3.2\times 10^{14}$ M$_\odot$.

\subsubsection{eMACSJ0840.2$+$4421}

While not fully relaxed (note the absence of a pronounced X-ray peak and the non-circular geometry apparent in Fig.~B11 
and Fig.~\ref{fig:top10}), eMACSJ0840.2 is one of the least disturbed clusters in this subset of the most X-ray luminous eMACS clusters. Both the excellent X-ray/optical alignment and the substantial but (for a cluster of this size) modest ICM temperature of $(7.5\pm 0.7)$ keV suggest that eMACSJ0840.2 is observed well past its most recent merger event. Two spectroscopically confirmed multiple-image systems constrain the mass in the cluster core to $M(<80\,{\rm kpc})=4\times 10^{13}$ M$_\odot$. Since the velocity dispersion of  $(1320\pm 120)$ km $s^{-1}$ (based on 39 redshifts) of eMACSJ0840.2 is unlikely to be inflated by ongoing merging activity along the line of sight, we can use the virial scaling relation to compute an estimate of the dynamical mass of the system \citep{2008ApJ...672..122E,2013ApJ...772...47S}. The result, $M_{\rm dyn}=(2.6\pm 0.1)\times 10^{15}$ M$_\odot$, places  
eMACSJ0840.2 in the top tier of massive clusters at any redshift (see Fig~\ref{fig:lxmz}).

\subsubsection{eMACSJ0935.1$+$0614}

Exceptionally X-ray luminous, in particular given its high redshift of $z=0.78$, eMACSJ0935.1 is one of two eMACS clusters in our Top Ten list that do not currently have a lens model. The {\it CXO} data for this system show neither a compact core nor pronounced substructure in the ICM (Fig.~\ref{fig:top10}); however, markedly elliptical X-ray surface-brightness contours, as well as a clear offset of the X-ray peak from the BCG, still identify eMACSJ0935.1 as a dynamically disturbed cluster (Fig.~B14). 
The cluster core features neither an extremely luminous BCG nor signs of strong gravitational lensing, suggesting a shallow mass profile that may be the result of mergers of several clusters of moderate mass.

\subsubsection{eMACSJ1353.7$+$4329}

eMACSJ1353.7, previously discussed in Section~\ref{sec:results-SL}, has one of the highest velocity dispersions of any eMACS cluster (see Section~\ref{sec:results-radvel}); however, closer inspection of all available data (Fig.~B27) 
suggests that the measured value of $1600^{+110}_{-160}$ km s$^{-1}$ is inflated by an infalling group. The cluster exhibits the textbook signature (a single X-ray peak between two BCGs, Fig.~\ref{fig:top10}) of what is probably a late-stage merger close to the plane of the sky. The two BCGs are at essentially the same redshift ($z=0.736$ and $z=0.739$); the looser group of galaxies to the East, however, features redshifts close to $z=0.75$, adding a pronounced and narrow peak to the radial-velocity histogram that we interpret as evidence of infall of a foreground group of galaxies. The lens model for eMACSJ1353.7, well constrained by three multiple-image systems, one of which is spectroscopically confirmed, yields a very high enclosed mass of $M(<250\,{\rm kpc})=4.3\times 10^{14}$ M$_\odot$ for this powerful gravitational lens. The not exceptionally high ICM temperature of ($7.9\pm 1.2$)\,kev measured by us suggests that this ongoing merger is observed well after the temporary temperature boost triggered by the initial collision. 

\subsubsection{eMACSJ1414.7$+$5446}

Like for eMACSJ1353.7, the extreme velocity dispersion measured for eMACSJ1414.7 ($1700^{+180}_{-210}$ km s$^{-1}$) is likely inflated by peculiar velocities from an ongoing merger proceeding along an axis that lies close to our line of sight. Although strong-lensing features are clearly discernible near the cluster core (Fig.~B28), 
we currently have no spectroscopically confirmed multiple-image system that could anchor a lens model for eMACSJ1414.7 The fact that the X-ray peak of eMACSJ1414.7 is offset to the West of the BCG (Fig.~\ref{fig:top10}), with no obvious second cluster core apparent in this direction in the galaxy distribution, might indicate a collision at a significant impact parameter. The gap in the radial-velocity histogram and the high global ICM temperature of 10\,keV suggest that the eMACSJ1414.7 merger is either observed shortly after the first core passage or prior to a second collision with an infalling group of galaxies.

\subsubsection{eMACSJ1527.6$+$2044}
eMACSJ1527.6 is another eMACS cluster with an exceptionally high velocity dispersion, $1580^{+190}_{-230}$ km s$^{-1}$, but, in contrast to eMACSJ1353.7 and eMACSJ1414.7, without clear signs of line-of-sight substructure in the radial-velocity histogram (Fig.~B32). 
Since the ICM temperature of ($8.7^{+1.9}_{-1.3}$)\,keV is above average, even for massive clusters, we tentatively conclude that eMACSJ1527.6 is an active merger observed during or shortly after first core passage. Although we cannot rule out that the absence of multiple peaks in the X-ray surface brightness distribution (Fig.~\ref{fig:top10}) is primarily the result of the very short {\it CXO}/ACIS-I exposure time of only 25\,ks, the presence of several BCG candidates in close proximity in the cluster core suggests, in conjunction with the very velocity dispersion, that the merging clusters are on trajectories that would make them difficult to separate even in a much deeper X-ray observation. Strong-lensing constraints from three multiple-image systems (one of them spectroscopically confirmed) yield a lens model of eMACSJ1527.6 that is markedly more elongated than the ICM distribution (Fig.~B32). 
A discrepancy between the collisonal and collisionless mass components of this kind is a textbook characteristic of ongoing merger activity.

\subsubsection{eMACSJ1756.8$+$4008}
\label{sec:e1756}
eMACSJ1756.8 is an obvious large-separation merger and the most powerful cluster lens in the eMACS sample according to \textsc{AStroLens}. Although we invested substantial resources into spectroscopic follow-up observations of the strong-lensing features identified in this system (Figs.~\ref{fig:sl} and B35), 
the lack of multiple-image systems near the eastern cluster core prevents us from constraining the mass in the presumably dominant part of this highly extended system (see also Section~\ref{sec:results-SL}. As a result, the Einstein radius and enclosed mass listed for eMACSJ1756.8 in Table~\ref{tab:re} should be considered firm lower limits. A very high total mass is expected also from the X-ray estimate of  $5.4\times 10^{14}$ M$_\odot$ for the total cluster mass (Table~\ref{tab:cxo}) and the very high velocity dispersion of $1400^{+80}_{-90}$ km s$^{-1}$, which is based on 117 spectroscopic redshifts, the largest number available for any eMACS cluster. The wide (optical) separation of the three cluster cores in the plane of the sky (800 kpc altogether), the location of the X-ray peak between two of them, and the high ICM temperature of ($8.7\pm 0.4$) keV identify eMACSJ1756.8 unambiguously as a linear, post-collision merger of as many as three galaxy clusters.

\subsubsection{eMACSJ1831.1$+$6214}

\noindent
Although narrowly beaten to the title of ``most X-ray luminous eMACS cluster'' by eMACSJ2327.4 (see below), eMACSJ1831.1 ($L_{\rm X,CXO}=(3.7\pm 0.2)\times 10^{45}$ erg s$^{-1}$) stands out as one of only a handful exceptionally massive clusters known at $z>0.8$. As is apparent from Fig.~B38 
eMACSJ1831.1 (not to be confused with eMACSJ1831.9, Fig.~B39) 
is a highly complex system that consists of at least two components, as evinced by its very disturbed X-ray morphology and the presence of several optical cluster cores (Fig.~\ref{fig:top10}). Both X-ray peaks are shallow and far from the closest BCG candidate, indicative of a pronounced segregation of collisional and non-collisional matter that is only observed in active mergers. Like eMACSJ1756.8 (Sections~\ref{sec:results-SL} and \ref{sec:e1756}), eMACSJ1831.1 lacks prominent strong-lensing features near two of its apparent optical cores, causing the associated cluster-scale mass concentrations to be missing from the \textsc{Lenstool} model (the respective \textsc{Lenstool} results are marked as lower limits in Table~\ref{tab:re}). Spectroscopic confirmation of a faint straight arc that appears to mark a saddle point in the potential close to the NE X-ray peak (cf.~Fig.~B38) 
would greatly improve the current, incomplete lens model of this exceptional system. The complexity of eMACSJ1831.1 ($z=0.8207$) is reflected also in redshift space, where the discovery of a superimposed background cluster at $z=0.858$ suggests that the system is part of an even more extended large-scale structure.

\subsubsection{eMACSJ2327.4$-$0204}

The most X-ray luminous eMACS cluster, eMACSJ2327.4 (Fig.~B45), 
was known previously and published as RCS2\,J2327 by \citet{2011AJ....141...94G}. For completeness' sake we show in 
Fig.~\ref{fig:top10} (and in Fig.~B45) 
the archival {\it CXO} data and a simple lens model based on the multiple-image systems at $z=1.42$ and $z=2.98$ (Fig.~\ref{fig:sl}) identified by \citet{2015ApJ...814...21S}. Since RCS2\,J2327 is a well known powerful gravitational lens, we refer the reader to the literature for details of its physical properties.

\subsection{Cluster galaxy properties}
\label{sec:results-gals}

The spectroscopic follow-up observations of galaxies in the fields of eMACS (candidate) clusters (see Section~\ref{sec:obs-spec}) served multiple purposes, the most fundamental of which was the confirmation of the cluster ID and the establishment of a cluster redshift above the $z=0.5$ threshold set for eMACS. Subsequent spectroscopic observations targeted not only the clusters' galaxy population in general (to measure velocity dispersions and search for substructure along the line of sight) but also specific galaxies of interest, such as the brightest cluster galaxies (BCGs); possible cases of ongoing ram-pressure stripping of infalling galaxies; potential AGN within (or superimposed onto) the cluster flagged as X-ray point sources; and of course strong-lensing features (i.e., highly amplified background galaxies) whose redshifts are crucial to constrain the cluster mass distribution (Section~\ref{sec:SL})

Although an exhaustive discussion of the findings of all these studies is beyond the scope of this paper, we present brief summaries and highlights in the following sections.

\subsubsection{Velocity dispersion and line-of-sight substructure}
\label{sec:results-radvel}

\begin{figure*}
    \centering
    \hspace*{-2mm}\includegraphics[width=0.98\textwidth]{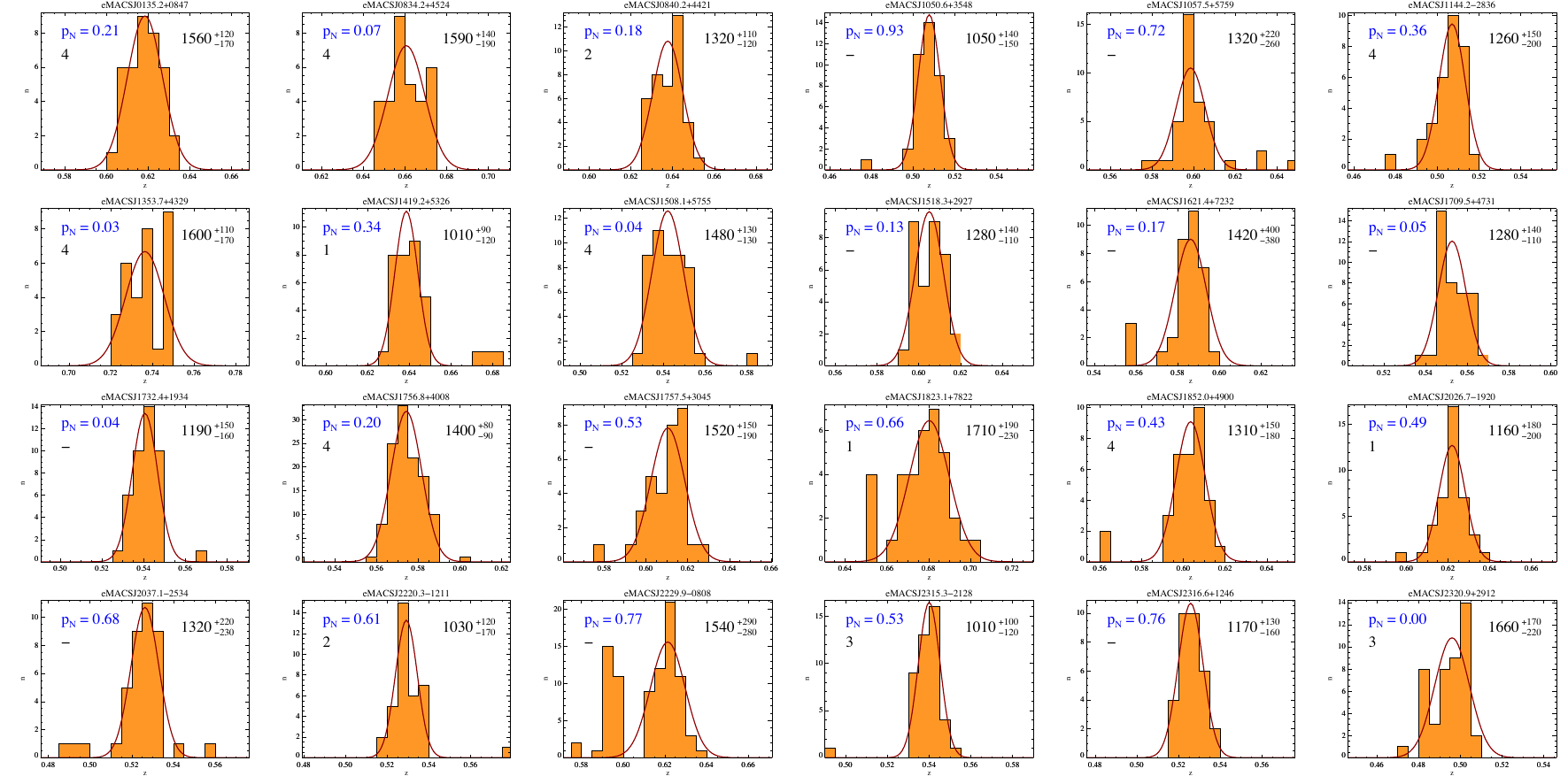}
    \caption{Histograms of the radial velocities for eMACS clusters with at least 30 spectroscopic redshifts. We list for each cluster the velocity dispersion in units of km s$^{-1}$ and, in the upper left corner of each panel, the probability of the redshift histogram within $\delta z = \pm 0.02$ of the systemic cluster redshift being drawn from a Gaussian distribution, as well as the morphological class of each cluster from Fig.~\ref{fig:cxocont}, where available.} 
    \label{fig:velhist}
\end{figure*}

While a robust systemic cluster redshift can be derived from a handful of concordant redshifts, a credible velocity dispersion requires much better statistics and should ideally also be computed from a stellar-mass-limited and spatially complete sample, in order to avoid biases introduced by an over-representation of galaxies near the cluster core or, conversely, the cluster outskirts. Since our selection of galaxies for spectroscopic follow-up was largely driven by technical constraints (spectrograph characteristics) and our desire to maximize efficiency by simultaneously advancing all science goals served by these observations, the set of cluster galaxies with spectroscopic redshifts is statistically not well defined. Specifically, in clusters with few spectroscopic redshifts (fewer than 20) the dense cluster cores are likely to be undersampled, leading to potentially systematically low velocity dispersions.

We show in Fig.~\ref{fig:velhist} the radial-velocity histograms of the 24 eMACS clusters for which at least 30 spectroscopic redshifts have been secured. At present eMACSJ2229.9 is the only eMACS cluster for which multiple components along the line of sight can be clearly resolved with the radial-velocity data in hand. Fig.~B42 
summarizes our knowledge of this system, a superposition of two clusters at $z=0.595$ and $z=0.625$. No {\it CXO} data have been obtained for this system to date; however, the presence of RASS\,6135 \citep{2017MNRAS.466..812V}, a luminous QSO that is in fact a member of the $z=0.625$ component of this double cluster, suggests a significant point-source contribution to the X-ray luminosity recorded in the RASS. A prominent fold arc at $z=3.375$ (dubbed ``The Scream'' by us, in homage to Edvard Munch's famous painting) with an obvious counterimage to the north, together with a second triple image at $z=5.166$, constrain a highly elongated \textsc{Lenstool} model\footnote{Our strong-lensing analysis supersedes the one presented by \citet{2021MNRAS.506.1595G} which is based on an erroneous redshift for The Scream.} that, however, does not fully reflect the mass distribution of both cluster components acting in projection along our line of sight (Section~\ref{sec:results-SL} and Table~\ref{tab:re}).

While the statistics of the radial-velocity data shown in Fig.~\ref{fig:velhist} are too poor to clearly reveal line-of-sight substructure in any other eMACS cluster, the probability \citep[as estimated by the Shapiro-Wilk statistic;][]{10.1093/biomet/52.3-4.591} of the observed set of redshifts having been drawn from a normal distribution is a useful diagnostic to complement the morphological classification based on a cluster's optical and X-ray appearance in projection on the sky (Section~\ref{sec:results-cxo} and Fig.~\ref{fig:cxocont}). As can be seen in Fig.~\ref{fig:velhist}, a disturbed morphology in the plane of the sky is not strongly correlated with significant substructure also along the line of sight: of the nine eMACS clusters in morphological class 3 or 4, only three have radial-velocity distributions that fail the Shapiro-Wilk normality test at greater than 95\% confidence ($p_{\rm N}<0.05$). Conversely, however, \textit{all} eMACS clusters that fail the test and have been observed with \textit{CXO} (eMACSJ1353.7, eMACSJ1508.1, and eMACSJ2320.9) are classified as heavily disturbed based on the X-ray / optical apperance, and the remaining two with $p_{\rm N}<0.05$ that have no \textit{CXO} data as of yet (eMACSJ1709.5, Fig.~B33, 
and eMACSJ1732.4, Fig.~B34) 
show the same highly complex galaxy distribution associated with a morphological classification of 3 or 4. We tentatively conclude that the collisions in progress in these systems proceed along axes that are oblique with respect to both our line of sight and the plane of the sky. For the remaining eMACS clusters in morphology class 3 and 4, the good statistical agreement of the observed radial-velocity distributions with a Gaussian suggests that the respective merger activity occurs predominantly in the plane of the sky. These numbers are roughly consistent with isotropy in the orientation of the merger axis. Interestingly, the complementary (if small) subsample of the five eMACS clusters with at least 30 spectroscopically confirmed cluster members and a relaxed or only mildly disturbed morphology (morphology class 1 or 2) passes the Shapiro-Wilk normality test without exception. 

Although encouraging, our insights into line-of-sight substructure remain qualitative even for the small subset of eMACS clusters targeted most heavily in our spectroscopic follow-up observations. Much more extensive redshift surveys of almost all eMACS clusters and their surroundings will be required to map the environment in which eMACS clusters are embedded and the pathways along which they accrete the matter that fuels their growth.

\subsubsection{Large-scale structure}
\label{sec:results-lss}

On much larger scales than typically probed by our spectroscopic follow-up observations of individual clusters (see Section~\ref{sec:obs-spec}), a glimpse of the cosmic web \citep{1996Natur.380..603B} is ironically provided with assistance from one of the non-clusters in the eMACS sample, eMACSJ0256.9. Although no diffuse X-ray emission is detected from eMACSJ0256.9 in our {\it CXO} follow-up observation (see Section~\ref{sec:misid}), an overdensity of galaxies is clearly discernible both in our {\it HST} images and in the histogram of radial velocities of galaxies in this field (Fig.~B6). 
We conclude that eMACSJ0256.9 is likely a filamentary structure aligned with our line of sight that masquerades as a three-dimensional overdensity only in this projection. Although eMACSJ0256.9 is thus not a three-dimensionally collapsed mass concentration, it appears to be a cluster in the process of formation, as indicated by the high velocity dispersion of $1170^{+140}_{-240}$ km s$^{-1}$ measured for galaxies in this structure. The presence of an actively evolving filament is of interest in the context of interconnected large-scale structure since eMACSJ0256.9$-$1631 ($z=0.867$) is only 6 Mpc (13\arcmin) away in projection from eMACSJ0256.7$-$1623 ($z=0.862$), an intrinsically massive cluster (Fig.~B5). 
This separation amounts to only about three times the virial radius of eMACSJ0256.7, rendering matter overdensities within the putative massive filament candidates for accretion onto the node of the cosmic web marked by eMACSJ0256.7 within the next few Gyr.

While (after evaluation of the {\it CXO} data) eMACSJ0256.9$-$1631 and eMACSJ0256.7$-$1623 remain a pair in the eMACS catalogue only by virtue of the commonality of a high velocity dispersion, eMACSJ1823.1 ($z=0.680$, Fig.~B37) 
is a true cluster pair, and the only obvious pre-collision double cluster in our sample. Separated by 865\,kpc in projection, both components appear dynamically relaxed, with the dominant north-eastern cluster exhibiting evidence of a cool core. Although the radial-velocity histograms show no sign of substructure along the line of sight (Fig.~B37), 
the very high velocity dispersion of the NE component ($\sigma > 1500$ km s$^{-1}$) suggests peculiar velocities (bulk motions) along the line of sight, a notion that is supported by the presence of a nearby foreground structure at $z=0.65$. eMACSJ1823.1 thus marks a node of the cosmic web that might be actively accreting matter from orthogonal directions.

\begin{figure*}
    \centering
    \includegraphics[width=0.392\textwidth]{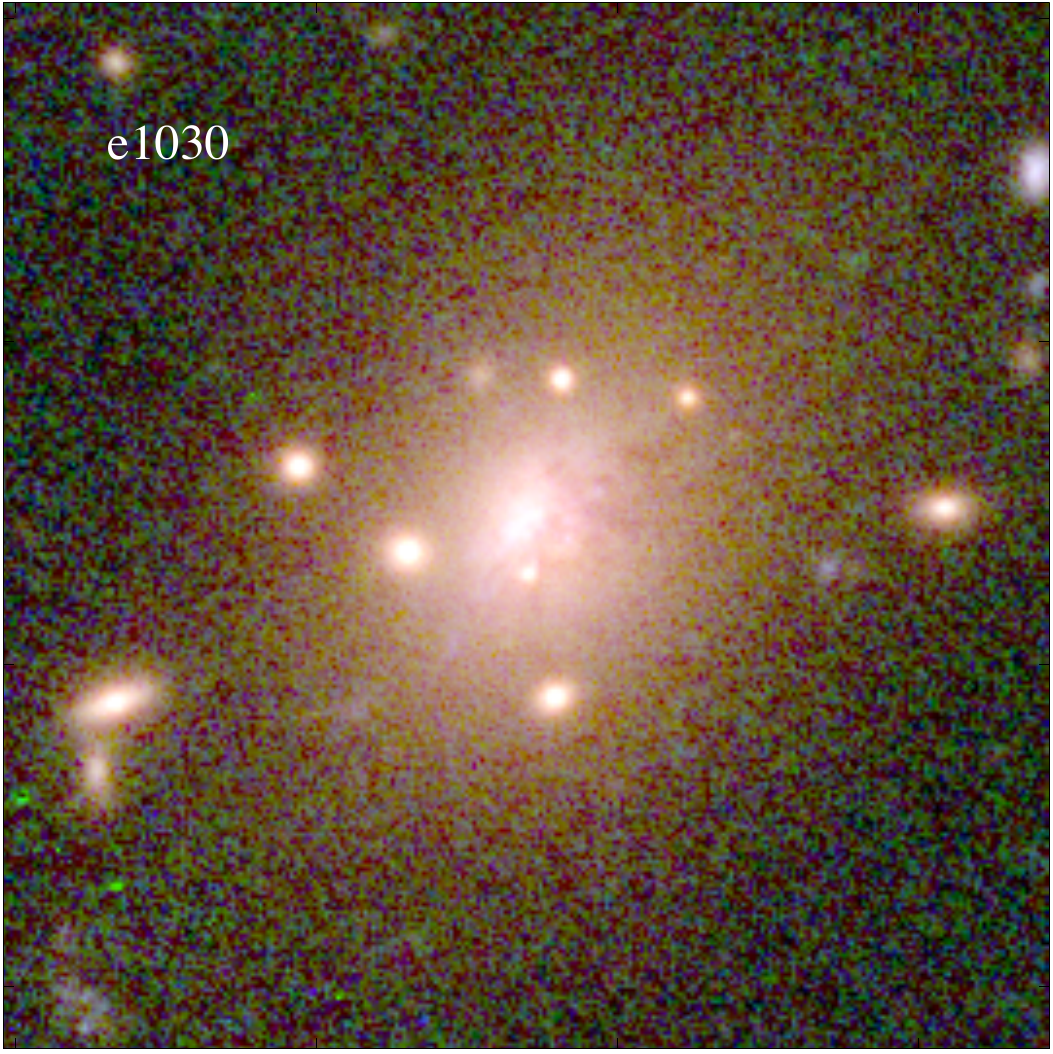}
    \includegraphics[width=0.591\textwidth]{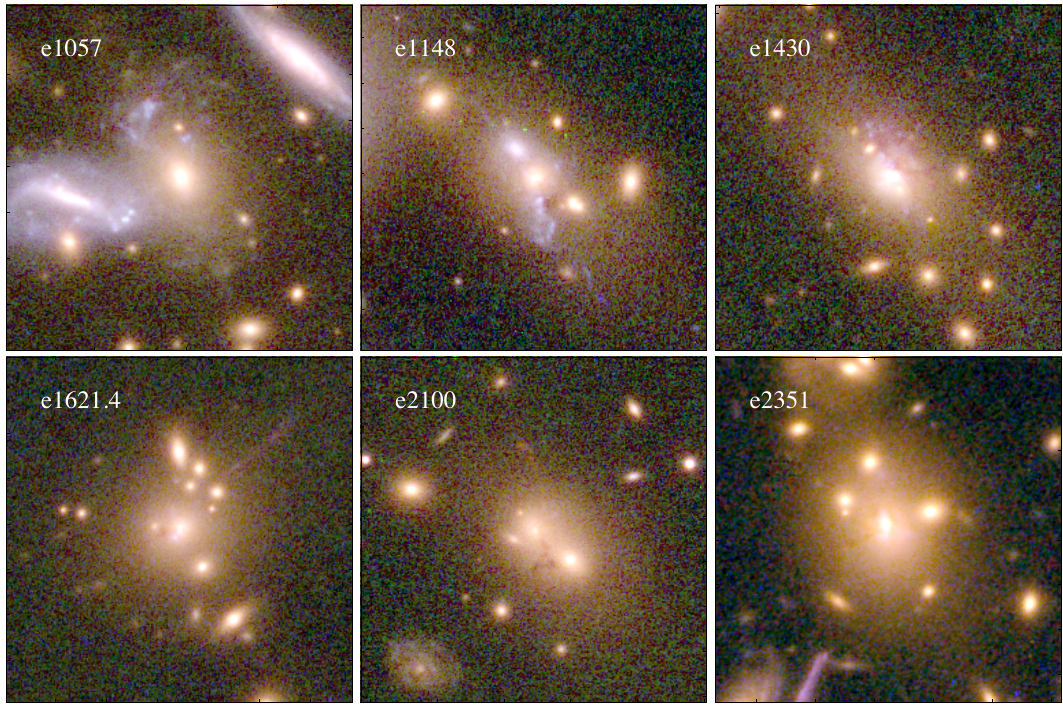}
    \caption{Close-up view (100\,kpc on a side) of notable eMACS BCGs, as seen with {\it HST}. Evidence of dust obscuration is found in the majority of these galaxies. eMACSJ1057.5 and  eMACSJ1148.0 show extended blue filaments to the north-east and south of the BCG, respectively, comparable to systems at lower redshift \citep{2010ApJ...719.1619O}. All are detected in either our JVLA radio observations or VLASS data.}
    \label{fig:bcgs}
\end{figure*}

\subsubsection{Brightest Cluster Galaxies}
\label{sec:bcg}

All X-ray selected samples of galaxy clusters contain clusters dominated by a BCG, these frequently display strong optical emission lines, star formation, and an associated reservoir of cold molecular gas \citep[e.g.,][]{1999MNRAS.306..857C,2016MNRAS.461..560G}. As demonstrated in the MUSE study of eMACSJ0252.4 \citep[][Fig.~\ref{fig:e0252}]{2021MNRAS.508.3663E}, such ``active'' BCGs are seen out to $z\sim 0.7$ and stand out in the optical regime as bluer than the expected colour of an evolved stellar population.

Our spectroscopic follow-up covers all BCGs in eMACS apart from those in eMACSJ0934.6, which has a published SDSS spectrum, and eMACSJ1852.0 which at present has no publicly available spectrum. For 17 of the 110 BCGs in the sample these spectra show significant [O{\sc ii}]$\lambda$3727 line emission, indicating star formation comparable to that seen in the most active BCGs at lower redshifts \citep{2016MNRAS.461..560G}, where H$\alpha$ line emission is typically used to gauge this activity. If we take the presence of an active BCG as an indicator of cluster relaxation, this spectroscopic estimate of the fraction of cool-core clusters of (15$\pm$4)\% is consistent with the percentage estimated from X-ray imaging (approximately 10\%) in Section~\ref{sec:results-cxo}. While the fraction of cool-core clusters is often found to be higher at lower redshifts, these more local studies benefit from both the accessibility of the H$\alpha$ line and more extensive X-ray observations. 

Radio emission from 28 BCGs in our sample is detected at $>5\sigma$ or 150$\mu$Jy in the JVLA observations at 5\,GHz. With 66 systems having been observed, this number corresponds to a detection rate of (42$\pm$8)\%, comparable to the detection rate at the same observed frequency reported by \citet{2015MNRAS.453.1201H} for the most X-ray luminous clusters at $z<0.3$. The range in radio spectral index and the mean radio spectral index of $\alpha=-0.95$, computed using available RACS, VLASS, TGSS, and LOFAR data for the detected sources, are also comparable to those found in previous studies. Of the remaining 45 clusters without a JVLA observation, 11 (i.e., 24\%) have a VLASS detection at 3\,GHz at the higher flux density limit of 0.5\,mJy reached in that all-sky survey, suggesting a comparable radio-power distribution for the sample as a whole.

Several BCGs in the eMACS sample in addition to eMACSJ0252.4 \citep[][Fig.~\ref{fig:e0252}]{2021MNRAS.508.3663E} are notable in their optical morphology. Specifically, eMACS1030.5 (Figs.~\ref{fig:bcgs}, B17 
harbors a spectacular case of an active BCG and is the brightest radio source in the eMACS sample, with properties comparable to those of compact, core-dominated sources such as 4C$+$55.16 \citep{2015MNRAS.453.1223H,2022MNRAS.509.2869R,2022A&A...668A..65T} and MACSJ0242.5 \citep{2023MNRAS.522.1118A}. Signs of active star formation, dust or multiple cores are discernible in the {\it HST} images of other eMACS BCGs, demonstrating that ``active'' BCGs are common at $z>0.5$; Fig.~\ref{fig:bcgs} shows 100\,kpc $\times$ 100\,kpc cutouts from our {\it HST} images for several examples.

\subsubsection{Ram-pressure Stripping}
\label{sec:rps}

In addition to being prime targets for investigations of phenomena and physics operating on scales of hundreds of kpc, massive galaxy clusters provide a high-density environment that shapes the evolution of the constituent galaxies on kpc scales. Two areas of galaxy evolution that are of particular interest are the interactions of BCGs and their active nuclei with the surrounding ICM (see Section~\ref{sec:bcg}) and the potential rapid morphological transformation of late-type field galaxies into early-type cluster members through ram-pressure stripping by the ICM \citep{1972ApJ...176....1G}. 

\begin{figure*}
    \centering
    \hspace*{2mm}\includegraphics[width=0.98\textwidth]{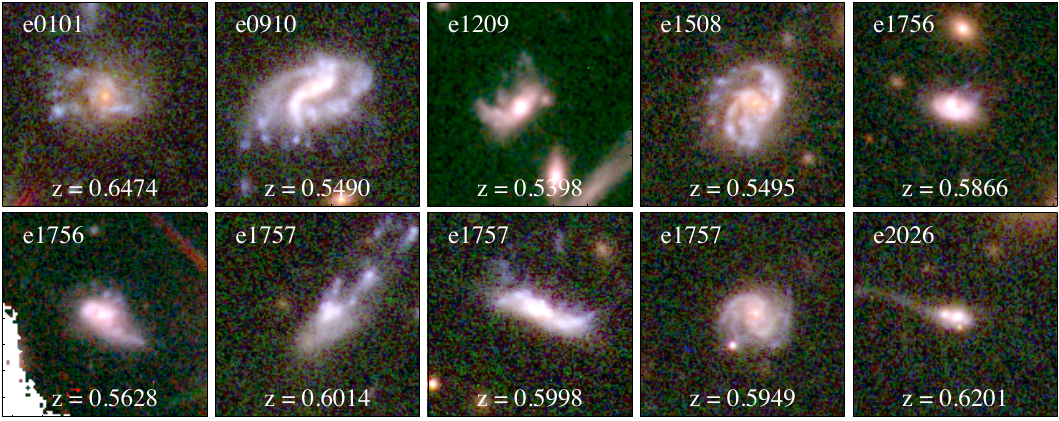}
    \caption{{\it HST} images (50 kpc on the side) of spectroscopically confirmed cluster members showing evidence of ram-pressure stripping by the ICM. Shortened names in the upper left corner of each panel indicate the respective host cluster. 
    \label{fig:jfgpanel}
    }
\end{figure*}

While the textbook ``jellyfish'' appearance of galaxies undergoing ram-pressure stripping \citep[e.g.,][]{2007MNRAS.376..157C,2012ApJ...750L..23O,ebeling2014} is less readily discernible at the high redshifts of eMACS clusters, the superb angular resolution of {\it HST} reveals many promising cases in the eMACS sample. We show in Fig.~\ref{fig:jfgpanel} examples of spectroscopically confirmed cluster members likely experiencing ram-pressure stripping. Note that we include spirals with an almost undisturbed morphology since, in the environment encountered within massive clusters, essentially all late-type galaxies are expected to be undergoing ram-pressure stripping to some degree and will ultimately be stripped of the majority of their atomic and molecular gas. Many other late-type galaxies with similarly intriguing morphology are found in {\it HST} images of eMACS clusters and await spectroscopic follow-up observations to establish cluster membership.

Although their faintness and small angular size renders these galaxies challenging targets for in-depth follow-up studies of their internal dynamics, the excitation state of the intragalactic medium, or the effects of ram-pressure stripping on central black holes \citep[all topics at the focus of ``jellyfish'' studies in the more nearby Universe out to about $z=0.3$: ][and references therein]{2019ApJ...887..158K}, the mere detection of debris trails or ICM-induced deformations of the galactic disk can provide valuable constraints on the direction of motion in the plane of the sky. The potential of ram-pressure stripping as a diagnostic of peculiar motions perpendicular to our line of sight was pointed out by \citet{2019ApJ...882..127E} and, in the eMACS sample, is exemplified by eMACSJ1757.5 (Fig.~B36) 
which illustrates that the tell-tale morphology of ``jellyfish'' galaxies can indeed reveal dominant axes of galaxy infall even at $z>0.5$. 

\section{Conclusions}
\label{sec:wrapup}

Based on RASS detections comprising as few as 6 photons, the extended Massive Cluster Survey (eMACS) has compiled a sample of over 100 very X-ray luminous galaxy clusters after searching over 20,000 square degrees of extragalactic sky. Limited to clusters at $z>0.5$, eMACS probes a population that was hitherto barely explored: the first generation of truly massive clusters ($M\gtrsim 10^{15}$ M$_\odot$). Extensive follow-up observations of eMACS cluster candidates with ground- and space-based observatories, crucial to overcome the poor photon statistics of the underlying RASS detections, confirmed the presence of clusters in the vast majority of the targeted systems but, so far, have also shown two to be non-clusters. As importantly, follow-up work from radio to X-ray wavelengths yielded valuable insights into the physical properties of many of these systems and allowed us to constrain their gravitational-lensing strength, evolutionary state, and ICM gas temperature. In addition, these observations revealed examples of highly active BCGs, powerful QSOs, and late-type galaxies undergoing ram-pressure stripping during their high-velocity passage through the dense ICM. 

The sample presented here includes the most distant galaxy cluster ever discovered in the RASS (eMACSJ0324.0 at $z=0.90$, Fig.~B7), 
extremely powerful gravitational lenses (e.g., eMACSJ1212.5, Fig.~B23; 
eMACSJ1341.9, Fig.~B26 
and \citealt{2018ApJ...852L...7E}; or eMACSJ1353.7, Fig.~B27), 
a newly discovered ``exotic'' hyperbolic umbilical lens (eMACSJ1248.2, Fig.~\ref{fig:e1437core}), exceptionally X-ray luminous cluster collisions in various phases of the merging process (e.g., eMACSJ1756.8, Fig.~B35; 
eMACSJ1831.1, Fig.~B38), 
extraordinarily ``active'' BCGs, e.g., in eMACSJ0252.4 \citep[][Fig.~\ref{fig:e0252}]{2021MNRAS.508.3663E} and eMACSJ1030.5 (Fig.~\ref{fig:bcgs}), and numerous systems whose potentially remarkable physical properties are only hinted at by the existing data (e.g., eMACSJ1144.2, eMACSJ1508.1, eMACSJ1757.5).

We highlight in particular the potential importance of the massive clusters at $z>0.5$ presented in this work as powerful gravitational telescopes to study faint sources in the very distant Universe (as already demonstrated, for example with the lensed sources at $z=3.375$ and $z=5.166$ in eMACSJ2229.9).  The main benefits of selected eMACS clusters as probes for {\it James Webb Space Telescope} ({\it JWST}) surveys of the very distant Universe, $z\gg 10$, are their high masses and high redshifts, which also yield a corresponding reduction in contamination from  intracluster light, allowing faint lensed features to be more easily identified in the high-amplification regions near the cluster centres. We anticipate that {\it JWST} surveys of these clusters will be as productive as the current exploitation of the MACS sample for studying faint galaxies and individual stars in the high-redshift Universe
\citep[e.g.,][]{Abdurrouf23,Atek23,Diego23,Asada24,Fujimoto24}.

While the current eMACS sample is not statistically complete, tentative conclusions about the physical properties of the population of very massive clusters at $z>0.5$ can be drawn from better defined subsets. Specifically, the morphology of the ICM of eMACS clusters in high-resolution imaging X-ray observations and the optical properties of eMACS BCGS from both imaging and spectroscopic follow-up work suggest that the cool-core fraction lies at or below $\sim$15\%, consistent with the findings of \citet{Mann2012} and expectations for a sample that approaches the redshift of cluster formation at $z \sim 1$. The analysis of all existing {\it CXO}/ACIS-I data for eMACS clusters shows that the power-law model of the $L_{\rm X}$--k$T$ relation established for less massive clusters at lower redshifts \citep[e.g.,][]{2013A&A...558A..75T} extends to the regime probed by eMACS (Fig.~\ref{fig:lxkt}). For the majority of eMACS clusters, velocity dispersions are very high, reaching and exceeding 1500 km s$^{-1}$ for several systems (Fig.~\ref{fig:velhist}).  While clear signs of line-of-sight substructure are not detectable at the current statistics, the highest velocity dispersions exceed reasonable virial estimates and suggest that inflation from peculiar (bulk) velocities from ongoing cluster growth is common in this cluster population.

As, over three decades after its completion, the RASS is being succeeded by eRASS1 \citep{2024A&A...682A..34M}, the second all-sky survey ever conducted with an imaging X-ray telescope, the samples of the most X-ray luminous clusters at $z>0.3$ compiled by the MACS and eMACS projects are being complemented by a wealth of discoveries down the mass and luminosity scale, all the way to groups of galaxies, protoclusters, and dense filaments. In addition, {\it eROSITA}'s much greater sensitivity will yield vastly better photon statistics, and at least estimated ICM temperatures, for all eMACS clusters that, as of yet, have not been targeted in dedicated follow-up observations with {\it CXO}. 

Although neither the compilation of the eMACS sample nor the analysis of the physical properties of all eMACS clusters are complete at this point, we here release all data and data products obtained and derived by us in the course of this program to avoid duplication and unnecessary investments of precious observing resources by the extragalactic community \citep{2023AstL...49..599Z}.  

\section*{Acknowledgements}

We thank Lukas Zalesky for permission to use unpublished AStroLens results in Section 5.2.2. of this paper. HE gratefully acknowledges financial support from grants administered by STScI and CXC. Specifically, support for programs GO-13671, -14098, -15132, -15466, -15608, -15843, -15844, and -16428 was provided by NASA through a grant from the Space Telescope Science Institute, which is operated by the Associations of Universities for Research in Astronomy, Incorporated, under NASA contract NAS5-26555. Support for this work was also provided by the National Aeronautics and Space Administration through {\it CXO} Award Number GO9-20120X, GO0-21125X, GO1-22119X, and GO2-23119X, issued by the {\it Chandra} X-ray Center, which is operated by the Smithsonian Astrophysical Observatory for and on behalf of the National Aeronautics Space Administration under contract NAS8-03060. Durham authors acknowledge support from STFC (ST/T000244/1, ST/X001075/1). 
This research has made use of data obtained from the {\it Chandra} Data Archive and the {\it Chandra} Source Catalogue, and software provided by the {\it Chandra} X-ray Center (CXC) in the application packages CIAO and Sherpa.

\section*{Data Availability}
The data underlying this article are available in the MAST, {\it CXO}, and Keck data archives.

\bibliography{paper_bib.bib}

\begin{thebibliography}{}
\makeatletter
\relax
\def\mn@urlcharsother{\let\do\@makeother \do\$\do\&\do\#\do\^\do\_\do\%\do\~}
\def\mn@doi{\begingroup\mn@urlcharsother \@ifnextchar [ {\mn@doi@}
  {\mn@doi@[]}}
\def\mn@doi@[#1]#2{\def\@tempa{#1}\ifx\@tempa\@empty \href
  {http://dx.doi.org/#2} {doi:#2}\else \href {http://dx.doi.org/#2} {#1}\fi
  \endgroup}
\def\mn@eprint#1#2{\mn@eprint@#1:#2::\@nil}
\def\mn@eprint@arXiv#1{\href {http://arxiv.org/abs/#1} {{\tt arXiv:#1}}}
\def\mn@eprint@dblp#1{\href {http://dblp.uni-trier.de/rec/bibtex/#1.xml}
  {dblp:#1}}
\def\mn@eprint@#1:#2:#3:#4\@nil{\def\@tempa {#1}\def\@tempb {#2}\def\@tempc
  {#3}\ifx \@tempc \@empty \let \@tempc \@tempb \let \@tempb \@tempa \fi \ifx
  \@tempb \@empty \def\@tempb {arXiv}\fi \@ifundefined
  {mn@eprint@\@tempb}{\@tempb:\@tempc}{\expandafter \expandafter \csname
  mn@eprint@\@tempb\endcsname \expandafter{\@tempc}}}

\bibitem[\protect\citeauthoryear{{Abdurro'uf} et~al.,}{{Abdurro'uf}
  et~al.}{2023}]{Abdurrouf23}
{Abdurro'uf} et~al., 2023, \mn@doi [\apj] {10.3847/1538-4357/acba06}, \href
  {https://ui.adsabs.harvard.edu/abs/2023ApJ...945..117A} {945, 117}

\bibitem[\protect\citeauthoryear{{Alavi} et~al.,}{{Alavi}
  et~al.}{2014}]{Alavi2014}
{Alavi} A.,  et~al., 2014, \mn@doi [\apj] {10.1088/0004-637X/780/2/143}, \href
  {https://ui.adsabs.harvard.edu/abs/2014ApJ...780..143A} {780, 143}

\bibitem[\protect\citeauthoryear{{Allen}, {Evrard}  \& {Mantz}}{{Allen}
  et~al.}{2011}]{Allen2011}
{Allen} S.~W.,  {Evrard} A.~E.,   {Mantz} A.~B.,  2011, \mn@doi [\araa]
  {10.1146/annurev-astro-081710-102514}, \href
  {https://ui.adsabs.harvard.edu/abs/2011ARA&A..49..409A} {49, 409}

\bibitem[\protect\citeauthoryear{{Allingham} et~al.,}{{Allingham}
  et~al.}{2023}]{2023MNRAS.522.1118A}
{Allingham} J. F.~V.,  et~al., 2023, \mn@doi [\mnras] {10.1093/mnras/stad917},
  \href {https://ui.adsabs.harvard.edu/abs/2023MNRAS.522.1118A} {522, 1118}

\bibitem[\protect\citeauthoryear{{Andrade-Santos} et~al.,}{{Andrade-Santos}
  et~al.}{2017}]{2017ApJ...843...76A}
{Andrade-Santos} F.,  et~al., 2017, \mn@doi [\apj] {10.3847/1538-4357/aa7461},
  \href {https://ui.adsabs.harvard.edu/abs/2017ApJ...843...76A} {843, 76}

\bibitem[\protect\citeauthoryear{{Applegate} et~al.,}{{Applegate}
  et~al.}{2014}]{2014MNRAS.439...48A}
{Applegate} D.~E.,  et~al., 2014, \mn@doi [\mnras] {10.1093/mnras/stt2129},
  \href {https://ui.adsabs.harvard.edu/abs/2014MNRAS.439...48A} {439, 48}

\bibitem[\protect\citeauthoryear{{Arnaud}, {Aghanim}  \& {Neumann}}{{Arnaud}
  et~al.}{2002}]{arnaud2002}
{Arnaud} M.,  {Aghanim} N.,   {Neumann} D.~M.,  2002, \mn@doi [\aap]
  {10.1051/0004-6361:20020378}, \href
  {https://ui.adsabs.harvard.edu/abs/2002A&A...389....1A} {389, 1}

\bibitem[\protect\citeauthoryear{{Asada} et~al.,}{{Asada}
  et~al.}{2024}]{Asada24}
{Asada} Y.,  et~al., 2024, \mn@doi [\mnras] {10.1093/mnras/stad3902}, \href
  {https://ui.adsabs.harvard.edu/abs/2024MNRAS.52711372A} {527, 11372}

\bibitem[\protect\citeauthoryear{{Asplund}, {Grevesse}, {Sauval}  \&
  {Scott}}{{Asplund} et~al.}{2009}]{asplund2009}
{Asplund} M.,  {Grevesse} N.,  {Sauval} A.~J.,   {Scott} P.,  2009, \mn@doi
  [\araa] {10.1146/annurev.astro.46.060407.145222}, \href
  {https://ui.adsabs.harvard.edu/abs/2009ARA&A..47..481A} {47, 481}

\bibitem[\protect\citeauthoryear{{Atek} et~al.,}{{Atek} et~al.}{2023}]{Atek23}
{Atek} H.,  et~al., 2023, \mn@doi [\mnras] {10.1093/mnras/stac3144}, \href
  {https://ui.adsabs.harvard.edu/abs/2023MNRAS.519.1201A} {519, 1201}

\bibitem[\protect\citeauthoryear{{Beauchesne} et~al.,}{{Beauchesne}
  et~al.}{2024}]{beauchesne2024}
{Beauchesne} B.,  et~al., 2024, \mn@doi [\mnras] {10.1093/mnras/stad3308},
  \href {https://ui.adsabs.harvard.edu/abs/2024MNRAS.527.3246B} {527, 3246}

\bibitem[\protect\citeauthoryear{{Beers}, {Flynn}  \& {Gebhardt}}{{Beers}
  et~al.}{1990}]{1990AJ....100...32B}
{Beers} T.~C.,  {Flynn} K.,   {Gebhardt} K.,  1990, \mn@doi [\aj]
  {10.1086/115487}, \href
  {https://ui.adsabs.harvard.edu/abs/1990AJ....100...32B} {100, 32}

\bibitem[\protect\citeauthoryear{{Bertin} \& {Arnouts}}{{Bertin} \&
  {Arnouts}}{1996}]{bertin1996}
{Bertin} E.,  {Arnouts} S.,  1996, \mn@doi [\aaps] {10.1051/aas:1996164}, \href
  {https://ui.adsabs.harvard.edu/abs/1996A&AS..117..393B} {117, 393}

\bibitem[\protect\citeauthoryear{{Bezanson} et~al.,}{{Bezanson}
  et~al.}{2022}]{Bezanson2022}
{Bezanson} R.,  et~al., 2022, \mn@doi [arXiv e-prints]
  {10.48550/arXiv.2212.04026}, \href
  {https://ui.adsabs.harvard.edu/abs/2022arXiv221204026B} {p. arXiv:2212.04026}

\bibitem[\protect\citeauthoryear{{Bleem} et~al.,}{{Bleem}
  et~al.}{2015}]{2015ApJS..216...27B}
{Bleem} L.~E.,  et~al., 2015, \mn@doi [\apjs] {10.1088/0067-0049/216/2/27},
  \href {https://ui.adsabs.harvard.edu/abs/2015ApJS..216...27B} {216, 27}

\bibitem[\protect\citeauthoryear{{Bleem} et~al.,}{{Bleem}
  et~al.}{2020}]{2020ApJS..247...25B}
{Bleem} L.~E.,  et~al., 2020, \mn@doi [\apjs] {10.3847/1538-4365/ab6993}, \href
  {https://ui.adsabs.harvard.edu/abs/2020ApJS..247...25B} {247, 25}

\bibitem[\protect\citeauthoryear{{Bocquet} et~al.,}{{Bocquet}
  et~al.}{2019}]{Bocquet2019}
{Bocquet} S.,  et~al., 2019, \mn@doi [\apj] {10.3847/1538-4357/ab1f10}, \href
  {https://ui.adsabs.harvard.edu/abs/2019ApJ...878...55B} {878, 55}

\bibitem[\protect\citeauthoryear{{B{\"o}hringer} et~al.,}{{B{\"o}hringer}
  et~al.}{2004}]{Bohringer2004}
{B{\"o}hringer} H.,  et~al., 2004, \mn@doi [\aap] {10.1051/0004-6361:20034484},
  \href {https://ui.adsabs.harvard.edu/abs/2004A&A...425..367B} {425, 367}

\bibitem[\protect\citeauthoryear{{B{\"o}hringer}, {Chon}, {Collins}, {Guzzo},
  {Nowak}  \& {Bobrovskyi}}{{B{\"o}hringer} et~al.}{2013}]{Bohringer2013}
{B{\"o}hringer} H.,  {Chon} G.,  {Collins} C.~A.,  {Guzzo} L.,  {Nowak} N.,
  {Bobrovskyi} S.,  2013, \mn@doi [\aap] {10.1051/0004-6361/201220722}, \href
  {https://ui.adsabs.harvard.edu/abs/2013A&A...555A..30B} {555, A30}

\bibitem[\protect\citeauthoryear{{Bond}, {Kofman}  \& {Pogosyan}}{{Bond}
  et~al.}{1996}]{1996Natur.380..603B}
{Bond} J.~R.,  {Kofman} L.,   {Pogosyan} D.,  1996, \mn@doi [\nat]
  {10.1038/380603a0}, \href
  {https://ui.adsabs.harvard.edu/abs/1996Natur.380..603B} {380, 603}

\bibitem[\protect\citeauthoryear{{Boselli}, {Fossati}  \& {Sun}}{{Boselli}
  et~al.}{2022}]{2022A&ARv..30....3B}
{Boselli} A.,  {Fossati} M.,   {Sun} M.,  2022, \mn@doi [\aapr]
  {10.1007/s00159-022-00140-3}, \href
  {https://ui.adsabs.harvard.edu/abs/2022A&ARv..30....3B} {30, 3}

\bibitem[\protect\citeauthoryear{{Bouwens}, {Illingworth}, {Ellis}, {Oesch}  \&
  {Stefanon}}{{Bouwens} et~al.}{2022}]{Bouwens2022}
{Bouwens} R.~J.,  {Illingworth} G.,  {Ellis} R.~S.,  {Oesch} P.,   {Stefanon}
  M.,  2022, \mn@doi [\apj] {10.3847/1538-4357/ac86d1}, \href
  {https://ui.adsabs.harvard.edu/abs/2022ApJ...940...55B} {940, 55}

\bibitem[\protect\citeauthoryear{{Bower}, {Lucey}  \& {Ellis}}{{Bower}
  et~al.}{1992}]{1992MNRAS.254..601B}
{Bower} R.~G.,  {Lucey} J.~R.,   {Ellis} R.~S.,  1992, \mn@doi [\mnras]
  {10.1093/mnras/254.4.601}, \href
  {https://ui.adsabs.harvard.edu/abs/1992MNRAS.254..601B} {254, 601}

\bibitem[\protect\citeauthoryear{{Brada{\v{c}}}, {Allen}, {Treu}, {Ebeling},
  {Massey}, {Morris}, {von der Linden}  \& {Applegate}}{{Brada{\v{c}}}
  et~al.}{2008}]{2008ApJ...687..959B}
{Brada{\v{c}}} M.,  {Allen} S.~W.,  {Treu} T.,  {Ebeling} H.,  {Massey} R.,
  {Morris} R.~G.,  {von der Linden} A.,   {Applegate} D.,  2008, \mn@doi [\apj]
  {10.1086/591246}, \href
  {https://ui.adsabs.harvard.edu/abs/2008ApJ...687..959B} {687, 959}

\bibitem[\protect\citeauthoryear{{Buchner} et~al.,}{{Buchner}
  et~al.}{2014}]{buchner2014}
{Buchner} J.,  et~al., 2014, \mn@doi [\aap] {10.1051/0004-6361/201322971},
  \href {https://ui.adsabs.harvard.edu/abs/2014A&A...564A.125B} {564, A125}

\bibitem[\protect\citeauthoryear{{Carlstrom} et~al.,}{{Carlstrom}
  et~al.}{2011}]{2011PASP..123..568C}
{Carlstrom} J.~E.,  et~al., 2011, \mn@doi [\pasp] {10.1086/659879}, \href
  {https://ui.adsabs.harvard.edu/abs/2011PASP..123..568C} {123, 568}

\bibitem[\protect\citeauthoryear{{Chambers} et~al.,}{{Chambers}
  et~al.}{2016}]{Chambers2016}
{Chambers} K.~C.,  et~al., 2016, \mn@doi [arXiv e-prints]
  {10.48550/arXiv.1612.05560}, \href
  {https://ui.adsabs.harvard.edu/abs/2016arXiv161205560C} {p. arXiv:1612.05560}

\bibitem[\protect\citeauthoryear{{Chiu}, {Klein}, {Mohr}  \& {Bocquet}}{{Chiu}
  et~al.}{2023}]{2023MNRAS.522.1601C}
{Chiu} I.~N.,  {Klein} M.,  {Mohr} J.,   {Bocquet} S.,  2023, \mn@doi [\mnras]
  {10.1093/mnras/stad957}, \href
  {https://ui.adsabs.harvard.edu/abs/2023MNRAS.522.1601C} {522, 1601}

\bibitem[\protect\citeauthoryear{{Chon} \& {B{\"o}hringer}}{{Chon} \&
  {B{\"o}hringer}}{2017}]{2017A&A...606L...4C}
{Chon} G.,  {B{\"o}hringer} H.,  2017, \mn@doi [\aap]
  {10.1051/0004-6361/201731854}, \href
  {https://ui.adsabs.harvard.edu/abs/2017A&A...606L...4C} {606, L4}

\bibitem[\protect\citeauthoryear{{Cooper}, {Newman}, {Davis}, {Finkbeiner}  \&
  {Gerke}}{{Cooper} et~al.}{2012}]{2012ascl.soft03003C}
{Cooper} M.~C.,  {Newman} J.~A.,  {Davis} M.,  {Finkbeiner} D.~P.,   {Gerke}
  B.~F.,  2012, {spec2d: DEEP2 DEIMOS Spectral Pipeline}, Astrophysics Source
  Code Library, record ascl:1203.003 (\mn@eprint {ascl} {1203.003})

\bibitem[\protect\citeauthoryear{{Cortese} et~al.,}{{Cortese}
  et~al.}{2007}]{2007MNRAS.376..157C}
{Cortese} L.,  et~al., 2007, \mn@doi [\mnras]
  {10.1111/j.1365-2966.2006.11369.x}, \href
  {https://ui.adsabs.harvard.edu/abs/2007MNRAS.376..157C} {376, 157}

\bibitem[\protect\citeauthoryear{{Crawford}, {Allen}, {Ebeling}, {Edge}  \&
  {Fabian}}{{Crawford} et~al.}{1999}]{1999MNRAS.306..857C}
{Crawford} C.~S.,  {Allen} S.~W.,  {Ebeling} H.,  {Edge} A.~C.,   {Fabian}
  A.~C.,  1999, \mn@doi [\mnras] {10.1046/j.1365-8711.1999.02583.x}, \href
  {https://ui.adsabs.harvard.edu/abs/1999MNRAS.306..857C} {306, 857}

\bibitem[\protect\citeauthoryear{{De Grandi}, {Molendi}, {B{\"o}hringer},
  {Chincarini}  \& {Voges}}{{De Grandi} et~al.}{1997}]{1997ApJ...486..738D}
{De Grandi} S.,  {Molendi} S.,  {B{\"o}hringer} H.,  {Chincarini} G.,   {Voges}
  W.,  1997, \mn@doi [\apj] {10.1086/304543}, \href
  {https://ui.adsabs.harvard.edu/abs/1997ApJ...486..738D} {486, 738}

\bibitem[\protect\citeauthoryear{{De Grandi} et~al.,}{{De Grandi}
  et~al.}{1999}]{degrandi1999}
{De Grandi} S.,  et~al., 1999, \mn@doi [\apj] {10.1086/306939}, \href
  {https://ui.adsabs.harvard.edu/abs/1999ApJ...514..148D} {514, 148}

\bibitem[\protect\citeauthoryear{{De Lucia} \& {Blaizot}}{{De Lucia} \&
  {Blaizot}}{2007}]{2007MNRAS.375....2D}
{De Lucia} G.,  {Blaizot} J.,  2007, \mn@doi [\mnras]
  {10.1111/j.1365-2966.2006.11287.x}, \href
  {https://ui.adsabs.harvard.edu/abs/2007MNRAS.375....2D} {375, 2}

\bibitem[\protect\citeauthoryear{{Diego} et~al.,}{{Diego}
  et~al.}{2023}]{Diego23}
{Diego} J.~M.,  et~al., 2023, \mn@doi [\aap] {10.1051/0004-6361/202347556},
  \href {https://ui.adsabs.harvard.edu/abs/2023A&A...679A..31D} {679, A31}

\bibitem[\protect\citeauthoryear{{Dubinski}}{{Dubinski}}{1998}]{1998ApJ...502..141D}
{Dubinski} J.,  1998, \mn@doi [\apj] {10.1086/305901}, \href
  {https://ui.adsabs.harvard.edu/abs/1998ApJ...502..141D} {502, 141}

\bibitem[\protect\citeauthoryear{{Ebeling} \& {Kalita}}{{Ebeling} \&
  {Kalita}}{2019}]{2019ApJ...882..127E}
{Ebeling} H.,  {Kalita} B.~S.,  2019, \mn@doi [\apj]
  {10.3847/1538-4357/ab35d6}, \href
  {https://ui.adsabs.harvard.edu/abs/2019ApJ...882..127E} {882, 127}

\bibitem[\protect\citeauthoryear{{Ebeling}, {Voges}, {Bohringer}, {Edge},
  {Huchra}  \& {Briel}}{{Ebeling} et~al.}{1996}]{1996MNRAS.281..799E}
{Ebeling} H.,  {Voges} W.,  {Bohringer} H.,  {Edge} A.~C.,  {Huchra} J.~P.,
  {Briel} U.~G.,  1996, \mn@doi [\mnras] {10.1093/mnras/281.3.799}, \href
  {https://ui.adsabs.harvard.edu/abs/1996MNRAS.281..799E} {281, 799}

\bibitem[\protect\citeauthoryear{{Ebeling}, {Edge}, {Fabian}, {Allen},
  {Crawford}  \& {B{\"o}hringer}}{{Ebeling} et~al.}{1997}]{Ebeling1997}
{Ebeling} H.,  {Edge} A.~C.,  {Fabian} A.~C.,  {Allen} S.~W.,  {Crawford}
  C.~S.,   {B{\"o}hringer} H.,  1997, \mn@doi [\apjl] {10.1086/310589}, \href
  {https://ui.adsabs.harvard.edu/abs/1997ApJ...479L.101E} {479, L101}

\bibitem[\protect\citeauthoryear{{Ebeling}, {Edge}, {Bohringer}, {Allen},
  {Crawford}, {Fabian}, {Voges}  \& {Huchra}}{{Ebeling}
  et~al.}{1998}]{1998MNRAS.301..881E}
{Ebeling} H.,  {Edge} A.~C.,  {Bohringer} H.,  {Allen} S.~W.,  {Crawford}
  C.~S.,  {Fabian} A.~C.,  {Voges} W.,   {Huchra} J.~P.,  1998, \mn@doi
  [\mnras] {10.1046/j.1365-8711.1998.01949.x}, \href
  {https://ui.adsabs.harvard.edu/abs/1998MNRAS.301..881E} {301, 881}

\bibitem[\protect\citeauthoryear{{Ebeling}, {Edge}, {Allen}, {Crawford},
  {Fabian}  \& {Huchra}}{{Ebeling} et~al.}{2000}]{Ebeling2000}
{Ebeling} H.,  {Edge} A.~C.,  {Allen} S.~W.,  {Crawford} C.~S.,  {Fabian}
  A.~C.,   {Huchra} J.~P.,  2000, \mn@doi [\mnras]
  {10.1046/j.1365-8711.2000.03549.x}, \href
  {https://ui.adsabs.harvard.edu/abs/2000MNRAS.318..333E} {318, 333}

\bibitem[\protect\citeauthoryear{{Ebeling}, {Edge}  \& {Henry}}{{Ebeling}
  et~al.}{2001}]{Ebeling2001}
{Ebeling} H.,  {Edge} A.~C.,   {Henry} J.~P.,  2001, \mn@doi [\apj]
  {10.1086/320958}, \href
  {https://ui.adsabs.harvard.edu/abs/2001ApJ...553..668E} {553, 668}

\bibitem[\protect\citeauthoryear{{Ebeling}, {Mullis}  \& {Tully}}{{Ebeling}
  et~al.}{2002}]{Ebeling2002}
{Ebeling} H.,  {Mullis} C.~R.,   {Tully} R.~B.,  2002, \mn@doi [\apj]
  {10.1086/343790}, \href
  {https://ui.adsabs.harvard.edu/abs/2002ApJ...580..774E} {580, 774}

\bibitem[\protect\citeauthoryear{{Ebeling}, {White}  \& {Rangarajan}}{{Ebeling}
  et~al.}{2006}]{2006MNRAS.368...65E}
{Ebeling} H.,  {White} D.~A.,   {Rangarajan} F.~V.~N.,  2006, \mn@doi [\mnras]
  {10.1111/j.1365-2966.2006.10135.x}, \href
  {https://ui.adsabs.harvard.edu/abs/2006MNRAS.368...65E} {368, 65}

\bibitem[\protect\citeauthoryear{{Ebeling}, {Barrett}, {Donovan}, {Ma}, {Edge}
  \& {van Speybroeck}}{{Ebeling} et~al.}{2007}]{Ebeling2007}
{Ebeling} H.,  {Barrett} E.,  {Donovan} D.,  {Ma} C.~J.,  {Edge} A.~C.,   {van
  Speybroeck} L.,  2007, \mn@doi [\apjl] {10.1086/518603}, \href
  {https://ui.adsabs.harvard.edu/abs/2007ApJ...661L..33E} {661, L33}

\bibitem[\protect\citeauthoryear{{Ebeling}, {Edge}, {Mantz}, {Barrett},
  {Henry}, {Ma}  \& {van Speybroeck}}{{Ebeling} et~al.}{2010}]{Ebeling2010}
{Ebeling} H.,  {Edge} A.~C.,  {Mantz} A.,  {Barrett} E.,  {Henry} J.~P.,  {Ma}
  C.~J.,   {van Speybroeck} L.,  2010, \mn@doi [\mnras]
  {10.1111/j.1365-2966.2010.16920.x}, \href
  {https://ui.adsabs.harvard.edu/abs/2010MNRAS.407...83E} {407, 83}

\bibitem[\protect\citeauthoryear{{Ebeling} et~al.,}{{Ebeling}
  et~al.}{2013}]{Ebeling2013}
{Ebeling} H.,  et~al., 2013, \mn@doi [\mnras] {10.1093/mnras/stt387}, \href
  {https://ui.adsabs.harvard.edu/abs/2013MNRAS.432...62E} {432, 62}

\bibitem[\protect\citeauthoryear{{Ebeling}, {Stephenson}  \& {Edge}}{{Ebeling}
  et~al.}{2014}]{ebeling2014}
{Ebeling} H.,  {Stephenson} L.~N.,   {Edge} A.~C.,  2014, \mn@doi [\apjl]
  {10.1088/2041-8205/781/2/L40}, \href
  {https://ui.adsabs.harvard.edu/abs/2014ApJ...781L..40E} {781, L40}

\bibitem[\protect\citeauthoryear{{Ebeling}, {Qi}  \& {Richard}}{{Ebeling}
  et~al.}{2017}]{Ebeling2017}
{Ebeling} H.,  {Qi} J.,   {Richard} J.,  2017, \mn@doi [\mnras]
  {10.1093/mnras/stx1636}, \href
  {https://ui.adsabs.harvard.edu/abs/2017MNRAS.471.3305E} {471, 3305}

\bibitem[\protect\citeauthoryear{{Ebeling}, {Stockmann}, {Richard}, {Zabl},
  {Brammer}, {Toft}  \& {Man}}{{Ebeling} et~al.}{2018}]{2018ApJ...852L...7E}
{Ebeling} H.,  {Stockmann} M.,  {Richard} J.,  {Zabl} J.,  {Brammer} G.,
  {Toft} S.,   {Man} A.,  2018, \mn@doi [\apjl] {10.3847/2041-8213/aa9fee},
  \href {https://ui.adsabs.harvard.edu/abs/2018ApJ...852L...7E} {852, L7}

\bibitem[\protect\citeauthoryear{{Ebeling}, {Richard}, {Smail}, {Edge},
  {Koekemoer}  \& {Zalesky}}{{Ebeling} et~al.}{2021}]{2021MNRAS.508.3663E}
{Ebeling} H.,  {Richard} J.,  {Smail} I.,  {Edge} A.~C.,  {Koekemoer} A.~M.,
  {Zalesky} L.,  2021, \mn@doi [\mnras] {10.1093/mnras/stab2725}, \href
  {https://ui.adsabs.harvard.edu/abs/2021MNRAS.508.3663E} {508, 3663}

\bibitem[\protect\citeauthoryear{{Eckert}, {Finoguenov}, {Ghirardini},
  {Grandis}, {Kaefer}, {Sanders}  \& {Ramos-Ceja}}{{Eckert}
  et~al.}{2020}]{eckert2020}
{Eckert} D.,  {Finoguenov} A.,  {Ghirardini} V.,  {Grandis} S.,  {Kaefer} F.,
  {Sanders} J.,   {Ramos-Ceja} M.,  2020, \mn@doi [The Open Journal of
  Astrophysics] {10.21105/astro.2009.13944}, \href
  {https://ui.adsabs.harvard.edu/abs/2020OJAp....3E..12E} {3, 12}

\bibitem[\protect\citeauthoryear{{Ettori} et~al.,}{{Ettori}
  et~al.}{2019}]{2019A&A...621A..39E}
{Ettori} S.,  et~al., 2019, \mn@doi [\aap] {10.1051/0004-6361/201833323}, \href
  {https://ui.adsabs.harvard.edu/abs/2019A&A...621A..39E} {621, A39}

\bibitem[\protect\citeauthoryear{{Evrard} et~al.,}{{Evrard}
  et~al.}{2008}]{2008ApJ...672..122E}
{Evrard} A.~E.,  et~al., 2008, \mn@doi [\apj] {10.1086/521616}, \href
  {https://ui.adsabs.harvard.edu/abs/2008ApJ...672..122E} {672, 122}

\bibitem[\protect\citeauthoryear{{Faber} et~al.,}{{Faber}
  et~al.}{2003}]{DEIMOS2003}
{Faber} S.~M.,  et~al., 2003, in {Iye} M.,  {Moorwood} A. F.~M.,  eds,  Society
  of Photo-Optical Instrumentation Engineers (SPIE) Conference Series Vol.
  4841, Instrument Design and Performance for Optical/Infrared Ground-based
  Telescopes. pp 1657--1669, \mn@doi{10.1117/12.460346}

\bibitem[\protect\citeauthoryear{{Fedeli} \& {Bartelmann}}{{Fedeli} \&
  {Bartelmann}}{2007}]{2007A&A...474..355F}
{Fedeli} C.,  {Bartelmann} M.,  2007, \mn@doi [\aap]
  {10.1051/0004-6361:20077572}, \href
  {https://ui.adsabs.harvard.edu/abs/2007A&A...474..355F} {474, 355}

\bibitem[\protect\citeauthoryear{{Feroz}, {Hobson}, {Cameron}  \&
  {Pettitt}}{{Feroz} et~al.}{2019}]{feroz2019}
{Feroz} F.,  {Hobson} M.~P.,  {Cameron} E.,   {Pettitt} A.~N.,  2019, \mn@doi
  [The Open Journal of Astrophysics] {10.21105/astro.1306.2144}, \href
  {https://ui.adsabs.harvard.edu/abs/2019OJAp....2E..10F} {2, 10}

\bibitem[\protect\citeauthoryear{{Freeman}, {Kashyap}, {Rosner}  \&
  {Lamb}}{{Freeman} et~al.}{2002}]{freeman2002}
{Freeman} P.~E.,  {Kashyap} V.,  {Rosner} R.,   {Lamb} D.~Q.,  2002, \mn@doi
  [\apjs] {10.1086/324017}, \href
  {https://ui.adsabs.harvard.edu/abs/2002ApJS..138..185F} {138, 185}

\bibitem[\protect\citeauthoryear{{Fruchter} \& {et al.}}{{Fruchter} \& {et
  al.}}{2010}]{fruchter2010}
{Fruchter} A.~S.,  {et al.} 2010, in 2010 Space Telescope Science Institute
  Calibration Workshop. pp 382--387

\bibitem[\protect\citeauthoryear{{Fruscione} et~al.,}{{Fruscione}
  et~al.}{2006}]{fruscione2006}
{Fruscione} A.,  et~al., 2006, in {Silva} D.~R.,  {Doxsey} R.~E.,  eds,
  Society of Photo-Optical Instrumentation Engineers (SPIE) Conference Series
  Vol. 6270, Society of Photo-Optical Instrumentation Engineers (SPIE)
  Conference Series. p. 62701V, \mn@doi{10.1117/12.671760}

\bibitem[\protect\citeauthoryear{{Fujimoto} et~al.,}{{Fujimoto}
  et~al.}{2024}]{Fujimoto24}
{Fujimoto} S.,  et~al., 2024, \mn@doi [\apj] {10.3847/1538-4357/ad235c}, \href
  {https://ui.adsabs.harvard.edu/abs/2024ApJ...964..146F} {964, 146}

\bibitem[\protect\citeauthoryear{{Gilbank}, {Gladders}, {Yee}  \&
  {Hsieh}}{{Gilbank} et~al.}{2011}]{2011AJ....141...94G}
{Gilbank} D.~G.,  {Gladders} M.~D.,  {Yee} H.~K.~C.,   {Hsieh} B.~C.,  2011,
  \mn@doi [\aj] {10.1088/0004-6256/141/3/94}, \href
  {https://ui.adsabs.harvard.edu/abs/2011AJ....141...94G} {141, 94}

\bibitem[\protect\citeauthoryear{{Gobat} et~al.,}{{Gobat}
  et~al.}{2011}]{2011A&A...526A.133G}
{Gobat} R.,  et~al., 2011, \mn@doi [\aap] {10.1051/0004-6361/201016084}, \href
  {https://ui.adsabs.harvard.edu/abs/2011A&A...526A.133G} {526, A133}

\bibitem[\protect\citeauthoryear{{Gonzalez} et~al.,}{{Gonzalez}
  et~al.}{2019}]{2019ApJS..240...33G}
{Gonzalez} A.~H.,  et~al., 2019, \mn@doi [\apjs] {10.3847/1538-4365/aafad2},
  \href {https://ui.adsabs.harvard.edu/abs/2019ApJS..240...33G} {240, 33}

\bibitem[\protect\citeauthoryear{{Green} et~al.,}{{Green}
  et~al.}{2016}]{2016MNRAS.461..560G}
{Green} T.~S.,  et~al., 2016, \mn@doi [\mnras] {10.1093/mnras/stw1338}, \href
  {https://ui.adsabs.harvard.edu/abs/2016MNRAS.461..560G} {461, 560}

\bibitem[\protect\citeauthoryear{{Griffiths}, {Rudisel}, {Wagner}, {Hamilton},
  {Huang}  \& {Villforth}}{{Griffiths} et~al.}{2021}]{2021MNRAS.506.1595G}
{Griffiths} R.~E.,  {Rudisel} M.,  {Wagner} J.,  {Hamilton} T.,  {Huang} P.-C.,
    {Villforth} C.,  2021, \mn@doi [\mnras] {10.1093/mnras/stab1375}, \href
  {https://ui.adsabs.harvard.edu/abs/2021MNRAS.506.1595G} {506, 1595}

\bibitem[\protect\citeauthoryear{{Gunn} \& {Gott}}{{Gunn} \&
  {Gott}}{1972}]{1972ApJ...176....1G}
{Gunn} J.~E.,  {Gott} J.~Richard I.,  1972, \mn@doi [\apj] {10.1086/151605},
  \href {https://ui.adsabs.harvard.edu/abs/1972ApJ...176....1G} {176, 1}

\bibitem[\protect\citeauthoryear{{HI4PI Collaboration} et~al.,}{{HI4PI
  Collaboration} et~al.}{2016}]{HI4PI2016}
{HI4PI Collaboration} et~al., 2016, \mn@doi [\aap]
  {10.1051/0004-6361/201629178}, \href
  {https://ui.adsabs.harvard.edu/abs/2016A&A...594A.116H} {594, A116}

\bibitem[\protect\citeauthoryear{{Harvey}, {Massey}, {Kitching}, {Taylor}  \&
  {Tittley}}{{Harvey} et~al.}{2015}]{Harvey2015}
{Harvey} D.,  {Massey} R.,  {Kitching} T.,  {Taylor} A.,   {Tittley} E.,  2015,
  \mn@doi [Science] {10.1126/science.1261381}, \href
  {https://ui.adsabs.harvard.edu/abs/2015Sci...347.1462H} {347, 1462}

\bibitem[\protect\citeauthoryear{{Hilton} et~al.,}{{Hilton}
  et~al.}{2021}]{2021ApJS..253....3H}
{Hilton} M.,  et~al., 2021, \mn@doi [\apjs] {10.3847/1538-4365/abd023}, \href
  {https://ui.adsabs.harvard.edu/abs/2021ApJS..253....3H} {253, 3}

\bibitem[\protect\citeauthoryear{{Hogan} et~al.,}{{Hogan}
  et~al.}{2015a}]{2015MNRAS.453.1201H}
{Hogan} M.~T.,  et~al., 2015a, \mn@doi [\mnras] {10.1093/mnras/stv1517}, \href
  {https://ui.adsabs.harvard.edu/abs/2015MNRAS.453.1201H} {453, 1201}

\bibitem[\protect\citeauthoryear{{Hogan} et~al.,}{{Hogan}
  et~al.}{2015b}]{2015MNRAS.453.1223H}
{Hogan} M.~T.,  et~al., 2015b, \mn@doi [\mnras] {10.1093/mnras/stv1518}, \href
  {https://ui.adsabs.harvard.edu/abs/2015MNRAS.453.1223H} {453, 1223}

\bibitem[\protect\citeauthoryear{{Jauzac} et~al.,}{{Jauzac}
  et~al.}{2018}]{2018MNRAS.481.2901J}
{Jauzac} M.,  et~al., 2018, \mn@doi [\mnras] {10.1093/mnras/sty2366}, \href
  {https://ui.adsabs.harvard.edu/abs/2018MNRAS.481.2901J} {481, 2901}

\bibitem[\protect\citeauthoryear{{Jullo}, {Kneib}, {Limousin},
  {El{\'\i}asd{\'o}ttir}, {Marshall}  \& {Verdugo}}{{Jullo}
  et~al.}{2007}]{2007NJPh....9..447J}
{Jullo} E.,  {Kneib} J.~P.,  {Limousin} M.,  {El{\'\i}asd{\'o}ttir} {\'A}.,
  {Marshall} P.~J.,   {Verdugo} T.,  2007, \mn@doi [New Journal of Physics]
  {10.1088/1367-2630/9/12/447}, \href
  {https://ui.adsabs.harvard.edu/abs/2007NJPh....9..447J} {9, 447}

\bibitem[\protect\citeauthoryear{{Kaiser} et~al.,}{{Kaiser}
  et~al.}{2002}]{Kaiser2002}
{Kaiser} N.,  et~al., 2002, in {Tyson} J.~A.,  {Wolff} S.,  eds,  Society of
  Photo-Optical Instrumentation Engineers (SPIE) Conference Series Vol. 4836,
  Survey and Other Telescope Technologies and Discoveries. pp 154--164,
  \mn@doi{10.1117/12.457365}

\bibitem[\protect\citeauthoryear{{Kalita} \& {Ebeling}}{{Kalita} \&
  {Ebeling}}{2019}]{2019ApJ...887..158K}
{Kalita} B.~S.,  {Ebeling} H.,  2019, \mn@doi [\apj]
  {10.3847/1538-4357/ab5184}, \href
  {https://ui.adsabs.harvard.edu/abs/2019ApJ...887..158K} {887, 158}

\bibitem[\protect\citeauthoryear{{Kneib} \& {Natarajan}}{{Kneib} \&
  {Natarajan}}{2011}]{Kneib2011}
{Kneib} J.-P.,  {Natarajan} P.,  2011, \mn@doi [\aapr]
  {10.1007/s00159-011-0047-3}, \href
  {https://ui.adsabs.harvard.edu/abs/2011A&ARv..19...47K} {19, 47}

\bibitem[\protect\citeauthoryear{{LaRoque} et~al.,}{{LaRoque}
  et~al.}{2003}]{2003ApJ...583..559L}
{LaRoque} S.~J.,  et~al., 2003, \mn@doi [\apj] {10.1086/345500}, \href
  {https://ui.adsabs.harvard.edu/abs/2003ApJ...583..559L} {583, 559}

\bibitem[\protect\citeauthoryear{{Lagattuta} et~al.,}{{Lagattuta}
  et~al.}{2023}]{Lagattuta23}
{Lagattuta} D.~J.,  et~al., 2023, \mn@doi [\mnras] {10.1093/mnras/stad803},
  \href {https://ui.adsabs.harvard.edu/abs/2023MNRAS.522.1091L} {522, 1091}

\bibitem[\protect\citeauthoryear{{Limousin} et~al.,}{{Limousin}
  et~al.}{2012}]{2012A&A...544A..71L}
{Limousin} M.,  et~al., 2012, \mn@doi [\aap] {10.1051/0004-6361/201117921},
  \href {https://ui.adsabs.harvard.edu/abs/2012A&A...544A..71L} {544, A71}

\bibitem[\protect\citeauthoryear{{Lotz} et~al.,}{{Lotz}
  et~al.}{2017}]{Lotz2017}
{Lotz} J.~M.,  et~al., 2017, \mn@doi [\apj] {10.3847/1538-4357/837/1/97}, \href
  {https://ui.adsabs.harvard.edu/abs/2017ApJ...837...97L} {837, 97}

\bibitem[\protect\citeauthoryear{{Lovisari} et~al.,}{{Lovisari}
  et~al.}{2020}]{2020ApJ...892..102L}
{Lovisari} L.,  et~al., 2020, \mn@doi [\apj] {10.3847/1538-4357/ab7997}, \href
  {https://ui.adsabs.harvard.edu/abs/2020ApJ...892..102L} {892, 102}

\bibitem[\protect\citeauthoryear{{Mahler} et~al.,}{{Mahler}
  et~al.}{2023}]{mahler2023}
{Mahler} G.,  et~al., 2023, \mn@doi [\apj] {10.3847/1538-4357/acaea9}, \href
  {https://ui.adsabs.harvard.edu/abs/2023ApJ...945...49M} {945, 49}

\bibitem[\protect\citeauthoryear{{Malte Sch{\"a}fer} \& {Bartelmann}}{{Malte
  Sch{\"a}fer} \& {Bartelmann}}{2007}]{2007MNRAS.377..253M}
{Malte Sch{\"a}fer} B.,  {Bartelmann} M.,  2007, \mn@doi [\mnras]
  {10.1111/j.1365-2966.2007.11596.x}, \href
  {https://ui.adsabs.harvard.edu/abs/2007MNRAS.377..253M} {377, 253}

\bibitem[\protect\citeauthoryear{{Mann} \& {Ebeling}}{{Mann} \&
  {Ebeling}}{2012}]{Mann2012}
{Mann} A.~W.,  {Ebeling} H.,  2012, \mn@doi [\mnras]
  {10.1111/j.1365-2966.2011.20170.x}, \href
  {https://ui.adsabs.harvard.edu/abs/2012MNRAS.420.2120M} {420, 2120}

\bibitem[\protect\citeauthoryear{{Mantz} et~al.,}{{Mantz}
  et~al.}{2015}]{Mantz2015}
{Mantz} A.~B.,  et~al., 2015, \mn@doi [\mnras] {10.1093/mnras/stu2096}, \href
  {https://ui.adsabs.harvard.edu/abs/2015MNRAS.446.2205M} {446, 2205}

\bibitem[\protect\citeauthoryear{{Markevitch}, {Gonzalez}, {David},
  {Vikhlinin}, {Murray}, {Forman}, {Jones}  \& {Tucker}}{{Markevitch}
  et~al.}{2002}]{Markevitch2002}
{Markevitch} M.,  {Gonzalez} A.~H.,  {David} L.,  {Vikhlinin} A.,  {Murray} S.,
   {Forman} W.,  {Jones} C.,   {Tucker} W.,  2002, \mn@doi [\apjl]
  {10.1086/339619}, \href
  {https://ui.adsabs.harvard.edu/abs/2002ApJ...567L..27M} {567, L27}

\bibitem[\protect\citeauthoryear{{Masters} \& {Capak}}{{Masters} \&
  {Capak}}{2011}]{2011PASP..123..638M}
{Masters} D.,  {Capak} P.,  2011, \mn@doi [\pasp] {10.1086/660023}, \href
  {https://ui.adsabs.harvard.edu/abs/2011PASP..123..638M} {123, 638}

\bibitem[\protect\citeauthoryear{{McDonald} et~al.,}{{McDonald}
  et~al.}{2016}]{2016ApJ...817...86M}
{McDonald} M.,  et~al., 2016, \mn@doi [\apj] {10.3847/0004-637X/817/2/86},
  \href {https://ui.adsabs.harvard.edu/abs/2016ApJ...817...86M} {817, 86}

\bibitem[\protect\citeauthoryear{{McLean} et~al.,}{{McLean}
  et~al.}{2010}]{MOSFIRE2010}
{McLean} I.~S.,  et~al., 2010, in {McLean} I.~S.,  {Ramsay} S.~K.,   {Takami}
  H.,  eds,  Society of Photo-Optical Instrumentation Engineers (SPIE)
  Conference Series Vol. 7735, Ground-based and Airborne Instrumentation for
  Astronomy III. p. 77351E, \mn@doi{10.1117/12.856715}

\bibitem[\protect\citeauthoryear{{McLean} et~al.,}{{McLean}
  et~al.}{2012}]{MOSFIRE2012}
{McLean} I.~S.,  et~al., 2012, in {McLean} I.~S.,  {Ramsay} S.~K.,   {Takami}
  H.,  eds,  Society of Photo-Optical Instrumentation Engineers (SPIE)
  Conference Series Vol. 8446, Ground-based and Airborne Instrumentation for
  Astronomy IV. p. 84460J, \mn@doi{10.1117/12.924794}

\bibitem[\protect\citeauthoryear{{Meneghetti} et~al.,}{{Meneghetti}
  et~al.}{2023}]{2023A&A...678L...2M}
{Meneghetti} M.,  et~al., 2023, \mn@doi [\aap] {10.1051/0004-6361/202346975},
  \href {https://ui.adsabs.harvard.edu/abs/2023A&A...678L...2M} {678, L2}

\bibitem[\protect\citeauthoryear{{Merloni} et~al.,}{{Merloni}
  et~al.}{2024}]{2024A&A...682A..34M}
{Merloni} A.,  et~al., 2024, \mn@doi [\aap] {10.1051/0004-6361/202347165},
  \href {https://ui.adsabs.harvard.edu/abs/2024A&A...682A..34M} {682, A34}

\bibitem[\protect\citeauthoryear{{Merritt}}{{Merritt}}{1985}]{1985ApJ...289...18M}
{Merritt} D.,  1985, \mn@doi [\apj] {10.1086/162860}, \href
  {https://ui.adsabs.harvard.edu/abs/1985ApJ...289...18M} {289, 18}

\bibitem[\protect\citeauthoryear{{Morishita} et~al.,}{{Morishita}
  et~al.}{2023}]{2023ApJ...947L..24M}
{Morishita} T.,  et~al., 2023, \mn@doi [\apjl] {10.3847/2041-8213/acb99e},
  \href {https://ui.adsabs.harvard.edu/abs/2023ApJ...947L..24M} {947, L24}

\bibitem[\protect\citeauthoryear{{Navarro}, {Frenk}  \& {White}}{{Navarro}
  et~al.}{1997}]{1997ApJ...490..493N}
{Navarro} J.~F.,  {Frenk} C.~S.,   {White} S. D.~M.,  1997, \mn@doi [\apj]
  {10.1086/304888}, \href
  {https://ui.adsabs.harvard.edu/abs/1997ApJ...490..493N} {490, 493}

\bibitem[\protect\citeauthoryear{{Newman}, {Treu}, {Ellis}  \& {Sand
  }}{{Newman} et~al.}{2013}]{2013ApJ...765...25N}
{Newman} A.~B.,  {Treu} T.,  {Ellis} R.~S.,   {Sand } D.~J.,  2013, \mn@doi
  [\apj] {10.1088/0004-637X/765/1/25}, \href
  {https://ui.adsabs.harvard.edu/abs/2013ApJ...765...25N} {765, 25}

\bibitem[\protect\citeauthoryear{{Newman}, {Ellis}, {Andreon}, {Treu},
  {Raichoor}  \& {Trinchieri}}{{Newman} et~al.}{2014}]{2014ApJ...788...51N}
{Newman} A.~B.,  {Ellis} R.~S.,  {Andreon} S.,  {Treu} T.,  {Raichoor} A.,
  {Trinchieri} G.,  2014, \mn@doi [\apj] {10.1088/0004-637X/788/1/51}, \href
  {https://ui.adsabs.harvard.edu/abs/2014ApJ...788...51N} {788, 51}

\bibitem[\protect\citeauthoryear{{O'Dea} et~al.,}{{O'Dea}
  et~al.}{2010}]{2010ApJ...719.1619O}
{O'Dea} K.~P.,  et~al., 2010, \mn@doi [\apj] {10.1088/0004-637X/719/2/1619},
  \href {https://ui.adsabs.harvard.edu/abs/2010ApJ...719.1619O} {719, 1619}

\bibitem[\protect\citeauthoryear{{Oemler}}{{Oemler}}{1976}]{1976ApJ...209..693O}
{Oemler} A. J.,  1976, \mn@doi [\apj] {10.1086/154769}, \href
  {https://ui.adsabs.harvard.edu/abs/1976ApJ...209..693O} {209, 693}

\bibitem[\protect\citeauthoryear{{Ogrean} et~al.,}{{Ogrean}
  et~al.}{2016}]{2016ApJ...819..113O}
{Ogrean} G.~A.,  et~al., 2016, \mn@doi [\apj] {10.3847/0004-637X/819/2/113},
  \href {https://ui.adsabs.harvard.edu/abs/2016ApJ...819..113O} {819, 113}

\bibitem[\protect\citeauthoryear{{Oke} et~al.,}{{Oke} et~al.}{1995}]{LRIS1995}
{Oke} J.~B.,  et~al., 1995, \mn@doi [\pasp] {10.1086/133562}, \href
  {https://ui.adsabs.harvard.edu/abs/1995PASP..107..375O} {107, 375}

\bibitem[\protect\citeauthoryear{{Owers}, {Couch}, {Nulsen}  \&
  {Randall}}{{Owers} et~al.}{2012}]{2012ApJ...750L..23O}
{Owers} M.~S.,  {Couch} W.~J.,  {Nulsen} P. E.~J.,   {Randall} S.~W.,  2012,
  \mn@doi [\apjl] {10.1088/2041-8205/750/1/L23}, \href
  {https://ui.adsabs.harvard.edu/abs/2012ApJ...750L..23O} {750, L23}

\bibitem[\protect\citeauthoryear{{Papovich} et~al.,}{{Papovich}
  et~al.}{2010}]{2010ApJ...716.1503P}
{Papovich} C.,  et~al., 2010, \mn@doi [\apj] {10.1088/0004-637X/716/2/1503},
  \href {https://ui.adsabs.harvard.edu/abs/2010ApJ...716.1503P} {716, 1503}

\bibitem[\protect\citeauthoryear{{Planck Collaboration} et~al.,}{{Planck
  Collaboration} et~al.}{2014}]{Planck2014}
{Planck Collaboration} et~al., 2014, \mn@doi [\aap]
  {10.1051/0004-6361/201321521}, \href
  {https://ui.adsabs.harvard.edu/abs/2014A&A...571A..20P} {571, A20}

\bibitem[\protect\citeauthoryear{{Pointecouteau}, {Arnaud}  \&
  {Pratt}}{{Pointecouteau} et~al.}{2005}]{2005A&A...435....1P}
{Pointecouteau} E.,  {Arnaud} M.,   {Pratt} G.~W.,  2005, \mn@doi [\aap]
  {10.1051/0004-6361:20042569}, \href
  {https://ui.adsabs.harvard.edu/abs/2005A&A...435....1P} {435, 1}

\bibitem[\protect\citeauthoryear{Prochaska et~al.,}{Prochaska
  et~al.}{2020}]{pypeit:joss_pub}
Prochaska J.~X.,  et~al., 2020, \mn@doi [Journal of Open Source Software]
  {10.21105/joss.02308}, 5, 2308

\bibitem[\protect\citeauthoryear{{Randall}, {Sarazin}  \& {Ricker}}{{Randall}
  et~al.}{2002}]{2002ApJ...577..579R}
{Randall} S.~W.,  {Sarazin} C.~L.,   {Ricker} P.~M.,  2002, \mn@doi [\apj]
  {10.1086/342239}, \href
  {https://ui.adsabs.harvard.edu/abs/2002ApJ...577..579R} {577, 579}

\bibitem[\protect\citeauthoryear{{Reiprich}}{{Reiprich}}{2001}]{Reiprich2001}
{Reiprich} T.~H.,  2001, PhD thesis, Max-Planck-Institute for Extraterrestrial
  Physics, Garching

\bibitem[\protect\citeauthoryear{{Repp} \& {Ebeling}}{{Repp} \&
  {Ebeling}}{2018}]{Repp2018}
{Repp} A.,  {Ebeling} H.,  2018, \mn@doi [\mnras] {10.1093/mnras/sty1489},
  \href {https://ui.adsabs.harvard.edu/abs/2018MNRAS.479..844R} {479, 844}

\bibitem[\protect\citeauthoryear{{Richard} et~al.,}{{Richard}
  et~al.}{2010}]{Richard2010}
{Richard} J.,  et~al., 2010, \mn@doi [\mnras]
  {10.1111/j.1365-2966.2009.16274.x}, \href
  {https://ui.adsabs.harvard.edu/abs/2010MNRAS.404..325R} {404, 325}

\bibitem[\protect\citeauthoryear{{Richard} et~al.,}{{Richard}
  et~al.}{2021}]{Richard2021}
{Richard} J.,  et~al., 2021, \mn@doi [\aap] {10.1051/0004-6361/202039462},
  \href {https://ui.adsabs.harvard.edu/abs/2021A&A...646A..83R} {646, A83}

\bibitem[\protect\citeauthoryear{{Rose} et~al.,}{{Rose}
  et~al.}{2022}]{2022MNRAS.509.2869R}
{Rose} T.,  et~al., 2022, \mn@doi [\mnras] {10.1093/mnras/stab3217}, \href
  {https://ui.adsabs.harvard.edu/abs/2022MNRAS.509.2869R} {509, 2869}

\bibitem[\protect\citeauthoryear{{Rossetti}, {Gastaldello}, {Eckert}, {Della
  Torre}, {Pantiri}, {Cazzoletti}  \& {Molendi}}{{Rossetti}
  et~al.}{2017}]{2017MNRAS.468.1917R}
{Rossetti} M.,  {Gastaldello} F.,  {Eckert} D.,  {Della Torre} M.,  {Pantiri}
  G.,  {Cazzoletti} P.,   {Molendi} S.,  2017, \mn@doi [\mnras]
  {10.1093/mnras/stx493}, \href
  {https://ui.adsabs.harvard.edu/abs/2017MNRAS.468.1917R} {468, 1917}

\bibitem[\protect\citeauthoryear{{Saro}, {Mohr}, {Bazin}  \& {Dolag}}{{Saro}
  et~al.}{2013}]{2013ApJ...772...47S}
{Saro} A.,  {Mohr} J.~J.,  {Bazin} G.,   {Dolag} K.,  2013, \mn@doi [\apj]
  {10.1088/0004-637X/772/1/47}, \href
  {https://ui.adsabs.harvard.edu/abs/2013ApJ...772...47S} {772, 47}

\bibitem[\protect\citeauthoryear{Shapiro \& Wilk}{Shapiro \&
  Wilk}{1965}]{10.1093/biomet/52.3-4.591}
Shapiro S.~S.,  Wilk M.~B.,  1965, \mn@doi [Biometrika]
  {10.1093/biomet/52.3-4.591}, 52, 591

\bibitem[\protect\citeauthoryear{{Sharon} et~al.,}{{Sharon}
  et~al.}{2015}]{2015ApJ...814...21S}
{Sharon} K.,  et~al., 2015, \mn@doi [\apj] {10.1088/0004-637X/814/1/21}, \href
  {https://ui.adsabs.harvard.edu/abs/2015ApJ...814...21S} {814, 21}

\bibitem[\protect\citeauthoryear{{Steidel}, {Adelberger}, {Dickinson},
  {Giavalisco}, {Pettini}  \& {Kellogg}}{{Steidel}
  et~al.}{1998}]{1998ApJ...492..428S}
{Steidel} C.~C.,  {Adelberger} K.~L.,  {Dickinson} M.,  {Giavalisco} M.,
  {Pettini} M.,   {Kellogg} M.,  1998, \mn@doi [\apj] {10.1086/305073}, \href
  {https://ui.adsabs.harvard.edu/abs/1998ApJ...492..428S} {492, 428}

\bibitem[\protect\citeauthoryear{{Steidel} et~al.,}{{Steidel}
  et~al.}{2014}]{2014ApJ...795..165S}
{Steidel} C.~C.,  et~al., 2014, \mn@doi [\apj] {10.1088/0004-637X/795/2/165},
  \href {https://ui.adsabs.harvard.edu/abs/2014ApJ...795..165S} {795, 165}

\bibitem[\protect\citeauthoryear{{Sunyaev} \& {Zeldovich}}{{Sunyaev} \&
  {Zeldovich}}{1980}]{1980ARA&A..18..537S}
{Sunyaev} R.~A.,  {Zeldovich} I.~B.,  1980, \mn@doi [\araa]
  {10.1146/annurev.aa.18.090180.002541}, \href
  {https://ui.adsabs.harvard.edu/abs/1980ARA&A..18..537S} {18, 537}

\bibitem[\protect\citeauthoryear{{Swetz} et~al.,}{{Swetz}
  et~al.}{2011}]{2011ApJS..194...41S}
{Swetz} D.~S.,  et~al., 2011, \mn@doi [\apjs] {10.1088/0067-0049/194/2/41},
  \href {https://ui.adsabs.harvard.edu/abs/2011ApJS..194...41S} {194, 41}

\bibitem[\protect\citeauthoryear{{Takey}, {Schwope}  \& {Lamer}}{{Takey}
  et~al.}{2013}]{2013A&A...558A..75T}
{Takey} A.,  {Schwope} A.,   {Lamer} G.,  2013, \mn@doi [\aap]
  {10.1051/0004-6361/201220213}, \href
  {https://ui.adsabs.harvard.edu/abs/2013A&A...558A..75T} {558, A75}

\bibitem[\protect\citeauthoryear{{Timmerman}, {van Weeren}, {Botteon},
  {R{\"o}ttgering}, {McNamara}, {Sweijen}, {B{\^\i}rzan}  \&
  {Morabito}}{{Timmerman} et~al.}{2022}]{2022A&A...668A..65T}
{Timmerman} R.,  {van Weeren} R.~J.,  {Botteon} A.,  {R{\"o}ttgering} H.~J.~A.,
   {McNamara} B.~R.,  {Sweijen} F.,  {B{\^\i}rzan} L.,   {Morabito} L.~K.,
  2022, \mn@doi [\aap] {10.1051/0004-6361/202243936}, \href
  {https://ui.adsabs.harvard.edu/abs/2022A&A...668A..65T} {668, A65}

\bibitem[\protect\citeauthoryear{{Vikhlinin}, {Markevitch}  \&
  {Murray}}{{Vikhlinin} et~al.}{2001}]{Vikhlinin2001}
{Vikhlinin} A.,  {Markevitch} M.,   {Murray} S.~S.,  2001, \mn@doi [\apj]
  {10.1086/320078}, \href
  {https://ui.adsabs.harvard.edu/abs/2001ApJ...551..160V} {551, 160}

\bibitem[\protect\citeauthoryear{{Vikhlinin} et~al.,}{{Vikhlinin}
  et~al.}{2009}]{Vikhlinin2009}
{Vikhlinin} A.,  et~al., 2009, \mn@doi [\apj] {10.1088/0004-637X/692/2/1060},
  \href {https://ui.adsabs.harvard.edu/abs/2009ApJ...692.1060V} {692, 1060}

\bibitem[\protect\citeauthoryear{{Villforth} et~al.,}{{Villforth}
  et~al.}{2017}]{2017MNRAS.466..812V}
{Villforth} C.,  et~al., 2017, \mn@doi [\mnras] {10.1093/mnras/stw3037}, \href
  {https://ui.adsabs.harvard.edu/abs/2017MNRAS.466..812V} {466, 812}

\bibitem[\protect\citeauthoryear{{Visvanathan} \& {Sandage}}{{Visvanathan} \&
  {Sandage}}{1977}]{visvanathan1977}
{Visvanathan} N.,  {Sandage} A.,  1977, \mn@doi [\apj] {10.1086/155464}, \href
  {https://ui.adsabs.harvard.edu/abs/1977ApJ...216..214V} {216, 214}

\bibitem[\protect\citeauthoryear{{Voges} et~al.,}{{Voges}
  et~al.}{1999}]{Voges1999}
{Voges} W.,  et~al., 1999, \aap, \href
  {https://ui.adsabs.harvard.edu/abs/1999A&A...349..389V} {349, 389}

\bibitem[\protect\citeauthoryear{{Voges}, {Henry}, {Briel}, {B{\"o}hringer},
  {Mullis}, {Gioia}  \& {Huchra}}{{Voges} et~al.}{2001}]{2001ApJ...553L.119V}
{Voges} W.,  {Henry} J.~P.,  {Briel} U.~G.,  {B{\"o}hringer} H.,  {Mullis}
  C.~R.,  {Gioia} I.~M.,   {Huchra} J.~P.,  2001, \mn@doi [\apjl]
  {10.1086/320673}, \href
  {https://ui.adsabs.harvard.edu/abs/2001ApJ...553L.119V} {553, L119}

\bibitem[\protect\citeauthoryear{{Vollmer}, {Cayatte}, {Balkowski}  \&
  {Duschl}}{{Vollmer} et~al.}{2001}]{2001ApJ...561..708V}
{Vollmer} B.,  {Cayatte} V.,  {Balkowski} C.,   {Duschl} W.~J.,  2001, \mn@doi
  [\apj] {10.1086/323368}, \href
  {https://ui.adsabs.harvard.edu/abs/2001ApJ...561..708V} {561, 708}

\bibitem[\protect\citeauthoryear{{Wang} et~al.,}{{Wang}
  et~al.}{2016}]{2016ApJ...828...56W}
{Wang} T.,  et~al., 2016, \mn@doi [\apj] {10.3847/0004-637X/828/1/56}, \href
  {https://ui.adsabs.harvard.edu/abs/2016ApJ...828...56W} {828, 56}

\bibitem[\protect\citeauthoryear{{Wilber} et~al.,}{{Wilber}
  et~al.}{2018}]{2018MNRAS.476.3415W}
{Wilber} A.,  et~al., 2018, \mn@doi [\mnras] {10.1093/mnras/sty414}, \href
  {https://ui.adsabs.harvard.edu/abs/2018MNRAS.476.3415W} {476, 3415}

\bibitem[\protect\citeauthoryear{{Zalesky} \& {Ebeling}}{{Zalesky} \&
  {Ebeling}}{2020}]{Zalesky2020}
{Zalesky} L.,  {Ebeling} H.,  2020, \mn@doi [\mnras] {10.1093/mnras/staa2180},
  \href {https://ui.adsabs.harvard.edu/abs/2020MNRAS.498.1121Z} {498, 1121}

\bibitem[\protect\citeauthoryear{{Zaznobin} et~al.,}{{Zaznobin}
  et~al.}{2023}]{2023AstL...49..599Z}
{Zaznobin} I.~A.,  et~al., 2023, \mn@doi [Astronomy Letters]
  {10.1134/S1063773723110105}, \href
  {https://ui.adsabs.harvard.edu/abs/2023AstL...49..599Z} {49, 599}

\bibitem[\protect\citeauthoryear{{Zitrin}, {Broadhurst}, {Barkana}, {Rephaeli}
  \& {Ben{\'\i}tez}}{{Zitrin} et~al.}{2011}]{2011MNRAS.410.1939Z}
{Zitrin} A.,  {Broadhurst} T.,  {Barkana} R.,  {Rephaeli} Y.,   {Ben{\'\i}tez}
  N.,  2011, \mn@doi [\mnras] {10.1111/j.1365-2966.2010.17574.x}, \href
  {https://ui.adsabs.harvard.edu/abs/2011MNRAS.410.1939Z} {410, 1939}

\bibitem[\protect\citeauthoryear{{van Weeren}, {de Gasperin}, {Akamatsu},
  {Br{\"u}ggen}, {Feretti}, {Kang}, {Stroe}  \& {Zandanel}}{{van Weeren}
  et~al.}{2019}]{vanWeeren2019}
{van Weeren} R.~J.,  {de Gasperin} F.,  {Akamatsu} H.,  {Br{\"u}ggen} M.,
  {Feretti} L.,  {Kang} H.,  {Stroe} A.,   {Zandanel} F.,  2019, \mn@doi [\ssr]
  {10.1007/s11214-019-0584-z}, \href
  {https://ui.adsabs.harvard.edu/abs/2019SSRv..215...16V} {215, 16}

\makeatother
\end{thebibliography}
\bibliographystyle{mnras}

\appendix
\section{The compilation of cluster candidates from the RASS}
\label{sec:app-rass}

The eMACS project was launched in 2010, almost a decade before the launch of {\it eROSITA}, and was motivated by the realization that the RASS, the only existing all-sky survey with an imaging X-ray telescope, had not been fully exploited in search for massive clusters. In fact, there are thousands of X-ray sources below the MACS flux limit that can be examined in an even more ambitious cluster survey, and yet have never been tapped systematically. Our approach to using this valuable resource is detailed in the following.

\subsection{The RASS Faint-Source Catalogue}

Conducted in great circles through the ecliptic poles that result in a median exposure time of 370s, the RASS was, until 2020, the only existing all-sky survey conducted in soft X-rays (0.1--2.4 keV) and a powerful resource for X-ray studies that require extremely large areal coverage. Two official source catalogues (the Bright Source Catalogue, BSC, and the Faint Source Catalogue, FSC) as well as images and photon event tables for the entire survey are publicly available. The BSC contains only sources  comprising at least 15 net photons, and featuring count rates in excess of 0.05 ct s$^{-1}$; its defining criterion, however, is the maximum likelihood value assigned to all RASS sources by the official RASS source detection algorithm. The BSC comprises 18,802 sources above this preset detection-likelihood threshold. {\it MACS is based on sources selected from the BSC.} 
  
\begin{figure}
\hspace*{-3mm}\includegraphics[width=0.49\textwidth]{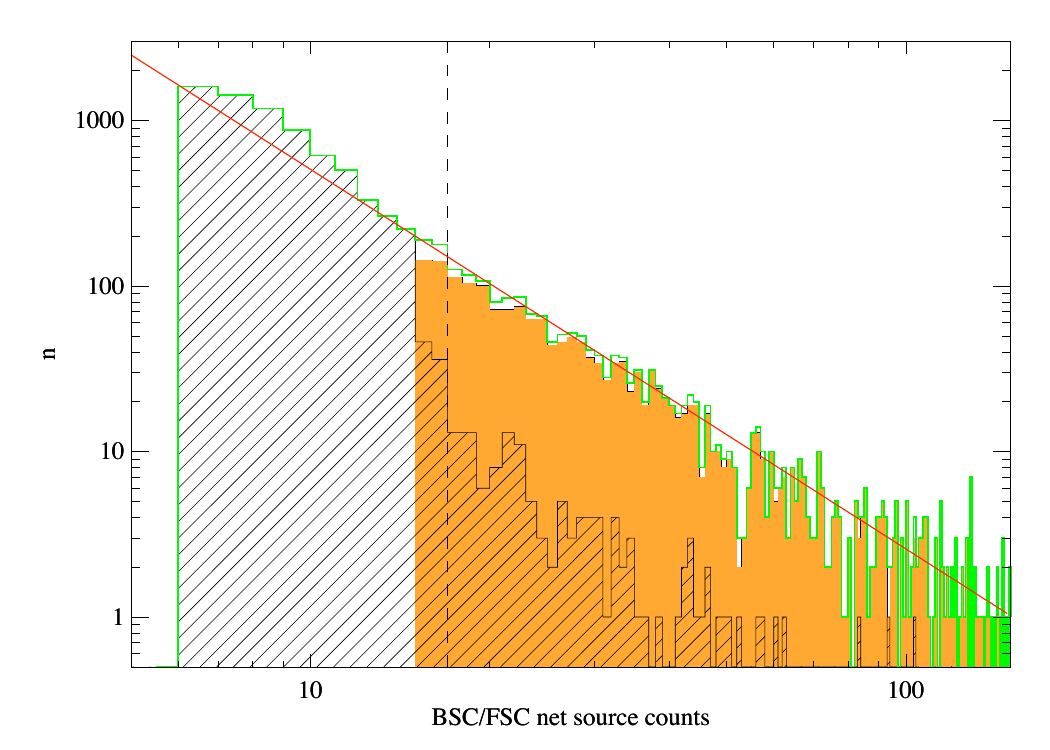}
\hspace*{-3mm}\includegraphics[width=0.49\textwidth]{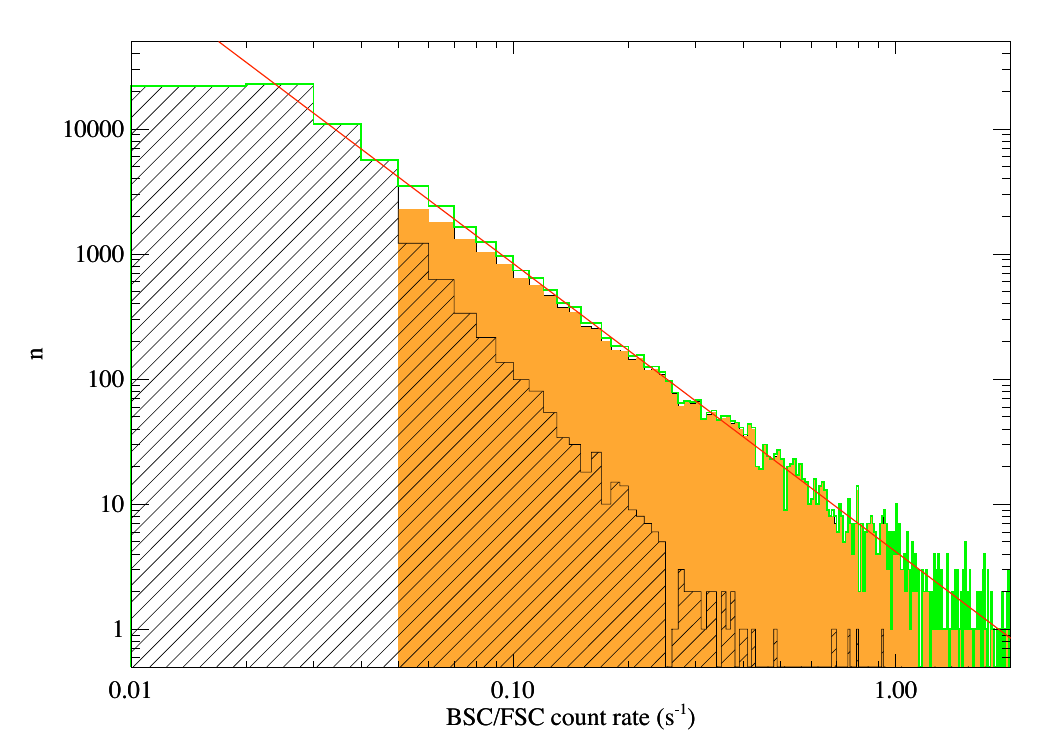}
\caption{Differential logN-logS distributions for net counts (top) and count rates (bottom) of BSC (orange) and FSC (shaded) sources at  $|b|\geq 20^\circ$. For the count distribution (top) only sources featuring exposure times between 200 and 300\,s were considered, thereby mitigating the impact of exposure time variations on the RASS detection efficiency. The source count limit of 17 photons adopted during the compilation of the MACS sample is marked by the dashed vertical line. The green histograms show the distributions that result when both catalogues are combined; best-fitting power laws are shown in red. 
} \label{fig:rass-ctcr}
\end{figure}

The much larger FSC, which lists over 100,000 additional sources comprising as few as 6 net photons, has seen very little use in scientific studies, due to its poorly known statistical properties. The enormous potential of the FSC becomes, however, immediately obvious when the distributions of net counts and count rate of FSC and BSC sources are inspected and compared. Following the approach taking for MACS \citep[for details, see][]{Ebeling2001}, we limit both data sets to the extragalactic sky ($|b|\geq 20^\circ$)  and the equatorial sky observable from Mauna Kea ($-40^\circ < \delta -80^\circ$) 
but apply no other cuts. The resulting histograms (proxies for the respective differential logN-logS distributions) for the BSC and FSC are plotted in Fig.~\ref{fig:rass-ctcr} and demonstrate that

\begin{enumerate}
\item in spite of its suggestive minimal source count value of 15, the BSC is not complete above this limit: 9\% of all RASS sources featuring more than 15 counts are in fact listed not in the BSC, but in the FSC;
\item for sources comprising between 6 and 11 photons, the FSC catalogue is about 20--30\% ``over-complete'', a well known phenomenon caused by an increasing number of spurious sources (many of them multiple detections of very bright or very extended sources) as the photon statistics approach the ultimate detection limit of, roughly, 5 to 6 net counts;
\item when both catalogues are combined, the differential logN--logS distributions for the resulting data sets follow a power law well below the count limit of 17 adopted for MACS, and well below the count-rate limit of 0.05 s$^{-1}$ of the BSC in general.   
\end{enumerate}

Figure~\ref{fig:rass-ctcr} demonstrates that RASS-based surveys can be successfully extended significantly beyond the limits of the BSC, a claim that is supported by the fact that the slopes of about $-2.3$ of the best-fitting power laws describing the merged FSC/BSC data sets  are in excellent agreement with the near-Euclidean value found for galaxy clusters and extragalactic RASS sources in general \citep{1998MNRAS.301..881E,Voges1999}. 

\begin{figure}
\hspace*{-1mm}\includegraphics[width=0.49\textwidth]{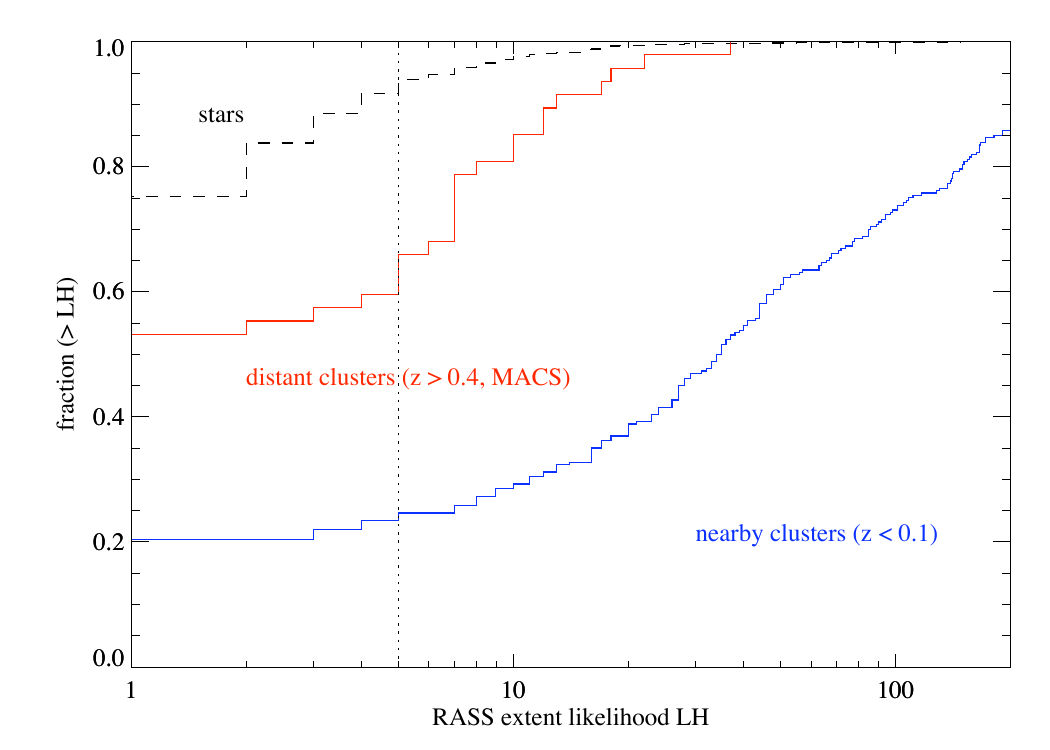}
\hspace*{-1mm}\includegraphics[width=0.49\textwidth]{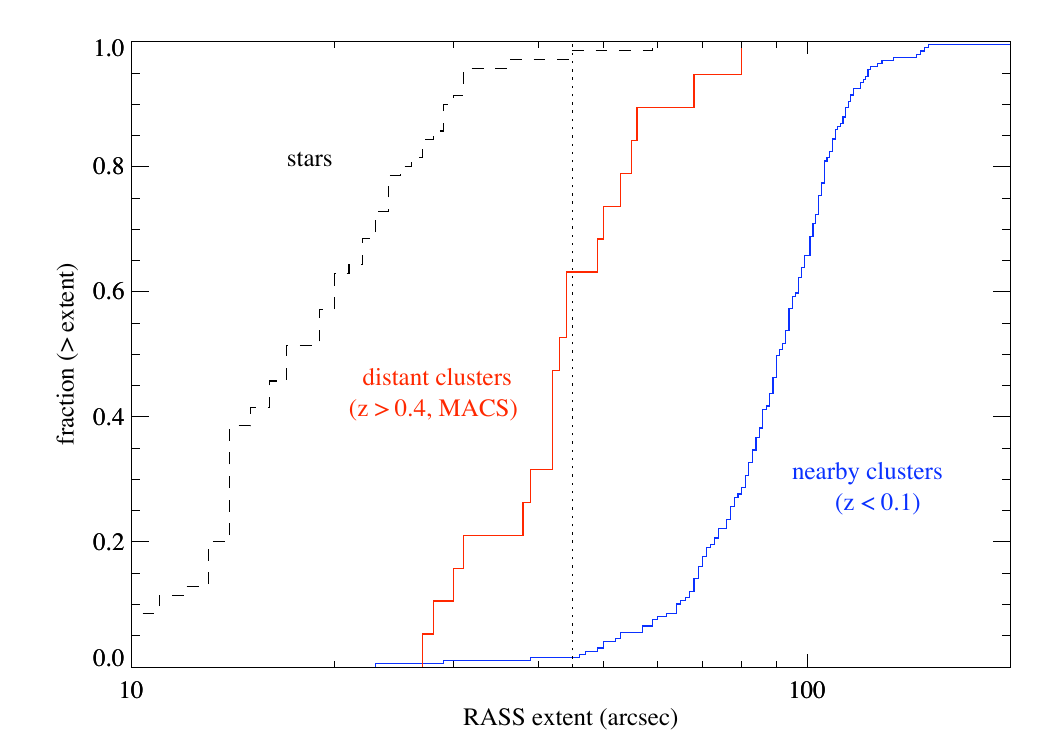}
\caption{{\it (Top)} Cumulative distribution of source extent likelihood in the RASS for three classes of objects: stars (as a reference for point sources), nearby clusters, and distant clusters. The extent likelihood distribution for stars illustrates that values below 5 (dotted vertical line) are not significant and often spurious. Note how over 60\% of all distant clusters are classified as point sources. {\it (Bottom)} Cumulative distribution of source extent for the same three object classes as shown on top, but using only sources with extent likelihood values of at least 5. The dotted vertical line marks the FWHM of the RASS point-spread function, which is also the pixel size of the RASS. Even for the minority of all distant clusters classified as extended (left panel) the assigned angular extent exceeds 1 pixel only for fewer than 40\%.}\label{fig:rass-ext}
\end{figure}

The conclusion that the RASS continues to hold great promise for {\it cluster}\/ surveys may appear premature though, given that our analysis so far used {\it all}\/ RASS sources in the extragalactic sky, i.e., a source population that is dominated by AGN and also includes a significant number of Galactic sources (primarily X-ray binaries and stars), all of which are X-ray point sources. Applying any insights gained from Fig.~\ref{fig:rass-ctcr} to clusters would indeed be impermissible if we aimed to find additional X-ray-faint systems at low redshift\footnote{and thus also of low X-ray luminosity}, where even galaxy groups exhibit significant angular extent in the RASS. {\it At $z>0.4$, however, galaxy clusters appear almost pointlike  at the angular resolution of the RASS} (Fig.~\ref{fig:rass-ext}). As a result, the background in the standard RASS detect cell is negligible for sources comprising at least 10 net photons, except for the deepest part of the RASS near the ecliptic poles where integration times approaching, and over a small solid angle in fact exceeding, 1000\,s are reached, resulting in increased background counts and thus a higher effective detection threshold in terms of net source counts \citep{2001ApJ...553L.119V}. 
  
\begin{figure}
\hspace*{-3mm}\includegraphics[width=0.52\textwidth]{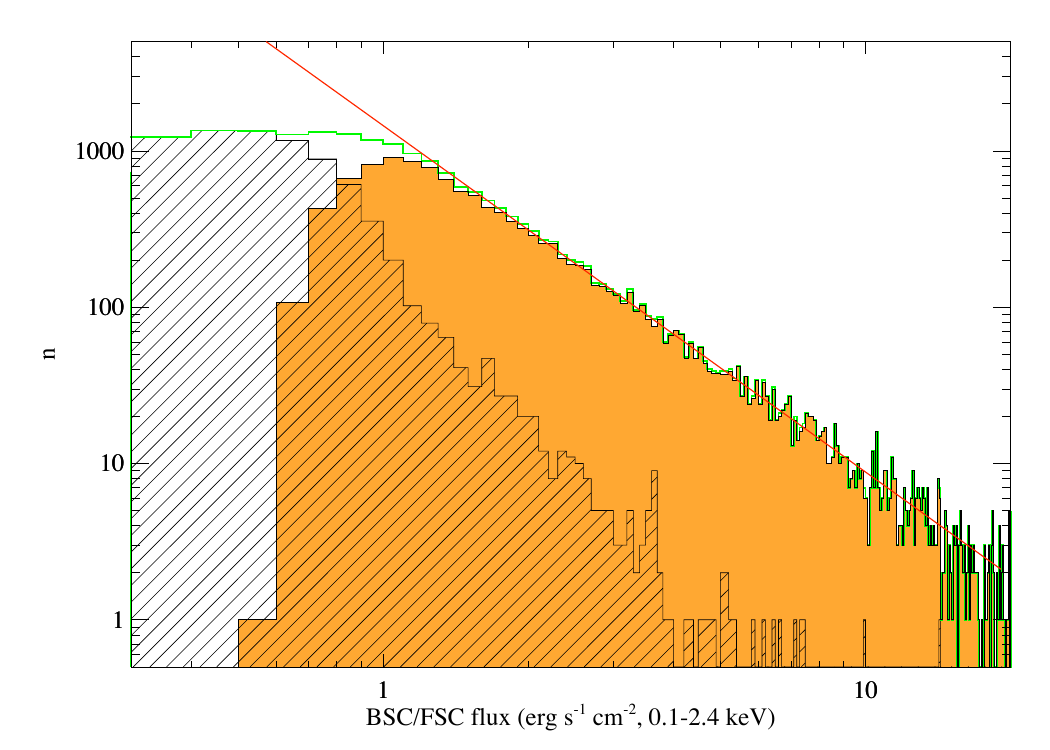}
\hspace*{-3mm}\includegraphics[width=0.52\textwidth]{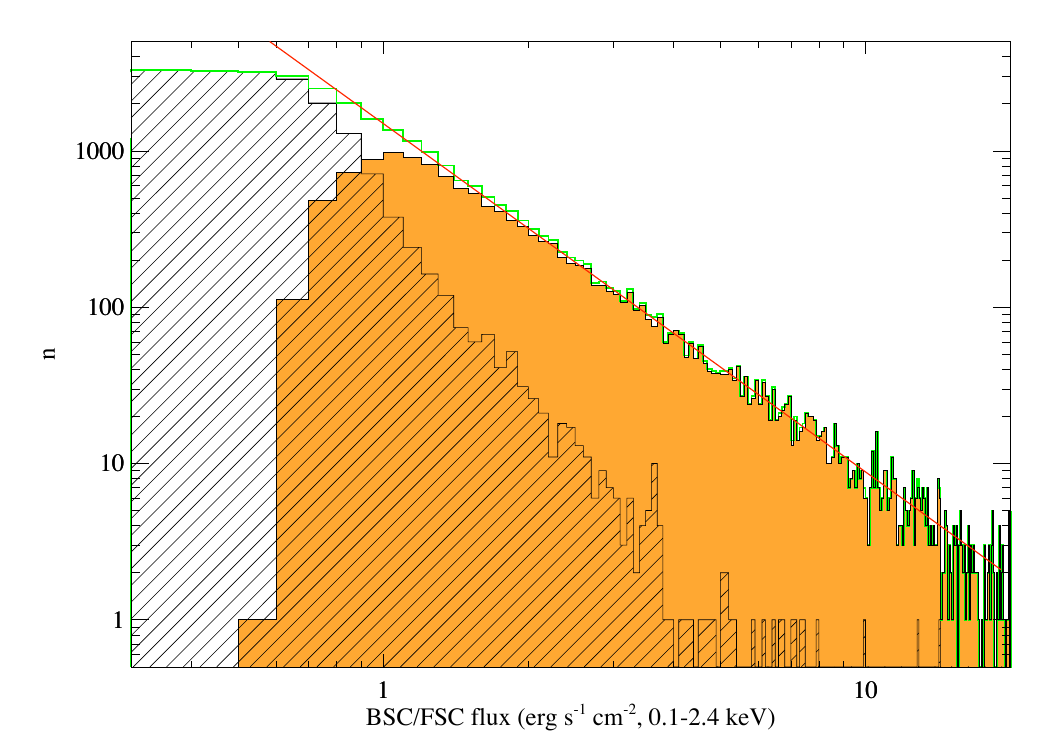}\mbox{}\\
\caption{Differential logN-logS distributions for BSC (orange) and FSC (shaded) sources at  $|b|\geq 20^\circ$. The plot on top uses only sources comprising at least 17 net photons (the threshold used by the MACS project and marked in Fig.~4); for the botttom plot this limit is lowered to 12 photons. Green histograms and red power-law models refer to the distributions resulting when the FSC and BSC catalogues are merged.\\*[-5mm]} \label{fig:rass-lnls}
\end{figure}

Although the efficiency of the RASS detection algorithm near the detection limit has to
be carefully tested as a function of RASS exposure time, the potential rewards make this effort very worthwhile. We show, in Fig.~\ref{fig:rass-lnls}, the differential log$N$-log$S$ distributions in terms of energy flux\footnote{For the conversion from count rate to flux we assumed a spectrum typical of very X-ray luminous clusters at moderately high redshift, characterized by $z=0.2$, k$T=8$ keV, $Z=0.2$, and the Galactic $n_{\rm H}$ column density in the direction of the source. Modifying these assumptions does not affect any of our conclusions noticeably.} when two different source count thresholds are applied: the top panel illustrates the impact of the FSC for sources with net counts in excess of 17 photons, the limit adopted for MACS. Not surprisingly, little can be gained above the MACS flux limit of $1\times 10^{-12}$ erg cm$^{-2}$ s$^{-1}$, but at fluxes down to  $5\times 10^{-13}$ erg cm$^{-2}$ s$^{-1}$ the inclusion of the FSC improves upon the MACS sample size by 66\%. If the count limit is lowered to 12 (bottom panel of Fig.~\ref{fig:rass-lnls}), easily achievable as shown in Fig.~\ref{fig:rass-ctcr}, the combined FSC/BSC data set increases the survey's effective solid angle noticeably already within the MACS regime (i.e., at $f_{\rm X}>1\times 10^{-12}$ erg cm$^{-2}$ s$^{-1}$), and allows the compilation of a sample of {\it more than twice the size of MACS}\/ at $f_{\rm X}>5\times 10^{-13}$ erg cm$^{-2}$ s$^{-1}$.

\subsection{Optical screening}

\begin{figure}
\hspace*{-3mm}\includegraphics[width=0.5\textwidth,clip,trim=15mm 60mm 10mm 40mm]{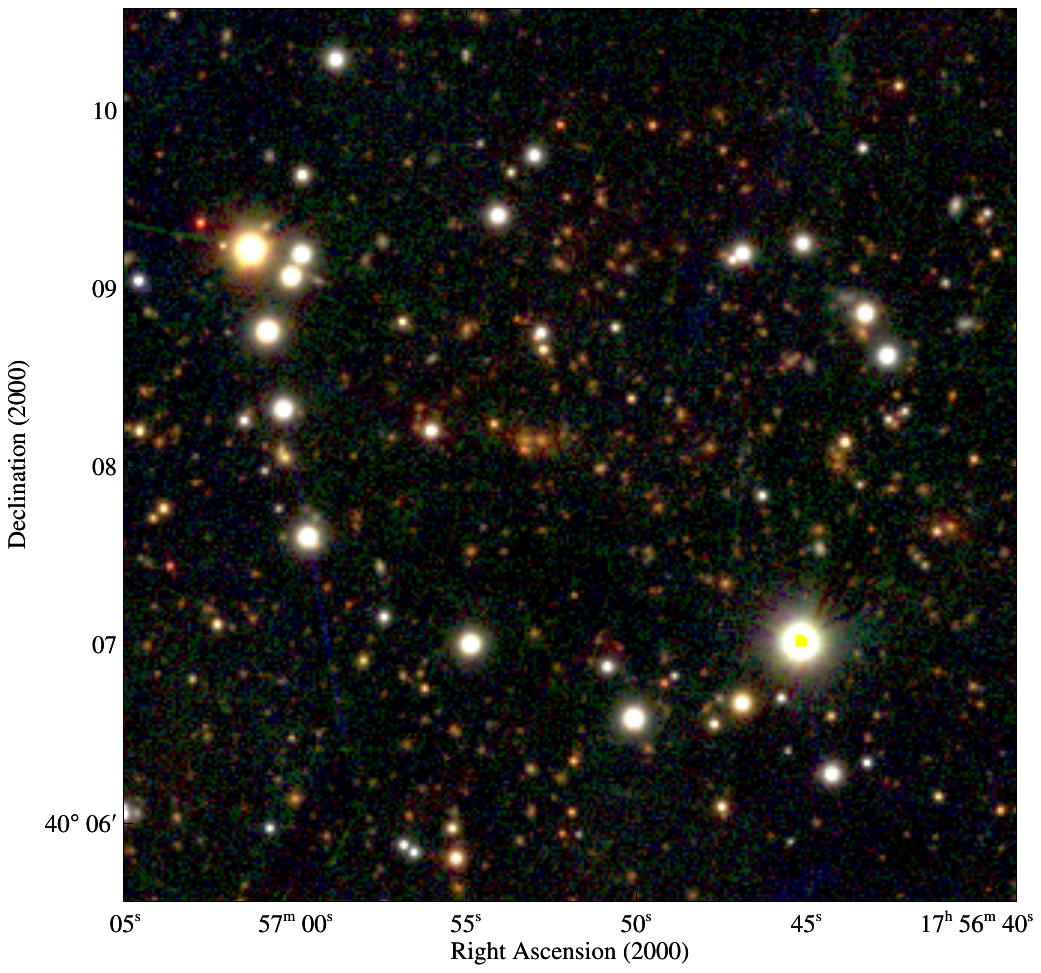}
\caption{PS1/3$\pi$ image ($gri$) of a $5\times 5$ arcmin$^{-2}$ region centred on the RASS source 1RXS\,J175651.8$+$400802. An overdensity of red galaxies near the image centre is immediately visible, as is a second apparent cluster core about 2\arcmin\ to the west. \label{fig:e1756-ps1}}
\end{figure}

Our strategy for the identification of distant galaxy clusters from these data sets is brute force: we select all X-ray sources listed in the  RASS BSC and FSC that fall within our study area and above the thresholds in X-ray flux and spectral hardness ratio, and then examine PS1 images in the $g$, $r$, $i$, and $z$ bands in a $5\times5$ arcmin$^2$ region around the X-ray source position.  Our goal, however, is not the identification of every X-ray source in the BSC and FSC; far from it. For each RASS source, we merely seek to answer the much simpler question: ``Based on the colour images provided by PS1, could the optical counterpart of this source conceivably be a massive, distant cluster?'' In essence this entails the visual identification of candidate clusters at $z\ga0.3$ as overdensities of faint, red galaxies, a simple approach that is motivated by the success of MACS, which demonstrated that distant RASS-detected clusters are readily discernible in relatively shallow optical images. As illustrated by Fig.~\ref{fig:e1756-ps1}, which shows eMACSJ1756.8 ($z=0.574$) as viewed by PS1 in the $3\pi$ survey (see also Section~\ref{sec:e1756}), this strategy remains successful for the population of X-ray fainter clusters probed by eMACS.

\begin{figure}
\hspace*{-3mm}\includegraphics[width=0.5\textwidth,clip,trim=15mm 60mm 10mm 40mm]{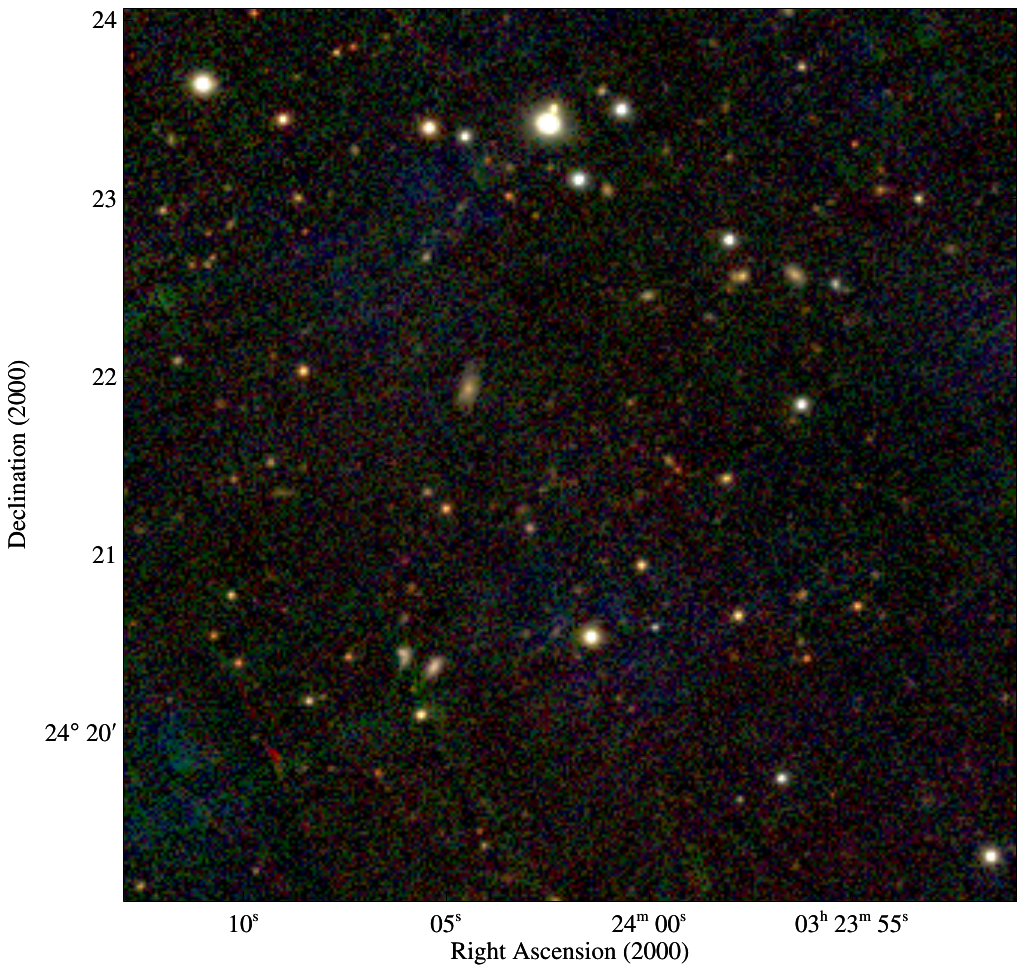}
\hspace*{-3mm}\includegraphics[width=0.5\textwidth,clip,trim=15mm 60mm 10mm 40mm]{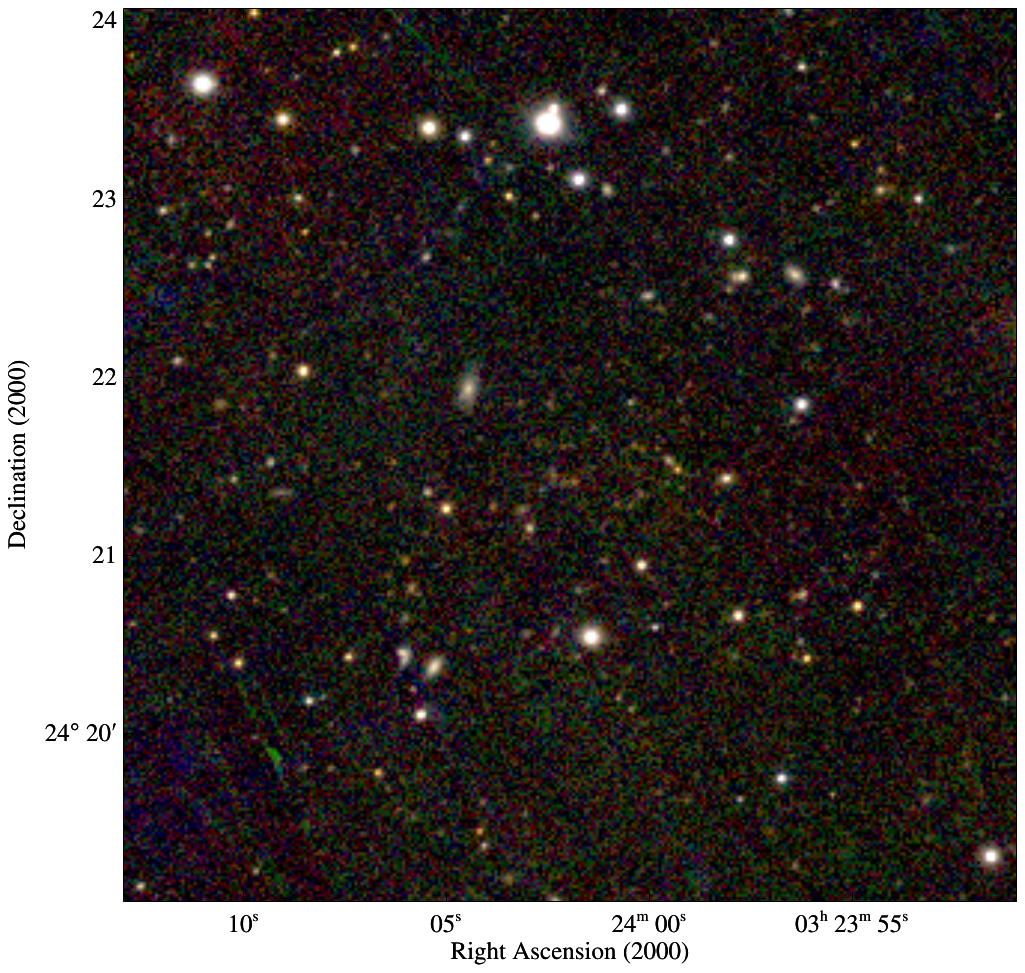}\mbox{}\\
\caption{PS1/3$\pi$ image of a $5\times 5$ arcmin$^{-2}$ region centred on the RASS source 1RXS\,J032401.8$+$242131. Requiring cluster candidates to show a pronounced galaxy overdensity in a  $gri$ image (top) would have led to the elimination of this source; even in the $riz$ image (bottom) the cluster galaxies are barely visible. Follow-up observations unambiguously established the presence of a massive galaxy cluster at $z=0.9$, the most distant cluster discovered to date in the RASS. \label{fig:e0324-ps1}}
\end{figure}

At the highest redshifts probed by eMACS, however, the Balmer/4000\AA\ break in cluster galaxies is getting close to moving out of the optical window. Care must thus be taken not to reject a RASS source based on the absence of an obvious galaxy overdensity in a $gri$ PS1 image; unless the presence of a bright star, a QSO candidate, or some other, known celestial object\footnote{In order to aid the identification of non-cluster sources, we also query NASA's Extragalactic Database (NED) for known celestial objects within 2 arcmin radius of the respective X-ray source. } suggest a non-cluster ID, a ``blank field'' in a PS1/3$\pi$ image is kept in our list of candidates and assigned high priority for follow-up observation. Although the associated follow-up observations resulted, in most cases, in the  rejection of such sources, several discoveries vindicate our conservative approach, as is illustrated by Fig.~\ref{fig:e0324-ps1} which shows an apparently blank field at the location of a RASS source. Owing to the lack of a plausible non-cluster ID we nonetheless included this source in our list of eMACS cluster candidates, a decision that led to the discovery of eMACSJ0324.0 ($z=0.90$) (see Section~\ref{sec:e0324}).

\section{eMACS one by one}
\label{sec:app-img}

We here provide an overview of the data available for all eMACS clusters mentioned in this paper.   We show only one example in this file and a complete version of the appendix is available on \url{http://www.astro.dur.ac.uk/~ace/eMACS.pdf}.

Specifically, we show colour images of the cluster core generated from {\it HST} data (where observations were performed in at least two passbands). Contours (in light grey) of the adaptively smoothed X-ray emission are overlaid for all eMACS clusters observed with {\it CXO} (see also Fig.~\ref{fig:cxocont}), while the critical lines for strong gravitational lensing of an object at $z=2$ are shown in yellow for systems for which a lens model is available (see Table~\ref{tab:re}). We mark multiple-image systems that are used to constrain these lens models and highlight strong-lensing features and (occasionally) cluster galaxies of special interest in insets placed around the main image.

We also show the redshift histogram in the vicinity of each cluster's systemic redshift and the velocity dispersion derived from it. The spatial distribution of all galaxies with spectroscopic redshifts measured in each field is also presented, where larger symbols mark cluster members (defined as featuring redshifts that fall within $3\sigma$ of the systemic cluster redshift in the radial-velocity histogram); in these plots, the area covered by the shown {\it HST} data is outlined by an orange polygon. In both the radial-velocity histograms and the maps showing the location of galaxies with spectroscopic redshifts we highlight in red the BCG (heavy line) and the second-brightest cluster galaxy (light line). 

Finally, we list all galaxies with spectroscopic redshifts (whether they are cluster members or not) and mark the BCG (B), any X-ray point sources detected in \textit{CXO} follow-up observations (X), and candidates for ram-pressure stripping (R).  
For the remainder of the sample, i.e., for all eMACS clusters that are listed in Table~\ref{tab:sample} but are not highlighted elsewhere in this paper, the same data summary is provided in an online-only extension of this appendix.

\begin{figure*}
    \centering
    \includegraphics[width=0.95\textwidth]{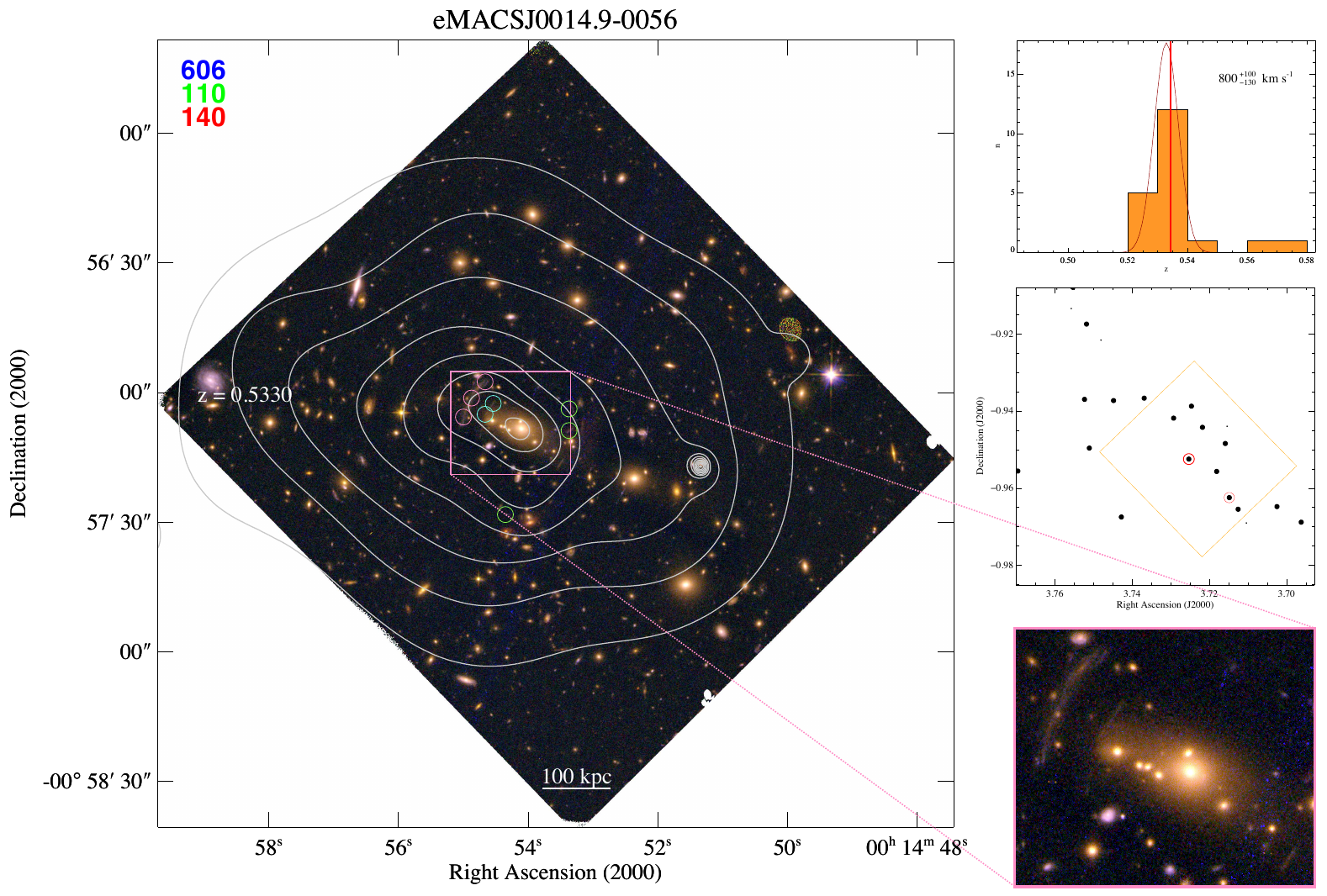}
    \caption{eMACSJ0014.9$-$0056 is an exceptionally X-ray luminous cluster (see Table~\ref{tab:cxo}) with a largely relaxed X-ray morphology. The modest number of radial velocities measured for cluster members to date precludes a more robust determination of the cluster velocity dispersion and an assessment of line-of-sight substructure. While strong-lensing features are clearly visible in the cluster core, their faintness makes follow-up spectroscopy challenging; as a result, we currently have no lens model for this system. }  
    \label{fig:e0014}
Coordinates (J2000) and spectroscopic redshifts of galaxies in the field of eMACSJ0014.9-0056.\\[2mm]
\centering
\begin{tabular}{c@{\hspace{0mm}}ccc}
 & R.A. (deg) & Dec (deg) & $z$  \\[1mm] \hline\\[-5mm]
 & & &  \\
  &   3.6932 & $-$0.9852 & 0.6368 \\
  &   3.6962 & $-$0.9689 & 0.5351 \\
  &   3.7025 & $-$0.9648 & 0.5283 \\
  &   3.7104 & $-$0.9691 & 0.4674 \\
  &   3.7125 & $-$0.9655 & 0.5276 \\
  &   3.7148 & $-$0.9625 & 0.5340 \\
  &   3.7154 & $-$0.9440 & 0.2769 \\
  &   3.7159 & $-$0.9484 & 0.5278 \\
  &   3.7181 & $-$0.9557 & 0.5349 \\
  &   3.7218 & $-$0.9442 & 0.5358 \\
  &   3.7246 & $-$0.9387 & 0.5321 \\
B &   3.7253 & $-$0.9525 & 0.5345 \\
  &   3.7292 & $-$0.9418 & 0.5372 \\
  &   3.7369 & $-$0.9367 & 0.5404 \\
  &   3.7428 & $-$0.9675 & 0.5293 \\
  &   3.7448 & $-$0.9373 & 0.5320 \\
  &   3.7481 & $-$0.9216 & 0.5707 \\
  &   3.7511 & $-$0.9496 & 0.5373 \\
  &   3.7518 & $-$0.9174 & 0.5278 \\
  &   3.7523 & $-$0.9370 & 0.5302 \\
  &   3.7553 & $-$0.9079 & 0.5330 \\
  &   3.7557 & $-$0.9134 & 0.5680 \\
  &   3.7595 & $-$0.9084 & 0.3697 \\
  &   3.7696 & $-$0.9556 & 0.5363 \\
 & & &  \\
\end{tabular}

\end{figure*}

\bsp
\label{lastpage}
\end{document}